\documentclass{article}

\usepackage{arxiv}

\usepackage[utf8]{inputenc} 
\usepackage[T1]{fontenc}    
\usepackage{url}            
\usepackage{hyperref}       
\usepackage{xcolor}
\hypersetup{
    colorlinks,
    linkcolor={red!50!black},
    citecolor={blue!50!black},
    urlcolor={blue!80!black}
}
\usepackage{booktabs}       
\usepackage{amsfonts}       
\usepackage{amsmath}        
\usepackage{nicefrac}       
\usepackage{microtype}      
\usepackage{array}          
\usepackage{geometry}       
\usepackage{subcaption}
\usepackage{caption} 
\usepackage{lipsum}
\usepackage{graphicx}
\usepackage{placeins}
\graphicspath{ {./figs/} }

\usepackage[natbib=true,backend=biber,sorting=nyt,style=apa]{biblatex}

\begin{document}

\begin{flushleft}
{\Large
\textbf\newline{Prediction of Maximum Temperature record in Spain 1960-2023 by the means of ERA5 atmospheric geopotentials}
}
\newline
\\
Elsa Barrio 1\textsuperscript{1,*},
Jesús Abaurrea \textsuperscript{1},
Jesús Asín \textsuperscript{1},
Jorge Castillo-Mateo \textsuperscript{1},
Ana Carmen Cebrián \textsuperscript{1},
Zeus Gracia \textsuperscript{1},
\\
\bigskip
\bf{1} Department of Statistical Methods, University of Zaragoza.
\\
\bigskip
* e.barrio@unizar.es

\end{flushleft}

\section*{Graphical Abstract}
\label{sec:gabs}

\begin{figure}[htb]
\includegraphics{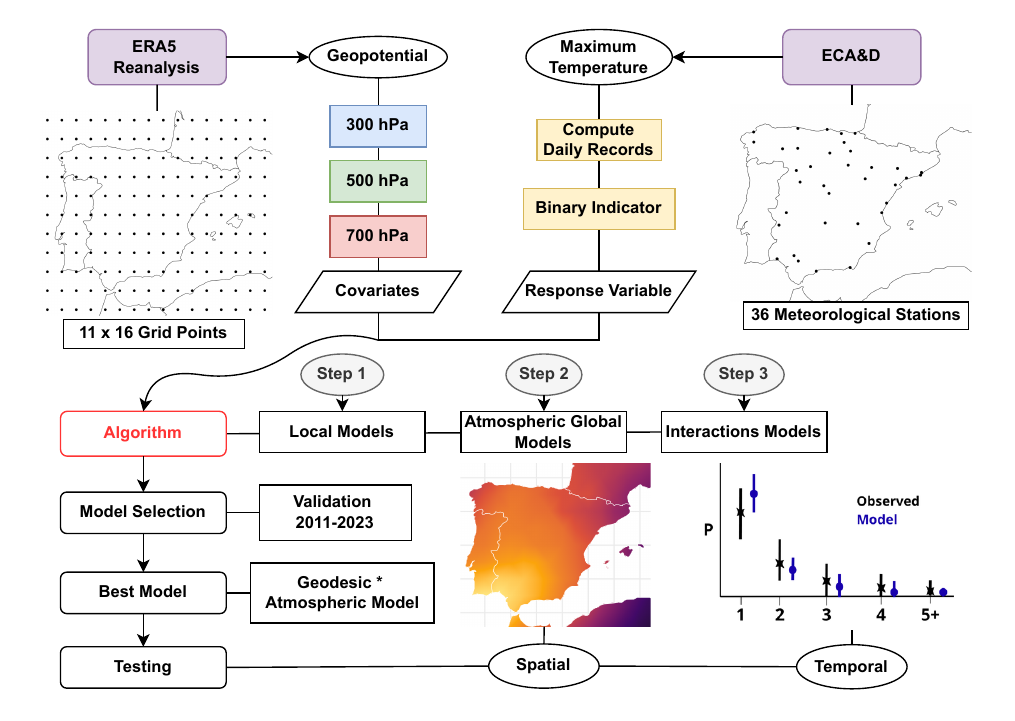}
\end{figure}

\newpage

\begin{abstract}

    The increasing frequency of extreme temperature events, such as daily maximum temperature ($T_x$) records, underscores the need for robust tools to understand their drivers and predict their occurrence. Although previous studies have identified increasing and non-stationary trends in $T_x$ records across the Iberian Peninsula \citep{cebrian2022record, castillo2023statistical}, particularly during summer, the literature directly exploring their connection with upper-level atmospheric covariates remains limited. This work develops and applies an innovative methodological framework to model the occurrence of $T_x$ records and their relationship with geopotential height fields. We used daily $T_x$ data from 36 Spanish stations (1960--2023) provided by ECA\&D and geopotential height data at 300, 500, and 700 hPa from ERA5. Exploratory analysis revealed a non-stationary trend in records, a higher frequency in the interior of the peninsula, and decreasing spatial co-occurrence with distance. 
    We designed a hierarchical spatio-temporal logistic regression algorithm prioritizing interpretability and high-dimensionality reduction. The approach involves: (1) fitting local models per station; (2) applying a spatial consensus filter based on statistical significance to reduce the initial 1620 covariates to 17 in a base model (M1); and (3) a controlled incorporation of interaction terms. Among the tested models, a global model (M2) that enhances M1 with geodetic interactions was selected for its optimal balance between predictive performance ($AUC$) and complexity.
    Model M2 demonstrates high predictive accuracy at interior stations and good performance at coastal stations. It also adequately reproduces key observed properties, including the persistence of record streaks and patterns of spatial co-occurrence. This study provides a novel tool for predicting upcoming record events with high accuracy while maintaining a concise and interpretable structure.
    
\end{abstract}


\section{Introduction}
\label{sec1}


There is an ongoing increase in the rate of extreme climatic events due to global warming. In particular, there is a rising frequency of temperature extremes such as heatwaves and record-breaking events \citep{zhang2025, paredes2023understanding, om2022,saddique2020}. These events have numerous consequences for ecosystems \citep{breshears2021underappreciated}, agriculture \citep{buntgen2024recent}, and public health systems \citep{roye2021effects}, among others. Europe, specifically, has experienced one of the most pronounced warming trends in recent decades compared to other continents \citep{di2020contribution,twardosz2021warming}, with Southern Europe and the Mediterranean sea considered climate \textit{hot-spots}. In these regions, summer temperatures have triggered devastating events such as wildfires and ``medicanes'' \citep{gonzalez2019potential,dupuy2020climate}.

Although extremes are commonly defined through maxima over a time period or exceedances over a high threshold, record-breaking temperatures, particularly calendar-day records, i.e., the highest value observed on a specific calendar day at a given station, offer a valuable alternative approach for analyzing extreme behavior. Many meteorological services use these events as climate indices. For example, the U.S. National Oceanic and Atmospheric Administration (NOAA) and its National Centers for Environmental Information (NCEI) provide daily updates on record events across various periods: see \url{https://www.ncdc.noaa.gov/cdo-web/datatools/records}. 
Similarly, the Copernicus Climate Change Service (C3S), via the European Centre for Medium-Range Weather Forecasts on behalf of the European Commission, includes these records in its annual European State of the Climate (ESOTC) report; specifically, the 2019 ESOTC report \citep{ESOTC2019}  analyzed the frequency of calendar-day records and compared it with theoretical expectations under a stationary climate. In Spain, the annual report of the national meteorological service \citep{aemet2023} includes a section (1.1.3)  dedicated to  the evolution of calendar-day records, distinct from Section 1.1.4, which focuses on heatwaves defined by local thresholds. Moreover, the frequency of record-breaking events is one of the three indicators used to summarize temperature trends in the summary for policymakers of the report.

From a methodological perspective, a major advantage of using record-breaking events is that, in a stationary climate (i.e., no global warming), their occurrence follows well-established, distribution-free probabilistic properties. This makes them a powerful metric for detecting deviations from stationarity and assessing the impacts of global warming \citep{cebrian2022record}.
Recent studies show that the frequency of calendar-day records observed across most of Peninsular Spain is inconsistent with a stationary climate. \textcite{castillo2023statistical} found that during the decade 2012–2021, the number of record-breaking temperature events nearly doubled compared to what would be expected under stationary conditions, with the increase being particularly notable during summer. \textcite{castillo2024spatio} further revealed that deviations from stationarity are neither temporally nor spatially homogeneous. Their findings include upward trends over time, seasonal variation, persistence effects, and spatial influences such as proximity to the coast. Their analysis did not consider potential influences from geopotential covariates.

The relationship between extreme temperatures and  different atmospheric covariates has been studied in various regions. \textcite{capozzi2025} identified large-scale circulation patterns associated to the occurrence of summer heat waves in the Apennines. \textcite{huang2025} found that abnormal North Atlantic tripole sea surface temperature (SST) anomalies and the Indo-Pacific zonal SST gradient (IPG) were potential origins of the record-breaking  extreme temperatures over northern China in October 2023. \textcite{qian2024rapid} investigated the June 2023 record-breaking heatwave in North China and found that insolation and adiabatic heating, linked to anomalous atmospheric circulation (a high-pressure ridge at 500 hPa), played a key role.  \textcite{ryu2023teleconnections} explored the influence of large-scale variability modes, such as the Pacific North America (PNA), North Atlantic Oscillation (NAO), and Pacific Decadal Oscillation (PDO), on extreme heatwaves in the U.S., using geopotential height and wind variables from the ERA5 database. These teleconnections significantly affected surface air temperatures during their negative phases, due to anomalous anticyclonic circulation patterns. In Spain, \textcite{garcia2015attributing} examined the impact of sea-level pressure, temperature, and geopotential variables at high atmospheric levels on extreme temperature events. In a similar  research line, \textcite{castillo2023statistical} analyzed record-breaking events at different pressure levels over the Iberian Peninsula but found no one-to-one correspondence between surface and upper-atmosphere records. However, to our knowledge, no previous studies have specifically assessed the potential relationship between record-breaking surface temperatures and geopotential covariates. 

Understanding and quantifying the influence of geopotential covariates on the occurrence of record-breaking temperatures is essential for conducting climate attribution studies and for developing statistical downscaling methods to project the future frequency of record-breaking temperatures under global warming scenarios. One might argue that such projections could be derived from daily temperature projections, but these often fail to accurately capture the behavior of temperature extremes and, in particular, record events, the most extreme observations. Therefore, generating reliable projections for record-breaking temperatures requires specific downscaling methods.

In this context, the objective of this work is to develop a spatio-temporal  model that can be used as a statistical downscaling tool for the occurrence of  record-breaking daily temperatures. Specifically, we aim to develop a logistic model capable of characterizing the occurrence of calendar-day records in daily maximum temperature ($T_x$), using geopotential covariates from the ERA5 reanalysis over a gridded study area. We will provide the tools to select the relevant covariates, fit the model, assess its performance, and draw conclusions  from it. We illustrate the procedure by analyzing a dataset of 36 daily temperature series across Peninsular Spain for the summer seasons from 1960 to 2023.

The structure of this work is as follows: Section \ref{sec:data} describes the dataset, including the daily $T_x$ series over Peninsular Spain and the geopotential covariates (700, 500, and 300 hPa) over the region. Section \ref{sec:eda} shows an exploratory data analysis of the current dataset. Section \ref{sec:methods} details the spatio-temporal models considered, including the model selection procedure. Section \ref{sec:results} summarizes the findings from the selected model, and Section \ref{sec:discussion} provides conclusions and outlines future research directions.

\section{Study area and dataset}
\label{sec:data}

The Spanish Iberian peninsula is a geographically diverse region. Its complex orography is defined by a central plateau, with the northern part averaging around 1000 meters above sea level and the southern part around 500 meters. The Ebro, Duero and Tajo river valleys, are situated within this plateau and the Guadalquivir valley lies in the south. The peninsula also features five main mountain ranges, with peaks exceeding 2000 meters. The Pyrenees form the northeastern border with France, the Cantabrian range lie in the northwest near the Atlantic ocean, the Iberian system occupies the northeast with a northwest–southeast orientation, the Central system divides the Central plateau, and the Baetic range is located in the southeast.

Despite its relatively small area, spanning latitudes 36 to 43.8$^\circ$ N (a north–south distance of approximately 1000 km) and longitudes 9.3$^\circ$ W to 3.$^\circ$ E, the Iberian peninsula exhibits significant climatic variability over short distances due to its complex topography. The main driver of maximum temperature  variation is seasonality, particularly the changes in potential daily solar radiation. During winter, atmospheric circulation over the Atlantic induces a succession of cyclonic and anticyclonic systems over the center and western parts of the peninsula. In contrast, summer is characterized by frequent anticyclonic conditions, leading to reduced cloud cover and increased solar radiation.

\begin{figure}
	\centering
	\includegraphics[width=0.7\textwidth]{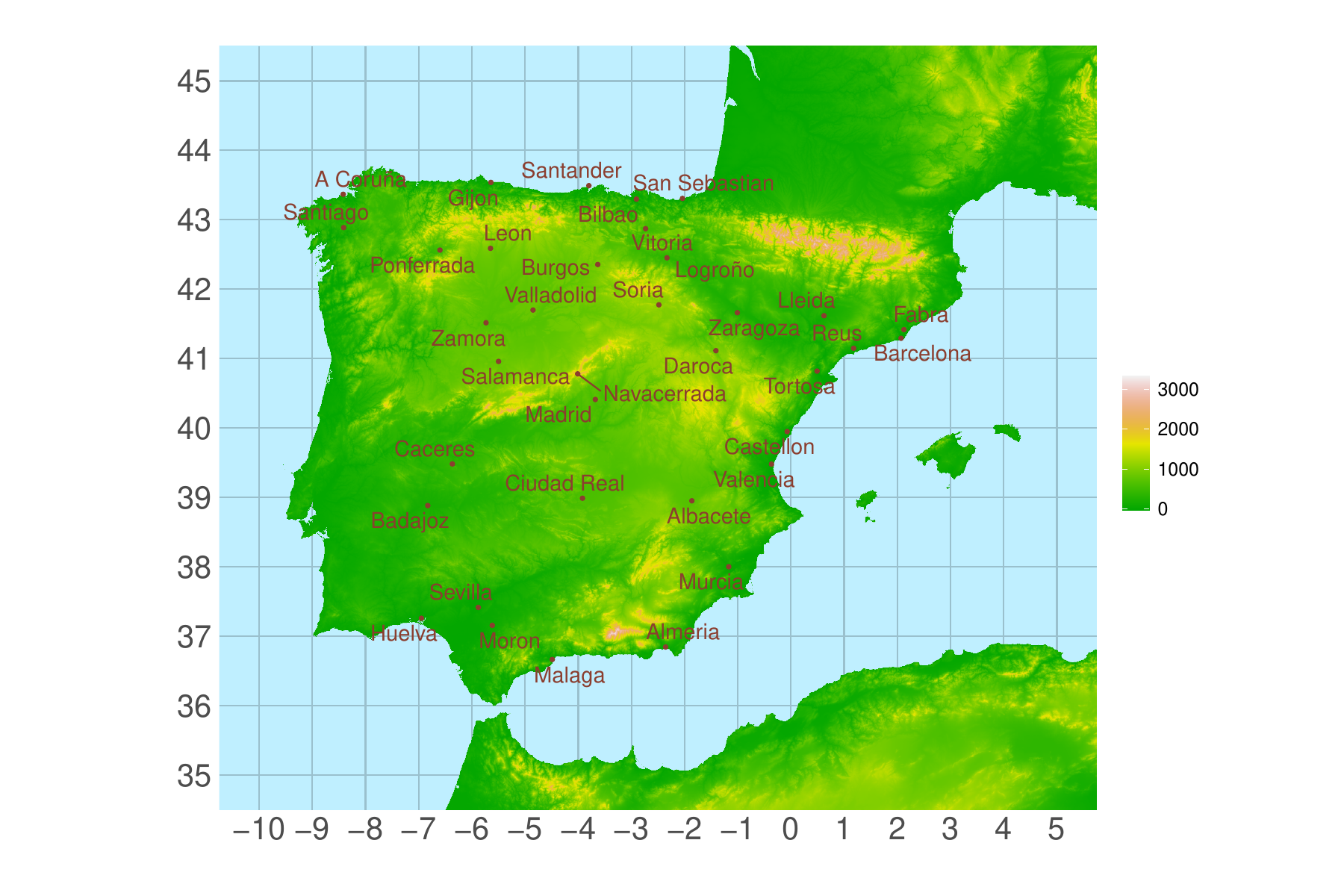}
	\caption{Map with  the 36  temperature stations  and the grid were geopotential variables are extracted.}
	\label{fig:stations-gridpoints}
\end{figure}

The temperature dataset, downloaded from the European Climate Assessment \& Dataset (ECA\&D;   \cite{klok2009updated}), contains 36 daily maximum temperature series $T_x$ across peninsular Spain for the period 1960--2023, see Figure \ref{fig:stations-gridpoints}. The selection criteria for these series included a percentage of missing values lower than 0.5\%, and ensuring a good representation of the climatic heterogeneity in the study area \citep{castillo2023statistical}. 
In this work, we  will focus on the calendar-day series derived from the original daily temperature series during the summer months, from June to August (JJA). This results in 92 time series (one per day) across years for each location. Notably, the dataset includes four centennial observing stations recognized by the World Meteorological Organization (WMO): Barcelona-Fabra, Madrid-Retiro, Daroca, and Tortosa (\url{https://wmo.int/activities/centennial-observing-stations}).

Given the 92 series of maximum daily temperature at each location, denoted by $Y_{t,\ell}(s_i)$ for $t \in \{1, \ldots, T\}$ and day $\ell \in \{1, \ldots, L\}$ at station $s_i$, the occurrence of calendar-day records is identified by the binary indicators
$$
I_{t,\ell}(s_i) = 
\begin{cases}
	1 & \text{if } Y_{t,\ell}(s_i) > \max \left\{ Y_{1,\ell}(s_i), \dots, Y_{t-1,\ell}(s_i) \right\} \\
	0 & \text{otherwise.}
\end{cases}
$$
 \noindent  For $t=1$, all  the observations are a record. 

The ERA5 series are downloaded from the \textit{ERA5 hourly data on pressure levels from 1940 to present} collection \citep{hersbach2023era5} at the Copernicus Climate Data Store (\url{https://cds.climate.copernicus.eu}). We consider daily geopotential height series at  time 12:00 (denoted as $G$), measured in m$^2$/s$^2$, at pressure levels of 700 hPa, 500 hPa, and 300 hPa for the period 1960--2023. The data were extracted at grid points  within the spatial domain [45$^\circ$N, 10$^\circ$W, 35$^\circ$N, 5$^\circ$E], covering the Iberian Peninsula, using a resolution of 1$^\circ \times$1$^\circ$, resulting in a total of $11 \times 16$ grid points,
see Figure \ref{fig:stations-gridpoints}. We denote the geopotential series at each grid point as $Gp_{[i,j]}$, where $p$ refers to the pressure level and the subscripts $i$ and $j$ represent the latitude and longitude coordinates of the grid point; for example, $G700_{41\text{N},1\text{W}}$ refers to the geopotential height at 700 hPa at the grid point located at 41$^\circ$N, 1$^\circ$W.


\section{Exploratory data analysis}
\label{sec:eda}

The main goal of this exploratory data analysis (EDA) is to characterize the occurrence of record-breaking temperatures in terms of their temporal evolution, persistence, and spatial dependence. Additionally, we aim to investigate the predictive effect of three geopotential height levels for the occurrence of daily maximum temperature records ($T_x$). Specifically, we will examine similarities in the underlying trends, patterns in record occurrences, and the distribution of geopotential heights conditional on whether a temperature record has occurred.

\subsection{Exploring record-breaking events in daily temperature}

\subsubsection{Non-stationary behavior}

\begin{figure}[htb]
 	\centering
 		\includegraphics[width=.4\textwidth]{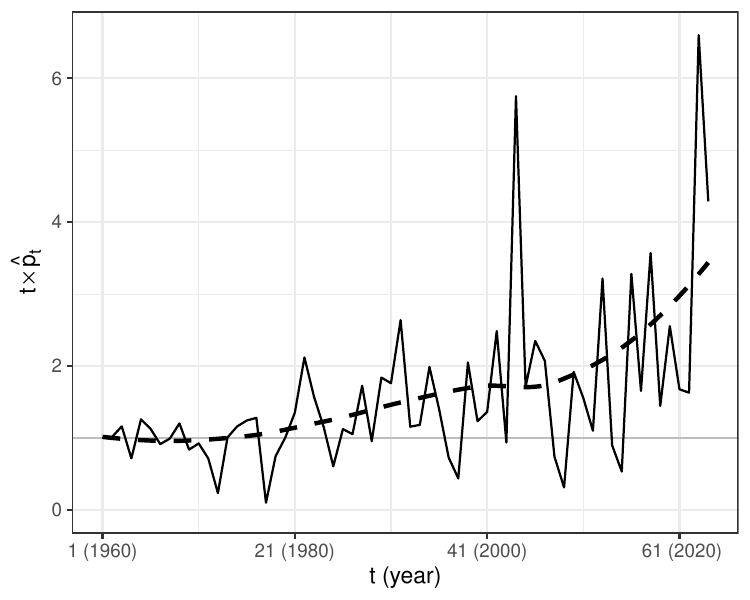} \\
        \includegraphics[width=.4\textwidth]{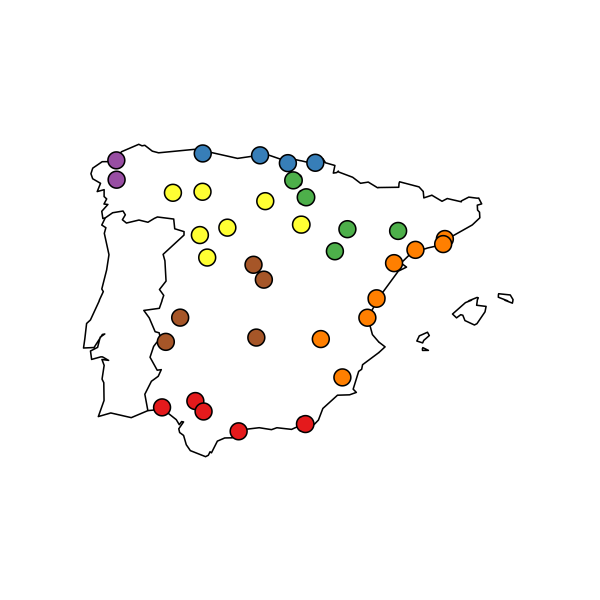}
	\includegraphics[width=.45\textwidth]{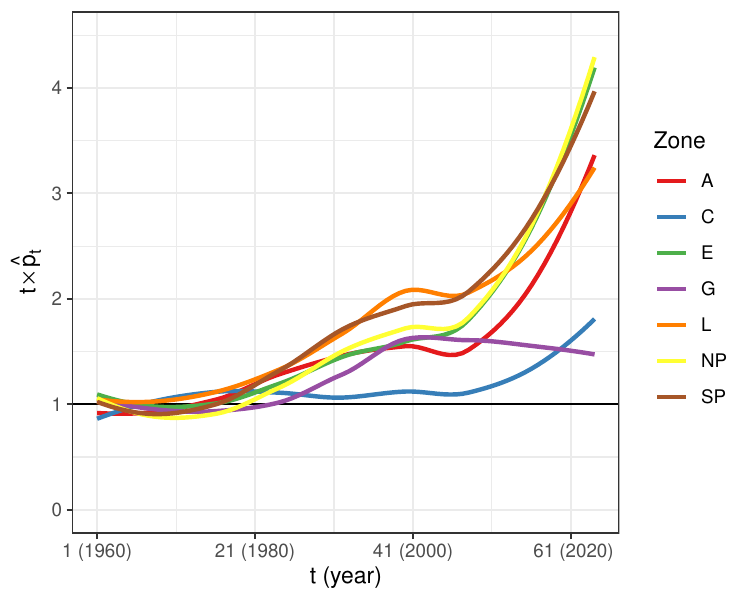}
 	\caption{Top: Evolution of $t \times \hat{p}_{t}$ over time, with LOESS smoother (dashed line). Value 1 represents the expected value under the assumption of stationarity. Bottom:  Map showing the locations colored by geographical area. Right: LOESS curves of $t \times \hat{p}_{t}$ by geographical area: Andalusia (A), Cantabrian Coast (C), Ebro Valley (E), Galicia (G), Levante (L), North Plateau (NP) and South Plateau (SP).}
 	\label{fig:ix_trends}
 \end{figure}

One of the advantages of using record events is that their occurrence in a stationary series (i.e., a sequence of independent and identically distributed random variables) follows a known, distribution-free probabilistic behavior. Specifically, in a stationary series, the probability of a record at time $t$ is given by $p_t = 1/t$. To assess whether this term adequately captures the temporal evolution of the probability of record, or to identify deviations from stationary behavior, Figure~\ref{fig:ix_trends} plots $t \times \hat{p}_t$ against $t$. Note that under stationarity, the expected value of this product is 1. The empirical estimate $\hat{p}_t$ is computed by averaging across space and days within each year, the indicator variables of records:
$$\hat{p}_{t} = \frac{1}{36} \frac{1}{92} \sum^{36}_{i=1} \sum^{92}_{\ell=1}  I_{t\ell}(s_i).$$
A LOESS smoother is included in the plot to highlight the underlying trend and deviations from stationarity. A clear non-stationary pattern emerges, with an increasing trend deviating from 1 starting around 1980.
 
To identify potential spatial differences in the trend, the previous approach is applied to specific geographical areas. Figure~\ref{fig:ix_trends} shows the selected regions and the corresponding LOESS curves of $t \times \hat{p}_{t}$. The results indicate that the northern coastal regions (Galicia and Cantabrian coast) exhibit the smallest increase, whereas inland areas have shown the largest in recent years.

This graphical analysis can be complemented using specific tests to study deviations from stationarity in the occurrence of records \citep{cebrian2022record, castillo2023recordtest}. \textcite{castillo2023statistical}, found that, during the summer months, the tests revealed significant non-stationary behavior in the occurrence of  records in the upper tail at most of the 36 stations analyzed in this study, with the exception of those located along the Galician and Cantabrian coasts.


\subsection{Persistence of record-breaking events}

\begin{figure}[htb]
    \centering
    \includegraphics[width=.35\textwidth]{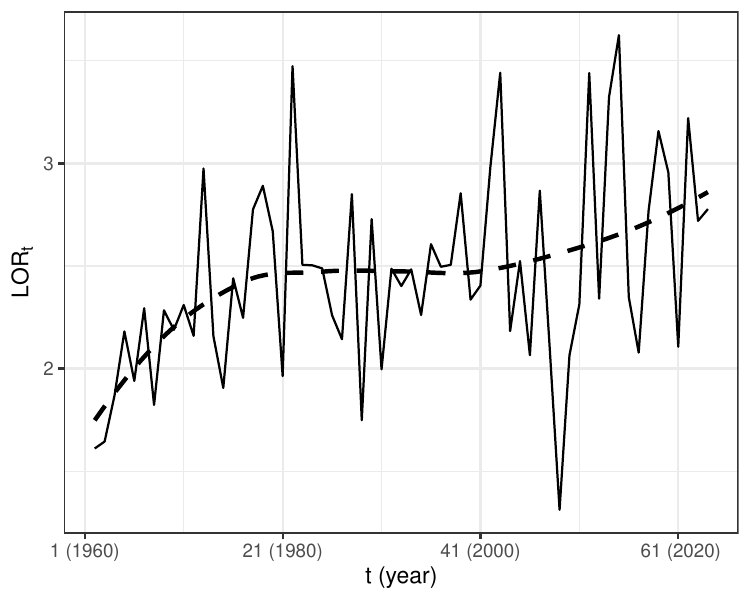}
    \caption{Log-odds ratio, $LOR_t$,  comparing the probability of  record given that the previous day was a record, to the probability given that it was not. LOESS smooth in dashed line.}
    \label{fig:lor}
\end{figure}

 Given the strong serial dependence of daily temperature, some persistence in the occurrence of records is to be expected, that is, the probability of a record on a given day may depend on whether the previous day was also a record. To explore this, we estimate the log-odds ratio which compares the probability of  record given that the previous day was a record,
$p_{t\ell}^{11} = P(I_{t\ell} = 1 \mid I_{t, \ell-1} = 1)$
to the probability given that it was not, $p_{t\ell}^{10} = P(I_{t\ell} = 1 \mid I_{t, \ell-1} = 0)$
using the expression:
$$
\log \left( \frac{p_{t\ell}^{11}/(1 - p_{t\ell}^{11})}{p_{t\ell}^{10}/(1 - p_{t\ell}^{10})} \right).
$$
The empirical log-odds ratio is computed for each year \( t \) and shown in Figure~\ref{fig:lor}. Values close to 0 indicate independence, while positive values suggest persistence. The estimates are clearly different from zero, increasing from nearly two at the beginning of the period to almost three by the end, indicating a strong and growing temporal persistence in record occurrences over the years.

\subsection{Co-occurrence of record-breaking events}

\begin{figure}[tb]
    \centering
    \includegraphics[width=0.64\textwidth]{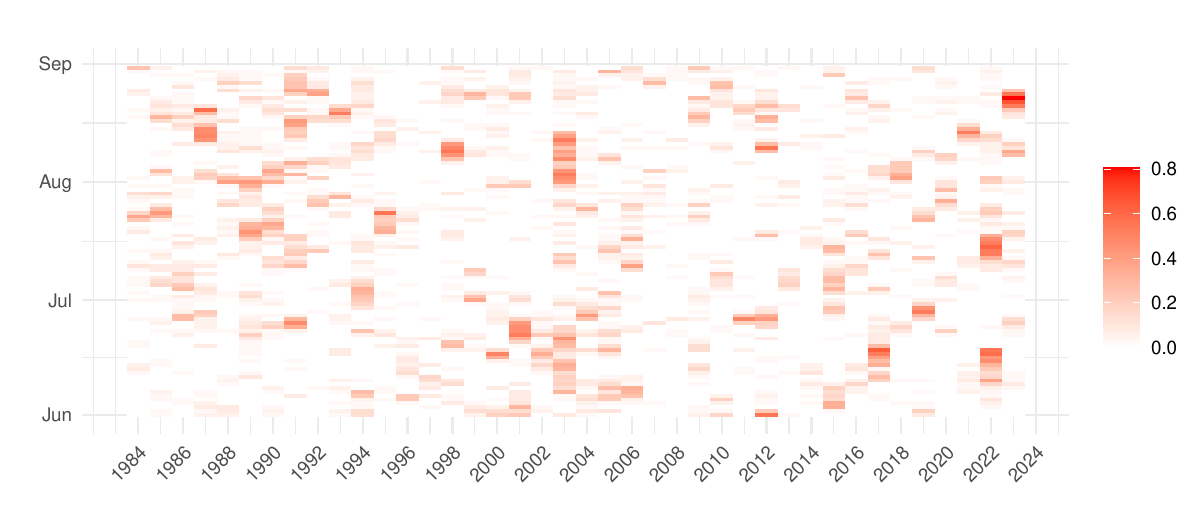}
        \includegraphics[width=0.35\textwidth]{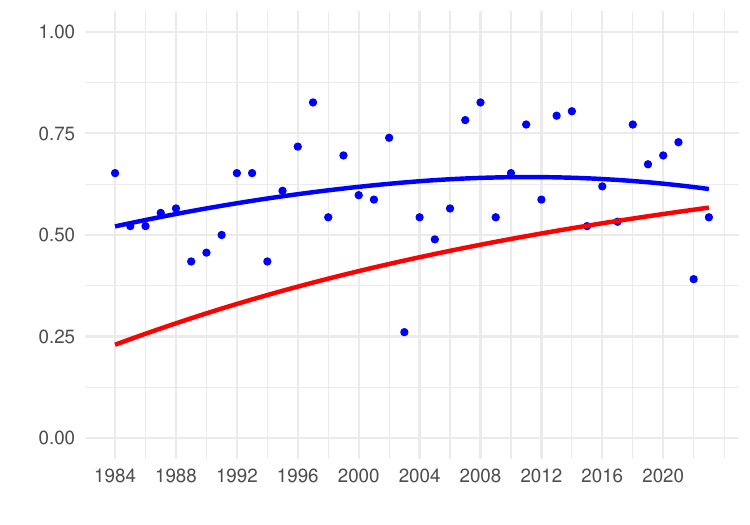}
    \caption{Left: Proportion of days with a temperature record for each day in JJA (rows) across the years in the 1984–2023 period (columns). Right: proportion of days in each year with $\sum_{i=1}^S I_{t\ell}(s_i) = 0$ and LOESS smoother (blue). Expected value under stationarity and independent stations (red) .}
    \label{fig:record.heatmap}
\end{figure}

The co-occurrence of record-breaking temperatures occurs when two or more locations experience a record temperatures on the same day  $\ell$ and  year $t$. To quantify the co-occurrence, we consider, for each day $\ell$,  the proportion of the 36 observatories having a record on the same day, 
$$PR_{t \ell}= \frac{1}{36} \sum^{36}_{i=1} I_{t \ell}(s_i).$$
Figure~\ref{fig:record.heatmap} shows these proportions  for each day in JJA (June, July, and August) during the period 1984--2023. 
The early years, during which the probability of record is relatively high, are omitted to highlight the increasing frequency of record-breaking events. The highest values, approaching 0.8, are observed in recent years, even though the probability of a record in a single series is expected to decrease over time; this suggests that co-occurrence is increasing over time.

To explore  this possibility, we will study the probability that no record occurs  at a given day, $t \ \ell$, in the $n=36$ stations, $P\left(\sum_{i=1}^n I_{t \ell}(s_i) = 0\right)$. Assuming a stationary climate and independent stations, the probability is,
$$
P\left(\sum_{i=1}^n I_{t \ell}(s_i) = 0\right) = \left(1 - \frac{1}{t}\right)^n.
$$
for any day. Of course, the 36 available time series are not independent even in a stationary climate due to spatial proximity, but this theoretical probability serves as a useful benchmark to evaluate changes in the frequency of simultaneous record events over time. Figure~\ref{fig:record.heatmap} shows the empirical estimators of these probabilities, calculated as the proportion of days $\ell$ in a year for which $\sum_{i=1}^n I_{t\ell}(s_i) = 0$, along with the benchmark under a stationary climate and independent stations.  Note that while the reference probability approaches 1 as $t \to \infty$, the estimated probability appears to level off, suggesting an increase in the co-occurrence of record events.

\textit{Jaccard index.} Another useful way to quantify co-occurrence between two series of records is by using the Jaccard index, a similarity coefficient for binary variables. The Jaccard index between two binary series  over time indexes $t$ and $\ell$, $I_{t \ell}(s_i)$ and $I_{t \ell}(s_k)$, is calculated as
$$
J\bigl(I_{t \ell}(s_i), I_{t \ell}(s_k)\bigr)
  = \frac{\sum_{t, \ell} I_{t \ell}(s_i)\,I_{t \ell}(s_k)}
         {\sum_{t,  \ell} \bigl(I_{t \ell}(s_i) + I_{t \ell}(s_k) - I_{t \ell}(s_i)\,I_{t \ell}(s_k)\bigr)}.
$$
It measures the similarity between the two binary series by comparing the number of observations during which both stations simultaneously register an event (a record, in this case) to the number of observations during which at least one of the stations registers an event. This index quantifies similarity while avoiding the influence of zero–zero coincidences, which do not represent meaningful similarity. The value ranges from $0$ (no simultaneous events) to $1$ (identical event patterns).

\begin{figure}[htb]
   \centering
 \includegraphics[width=9cm]{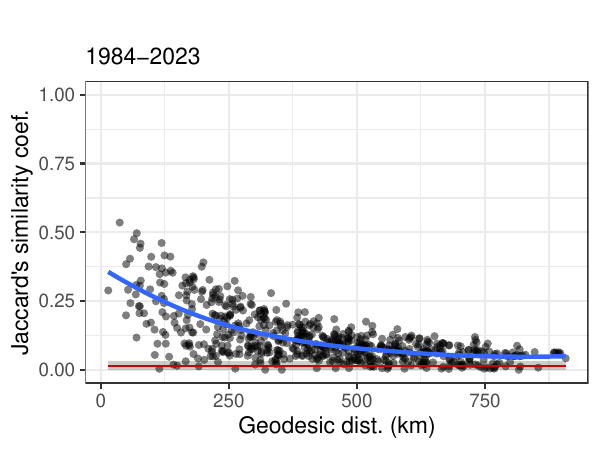}    \caption{
The Jaccard index between  record series of station pairs for the period 1984–2023  against the distance between stations. A LOESS smoother (in blue) illustrates the overall trend. The expected value and 95\% confidence intervals of the Jaccard index for two independent series under a stationary climate (in red and gray) are shown as a reference.}
    \label{fig:EDA:correlation:geodistance}
\end{figure}

Figure~\ref{fig:EDA:correlation:geodistance} displays the Jaccard indices for series of records from all pairs of stations during the period 1984–2023, plotted against the distance (in km) between the stations. The expected value and 95\% confidence intervals for the Jaccard index between two independent series under a stationary climate were estimated by simulation and are shown in the previous graph, with a mean estimate of 0.01. The plot reveals a strong spatial dependence that diminishes with distance: values are higher than 0.25 on average for distances less than around 100 km, and decrease as the distance increases. For the most distant locations, the binary series can generally be assumed to be independent. However, even among the farthest station pairs, most of the estimated Jaccard indices fall outside the confidence interval obtained under the assumption of records from independent and identically distributed  series, providing evidence of deviations from a stationary climate.

\subsection{Relationship between geopotential variables and daily maximum temperature}

\subsubsection{Comparing time trend}
 	
\begin{figure}[htb]
	\centering
	\includegraphics[width=.65\textwidth]{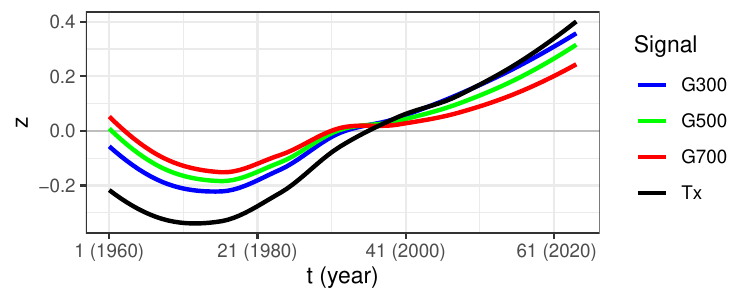}
	\caption{LOESS curves of the spatial averages of the standardized $T_x$ and geopotential variable series. }
	\label{fig:tx_trends}
\end{figure}

To compare the time evolution of the $T_x$ series and the three geopotential variables across the study region, we examine the average of the 36 $T_x$ series and the average of the geopotential variable series at the $11 \times 16$ grid points. These averages are computed from standardized signals, where standardization is based on the mean and standard deviation of each series during the 1981–2010 reference period. Figure \ref{fig:tx_trends} displays the LOESS trends of the resulting averages. A similar pattern, marked by an upward trend beginning just before the 1980s, is observed across all four variables, with $T_x$ displaying the steepest slope. Among the geopotential variables, G300 shows the most similar evolution to $T_x$, and the trend appears to weaken with decreasing height. 

\subsubsection{Comparing evolution in the occurrence of records}

\begin{figure}[htb]
    \centering
    \includegraphics[width=0.45\textwidth]{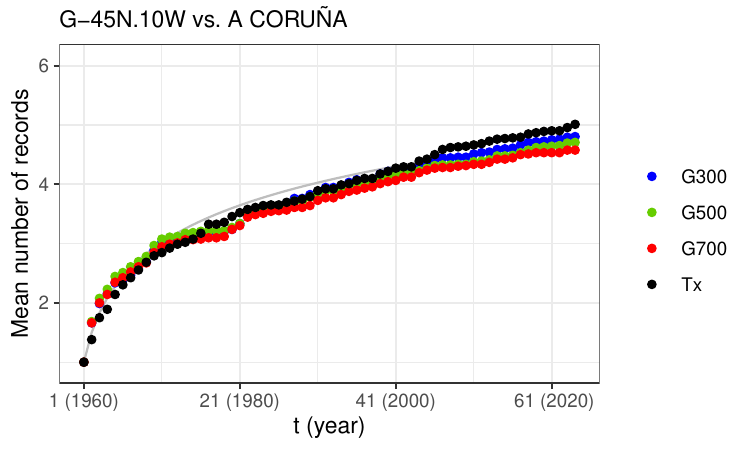}
    \includegraphics[width=0.45\textwidth]{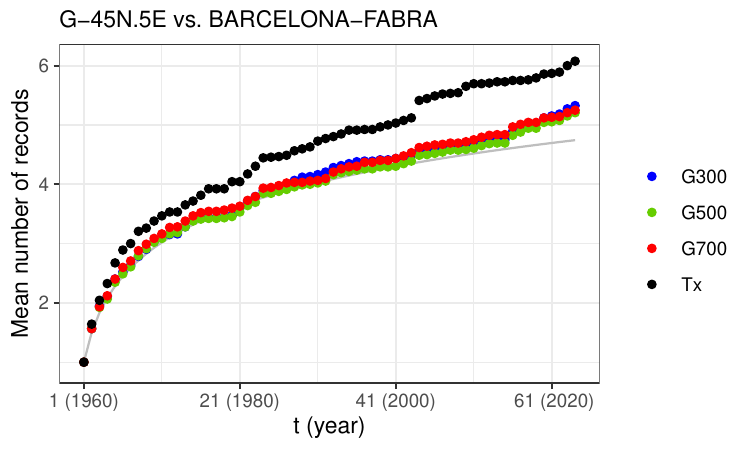}
    \includegraphics[width=0.45\textwidth]{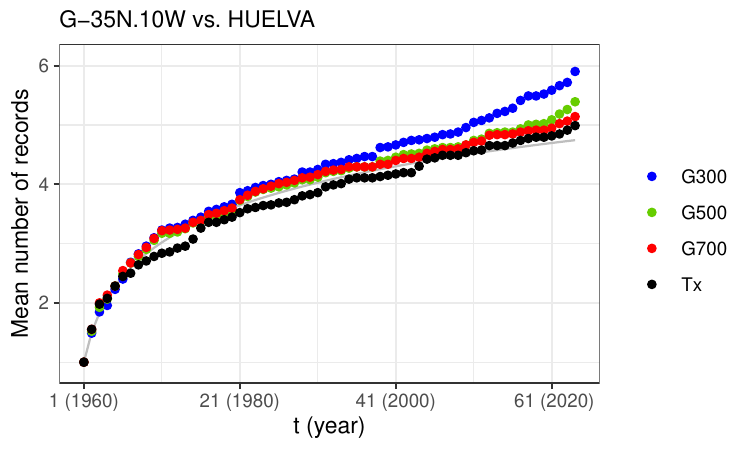}
    \includegraphics[width=0.45\textwidth]{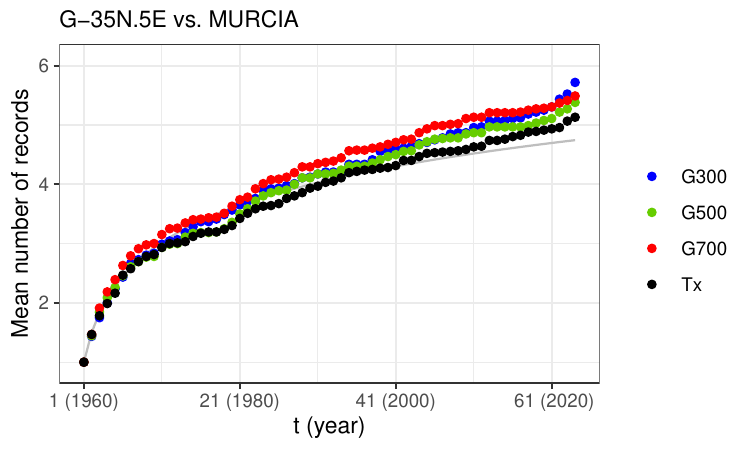}
    \caption{Mean number of records per year $t$ at each corner of the ERA5 grid for the geopotential in 700 (red), 500 (green), and 300 (blue) hPa. In black the closest station series of $T_x$ records. Vertical lines shows the estimated year of the change point.
    }
    \label{fig:record.g.trends}
\end{figure}

In this section, we compare the evolution of the cumulative number  of calendar-day records over time  in $T_x$  and the three geopotential  variables. As an illustration Figure~\ref{fig:record.g.trends} compares the series in the four corners of the grid covering the Iberian peninsula and the closest stations to those points: La Coruña, Barcelona-Fabra, Murcia and Huelva; together  with the expected behavior in a stationary situation, $\sum_{i=1}^t 1/i$. Supplementary Table~\ref{sup:tab:nttest} includes the $N_t$ statistics and p-values.

Although a similar temporal evolution is observed across the four variables in each plot, the patterns are not equivalent,  with  Huelva, and specially Barcelona showing the highest differences. Additionally, there is no homogeneous spatial behavior among the patterns of the four variables. For example, while the highest deviation from stationarity in Barcelona is observed in $T_x$, this variable shows the lowest deviation in Huelva and Murcia. The behavior is similarly inconsistent in the geopotential variables: in Huelva, the greatest deviation occurs in G300, whereas in A Coruña and Barcelona, G300 exhibits one of the lowest deviations. In summary, although there is a general relationship in the temporal evolution of record occurrences among the four signals, it is not a one-to-one correspondence.

\subsubsection{Influence of geopotential variables in the occurrence of record-breaking temperatures}
 
\begin{figure}[bt]
    \centering
    \includegraphics[width=0.32\linewidth]{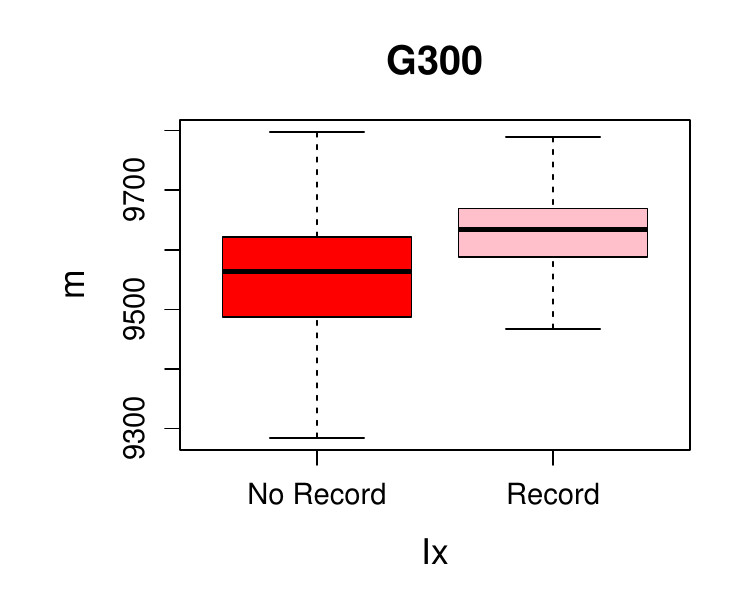}
    \includegraphics[width=0.32\linewidth]{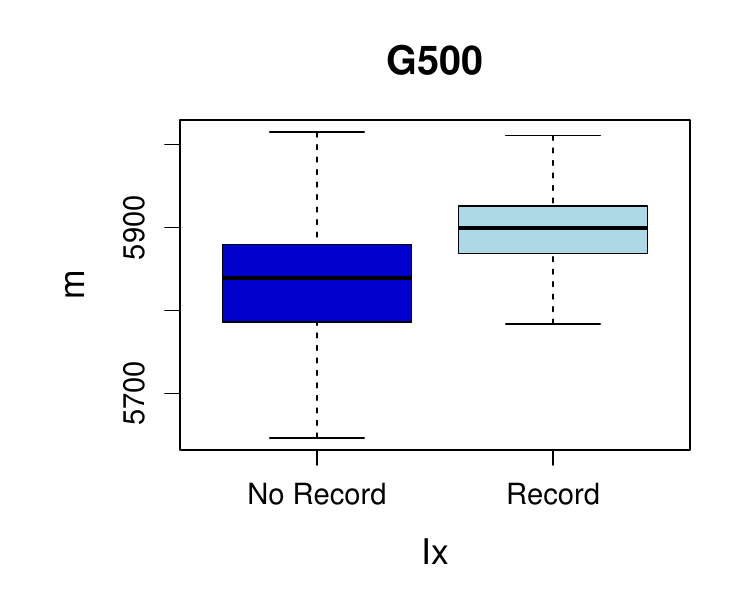}
    \includegraphics[width=0.32\linewidth]{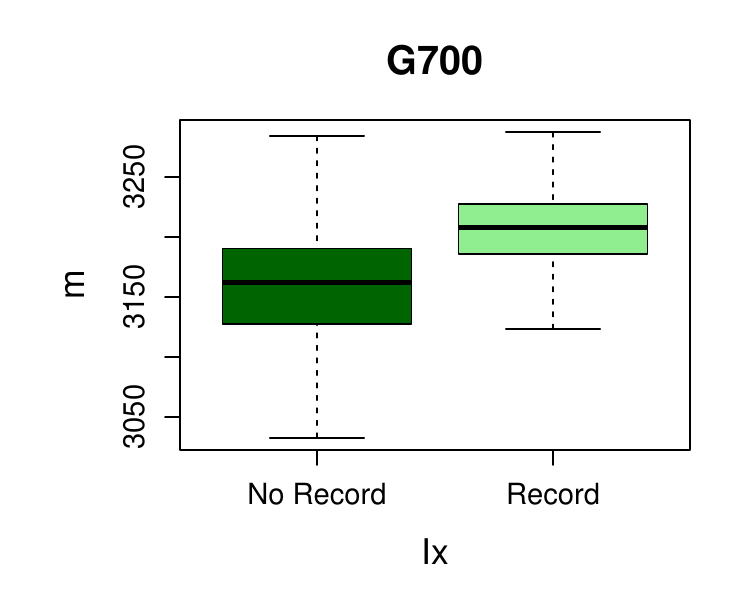}
    \caption{Boxplots (without the outliers) of geopotential variables at the 300, 500, and 700 hPa pressure levels (in geopotential high meters) during the  period 1984--2023, conditioned on the occurrence or non-occurrence of maximum temperature records.}
    \label{fig:boxplots:by:records}
\end{figure}

To study the  power of geopotential variables to predict the occurrence of  record-breaking temperatures, we compare the distribution of the variable given that a record has occurred in the maximum temperature or not. Figure \ref{fig:boxplots:by:records} shows the boxplots of the geopotential heights at the 300, 500, and 700 hPa pressure levels during the  period 1984--2023, conditioned on the occurrence or non-occurrence of maximum temperature records. The distribution of geopotential given the occurrence of a record  is clearly shifted upward, and its variability is smaller compared to when no record occurs. The difference between the medians for the two conditioning cases is slightly larger in the variables at lower atmospheric levels. 

\section{Methods}
\label{sec:methods}

\subsection{Algorithm}

\begin{figure}[htb]
    \centering
    \includegraphics[width=1\linewidth]{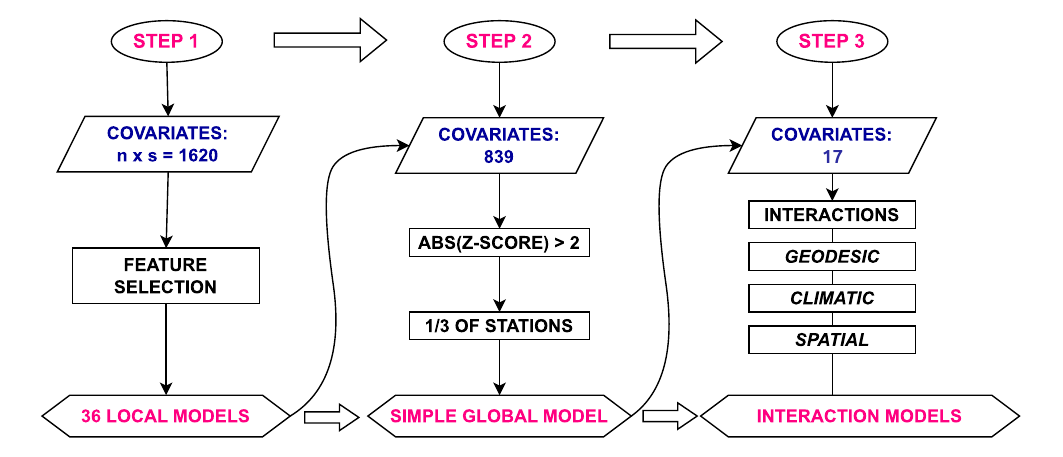}
    \caption{Diagram that summarizes the three steps of the algorithm.}
    \label{fig:algorithm}
\end{figure}

An algorithm was developed to construct a global spatio-temporal model for the occurrence of $T_x$ records across the Iberian Peninsula. The algorithm comprises three main steps. First, local models are fitted to each station by selecting the optimal combination of atmospheric covariates and their derivatives. Second, the most prevalent covariates from these local models are identified and used to build a global model driven solely by atmospheric data. Third, the resulting global model is enhanced by incorporating interactions with non-atmospheric covariates. These steps are summarized in Figure~\ref{fig:algorithm}.

\begin{figure}[htb]
    \centering
    \includegraphics[width=.45\textwidth]{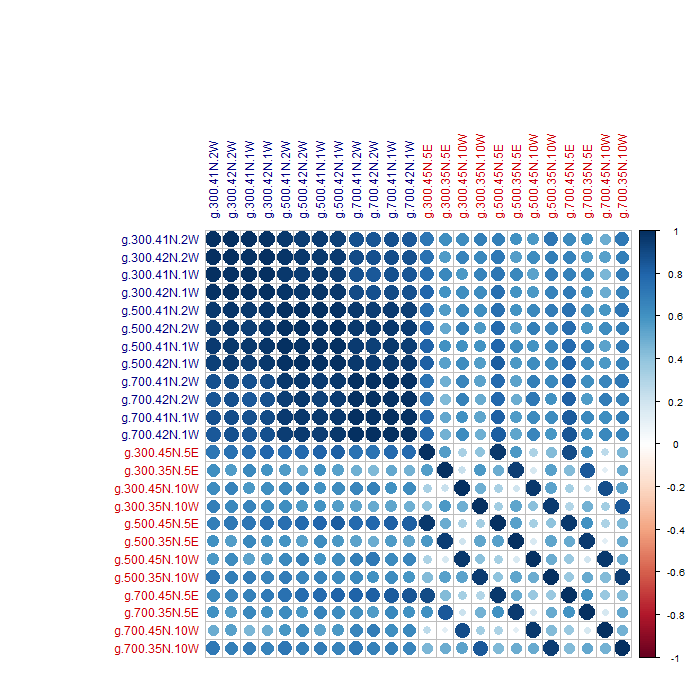}
    \includegraphics[width=.45\textwidth]{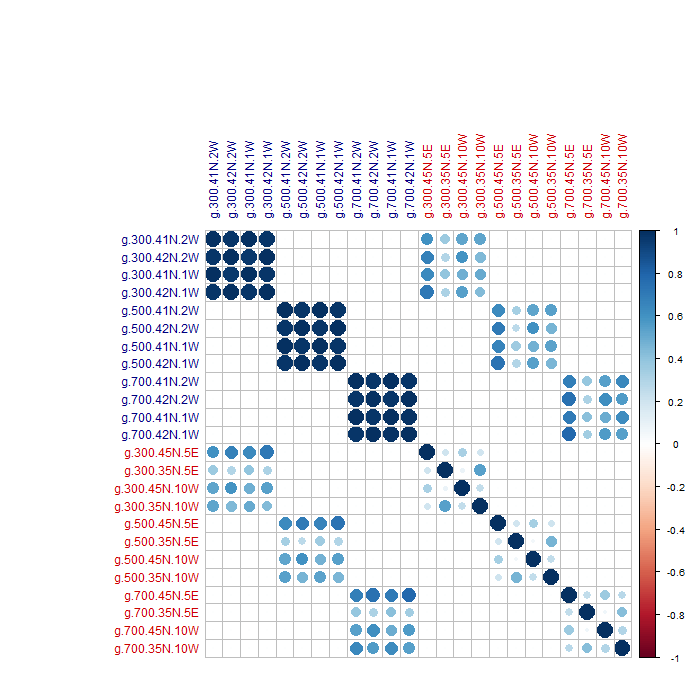}
    \caption{Pearson's (left) and intra-class (right) cross-correlation plots of the geopotential altitude covariates at the three pressure levels in the four nearest and furthest grid points of Zaragoza station (41N,1W).}
    \label{fig:supp:coricc}
\end{figure}

\textbf{Step 1}: A logistic regression model was optimized for each of the 36 stations, with the daily $T_x$ record indicator $I_{t,\ell}(s_i)$ as the target variable at station $s_i$. Predictors consisted of geopotential variables (G) from three pressure levels, including: (1) the nearest grid point, (2) its four surrounding grid corners, (3) their 1-day lagged terms, and (4) their second-order polynomial terms. Exploratory analyses justified this approach, revealing near-perfect correlations among adjacent grid points (Supplementary Figure~\ref{fig:supp:coricc}), which allowed the use of a single representative point. These analyses also identified significant persistence effects and nonlinear trends (Figures~\ref{fig:lor} and~\ref{fig:ix_trends}), motivating the inclusion of lagged and polynomial terms.
Hence, for each station, the algorithm evaluated 45 potential predictors via backward stepwise regression using the \textit{stepAIC} function (R package \texttt{MASS} v7.3-60.0.1, \cite{MASS4}) with $k = 2$ (default). Four model configurations were compared: (i) base G-terms only, (ii) G-terms plus lagged-1 terms, (iii) G-terms plus second-order polynomials, and (iv) G-terms with both lagged and polynomial terms. The configuration with the lowest AIC was selected as the optimal local model, automatically pruning unwarranted complexity when additional terms did not improve fit. This process produced 36 unique station-specific models.

\textbf{Step 2}: A global model was constructed using concatenated data from all stations. Predictor selection was based on geopotential covariates with absolute standardized coefficients ($|z|$-scores) exceeding 2 (approximately equivalent to a two-sided $p$-value $< 0.05$) in at least one-third of the 36 optimal local models (i.e., $\geq 12$ stations). These thresholds balanced predictor relevance with cross-station representativeness. The selected predictors were used to fit a global model, which underwent exhaustive feature selection via \textit{stepAIC} with bidirectional elimination and an adjusted penalty parameter ($k = 10.83$). This stringent threshold (corresponding to the 99.9th percentile of the $\chi^2$ distribution with 1 degree of freedom) mitigated potential overspecification that could arise from using the default $k = 2$ in large datasets. The resulting model (M1) contained only atmospheric predictors.

\textbf{Step 3}: The global modeling framework was extended by augmenting M1 with interaction terms while maintaining the stringent \textit{stepAIC} selection criteria ($k = 10.83$). Four enhanced models were developed: M2 introduced interactions between geopotential variables and geodetic covariates (latitude and longitude); M3 incorporated interactions with climatic covariates (1981--2010 $T_x$ mean and standard deviation); M4 included interactions with spatial covariates (altitude and coastal distance); and M5 combined all interaction types (geodetic, climatic, and spatial). All models preserved the rigorous feature selection protocol from Step 2 while systematically exploring distinct environmental dimensions, enabling the isolation of specific geophysical effects while controlling for potential overfitting.

\begin{table}[htb]
    \centering
    \caption{Global model summaries by their type of covariates included and their interactions.}
    \medskip
    \begin{tabular}{l}
     Global models description \\ [6pt]
     \hline  \\[0.1cm]
     \textbf{M1:} $I_{t,l}(s_i) \sim \text{Geopotentials}$ \\ [2pt]
     \textbf{M2:} $I_{t,l}(s_i) \sim \text{Geopotentials} * \text{Geodetic}$ \\[2pt]
     \textbf{M3:}  $I_{t,l}(s_i) \sim \text{Geopotentials} * \text{Climatic}$  \\[2pt]
     \textbf{M4:}  $I_{t,l}(s_i) \sim \text{Geopotentials} * \text{Spatial}$  \\[2pt]
    \textbf{M5:}  $I_{t,l}(s_i) \sim \text{Geopotentials} * (\text{Geodetic} + \text{Climatic} + \text{Spatial})$ 
    \end{tabular}
    \label{tab:models}
\end{table}

\subsection{Model comparison and validation}

To evaluate the models' predictive performance, the dataset was divided into training ($\approx$80\%) and testing subsets using a temporal split. The training set comprised the first 51 years of data, while the testing set contained the most recent 13 years. All model computation and feature selection procedures were performed exclusively on the training data. The resulting models were then applied to the testing set for prediction. Given the class imbalance in the record series (predominantly non-records), model validation employed Receiver Operating Characteristic (ROC) curve analysis. ROC curves plot the True Positive Rate (Sensitivity) against the False Positive Rate (1-Specificity) across all possible prediction thresholds. Model performance was quantified using the Area Under the Curve (AUC), which ranges from 0.5 (random guessing) to 1 (perfect classification), providing a comprehensive measure of predictive accuracy.

The primary model selection criterion was the highest global Area Under the Curve (AUC). Secondary considerations included: (1) model parsimony (lower number of predictors), (2) Akaike Information Criterion (AIC) values, and (3) AUC performance specifically for coastal stations (located within 50 km of the sea). This multi-faceted approach ensured optimal balance between predictive performance and model simplicity while addressing potential geographic biases.

To assess the model's capacity to reproduce record persistence, we compared observed and simulated probabilities of record run lengths during the validation period. For each station, we calculated the probability distribution of runs spanning: (1) single-day events, (2) two consecutive days, (3) three days, (4) four days, and (5) five or more days. Model predictions were evaluated against 10,000 randomly generated uniform probability series to produce binary event sequences. The analysis quantified persistence by computing station-averaged relative frequencies for each run-length category, comparing observed values against the simulated 95\% percentile intervals. This approach evaluated the model's ability to capture temporal clustering of extreme records.

The model's spatial performance was evaluated by comparing observed and simulated co-occurrence patterns of records across station pairs. We computed Jaccard Index (JI) scores for all station pairs during the validation period, comparing observed values against those derived from 10,000 randomly generated uniform probability series. 


\section{Results}
\label{sec:results}

\subsection{Models}

This section evaluates the performance of local versus global models. Model selection was based on (1) validation-set performance (assessed using standardized metrics) and (2) structural parsimony. We further highlight the algorithm's key advantages over conventional approaches.

\subsubsection{Local Models}

\begin{figure}
    \centering
    \includegraphics[width=0.7\linewidth]{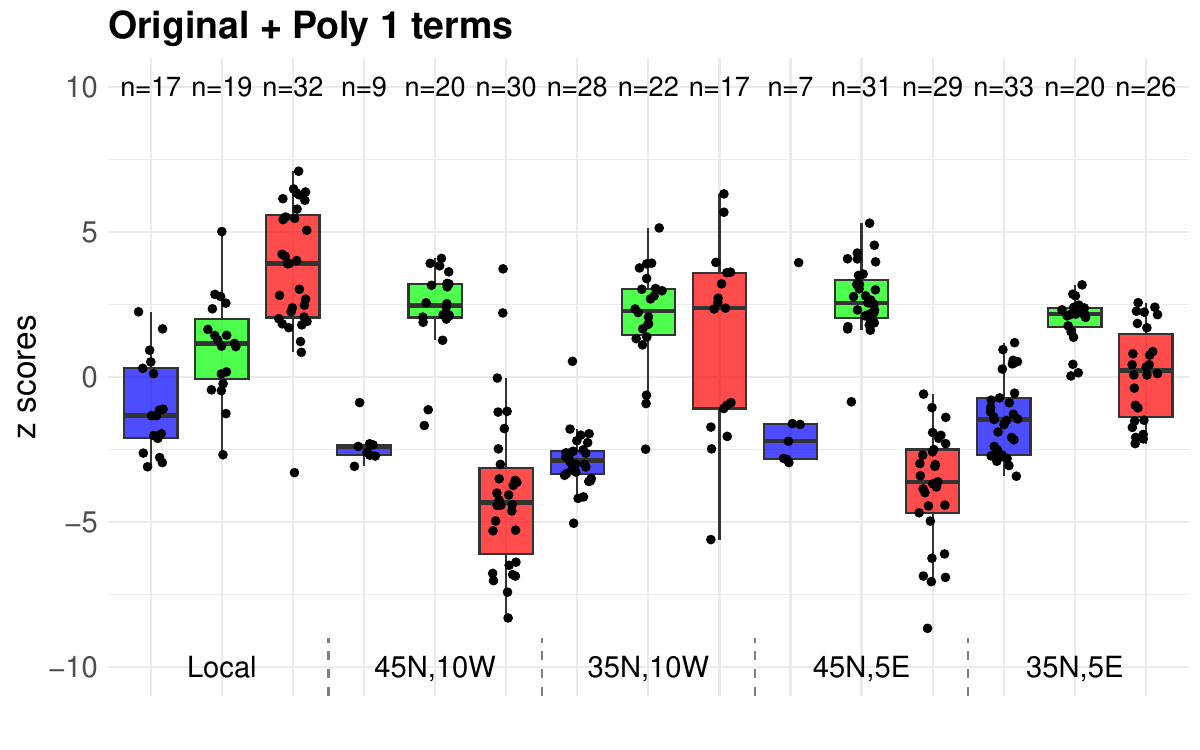}
    \includegraphics[width=0.7\linewidth]{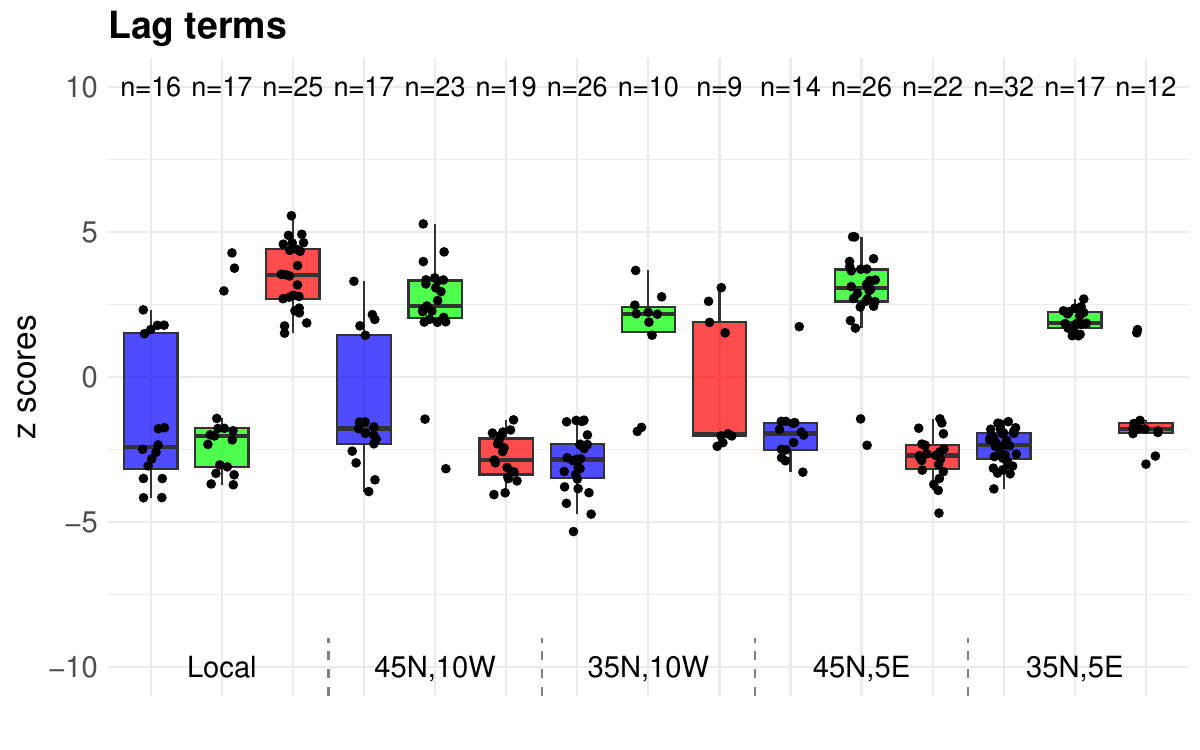}
    \includegraphics[width=0.7\linewidth]{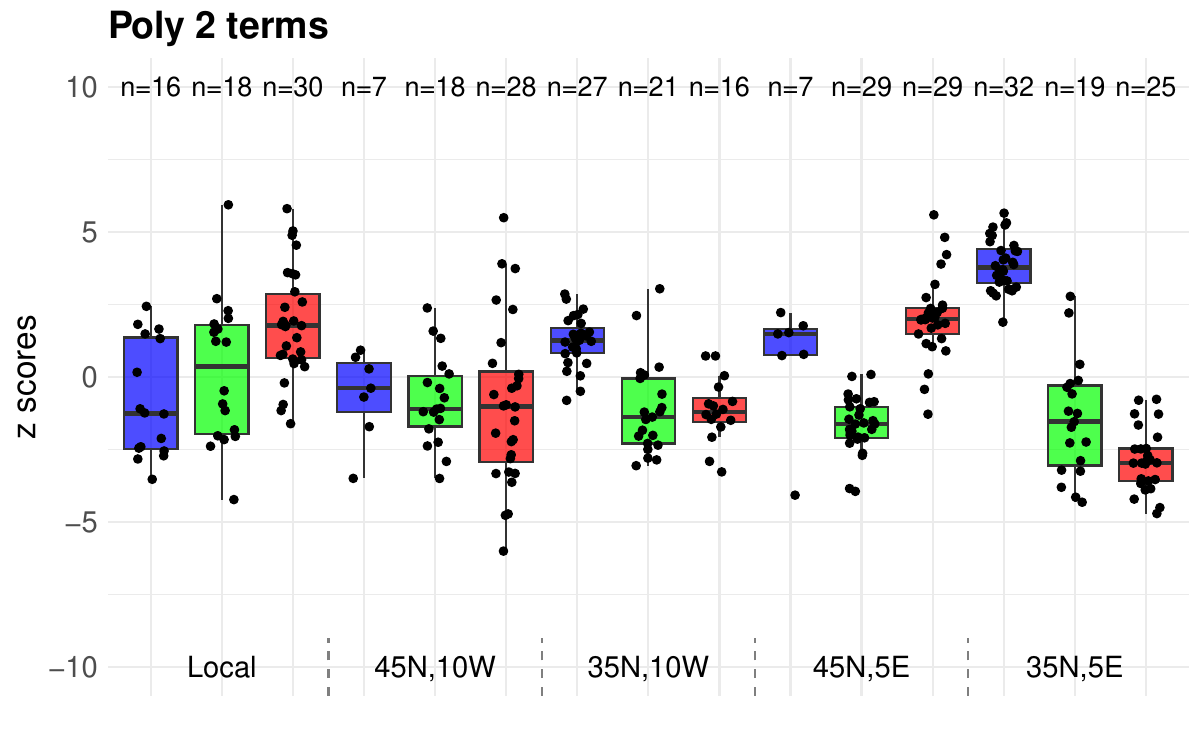}
    \caption{Jittered boxplot of the z-values of the beta coefficients from the optimal local models. Color indicates preassure level in hPa: 300 (blue), 500 (green), 700 (red). X-axis indicates the five grid points: nearest (Local), North-West (45N.10W), South-West (35N.10W), North-East (45N.5E), South-East (35N.5E). Covariates: single or poly-1 terms (top), 1-lagged terms (middle), poly-2 terms (bottom). '$n$' above each covariate indicates the number of stations scoring an absolute z-score higher than two.}
    \label{fig:local:z}
\end{figure}

Optimal local models were fitted for each station using stepwise regression. Figure~\ref{fig:local:z}
displays the z-scores of retained coefficients, categorized by grid location (nearest/corners), pressure level (300/500/700~hPa), and term type (single/lagged/polynomial). The nearest grid point exhibits the strongest positive effects at 700~hPa across all term types. In contrast, corner points show distinct patterns: single terms dominate at 500--700~hPa, while lagged terms alternate between positive (500~hPa) and negative effects (300/700~hPa). Polynomial terms generally contribute weakly, except at 300~hPa in the south-east corner. These results highlight systematic pressure-level dependencies linked to Iberian topography.

Additionally, Supplementary Table~\ref{sup:tab:local} summarizes the complexity and performance of each local model, reporting: (1)~the number of parameters ($k$), (2)~the area under the curve ($\text{AUC}$), and (3)~the Akaike Information Criterion ($\text{AIC}$) for all stations. The models achieve substantial dimensionality reduction, with $k$ values ranging from 15 to 35 (from an initial set of 45 potential predictors). $\text{AUC}$ was computed using independent validation data (2011--2023), which includes exceptionally warm years, providing a rigorous out-of-sample test. All models significantly outperform random chance ($\text{AUC} > 0.5$), with most stations exhibiting excellent discrimination ($\text{AUC} > 0.9$) and others showing robust skill ($\text{AUC} > 0.8$). Notably, Gijón's model has the weakest performance ($\text{AUC} = 0.67$), suggesting local atmospheric drivers may deviate from the regional pattern captured elsewhere.

\subsubsection{Global models}

Using a threshold of absolute z-scores $>2$ in at least one-third of stations (12 of 36), we identified 23 potential predictors for the global atmospheric model. Supplementary Table~\ref{sup:tab:local:thresholds} demonstrates how alternative selection thresholds would affect the final covariate count. The selected predictors were then used to fit a global model, which underwent exhaustive feature selection via \textit{stepAIC}, ultimately retaining 17 covariates. The stringent selection criteria ensure both model parsimony and atmospheric interpretability.

\begin{figure}[htb]
    \centering
    \includegraphics[width=1\linewidth]{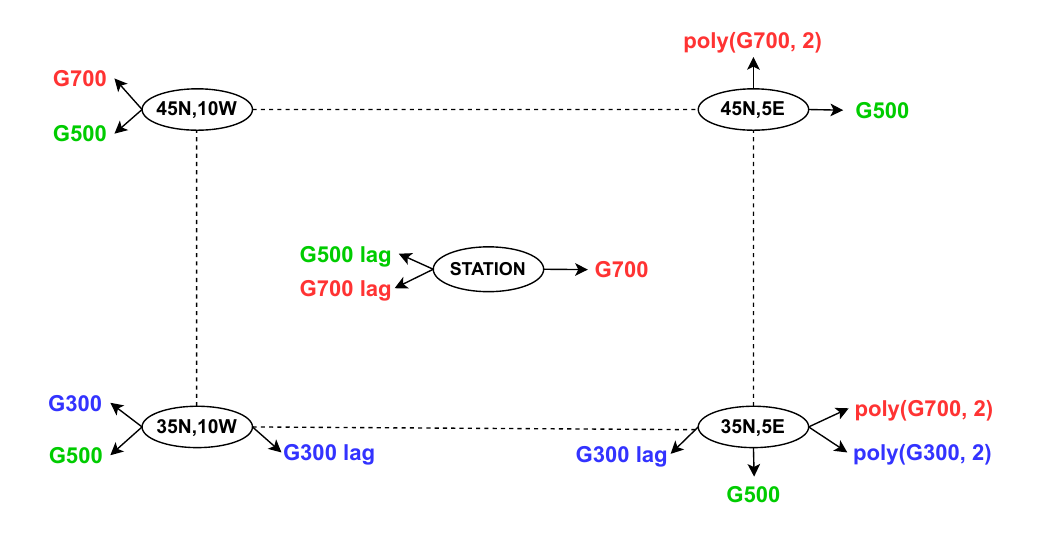}
    \includegraphics[width=1\linewidth]{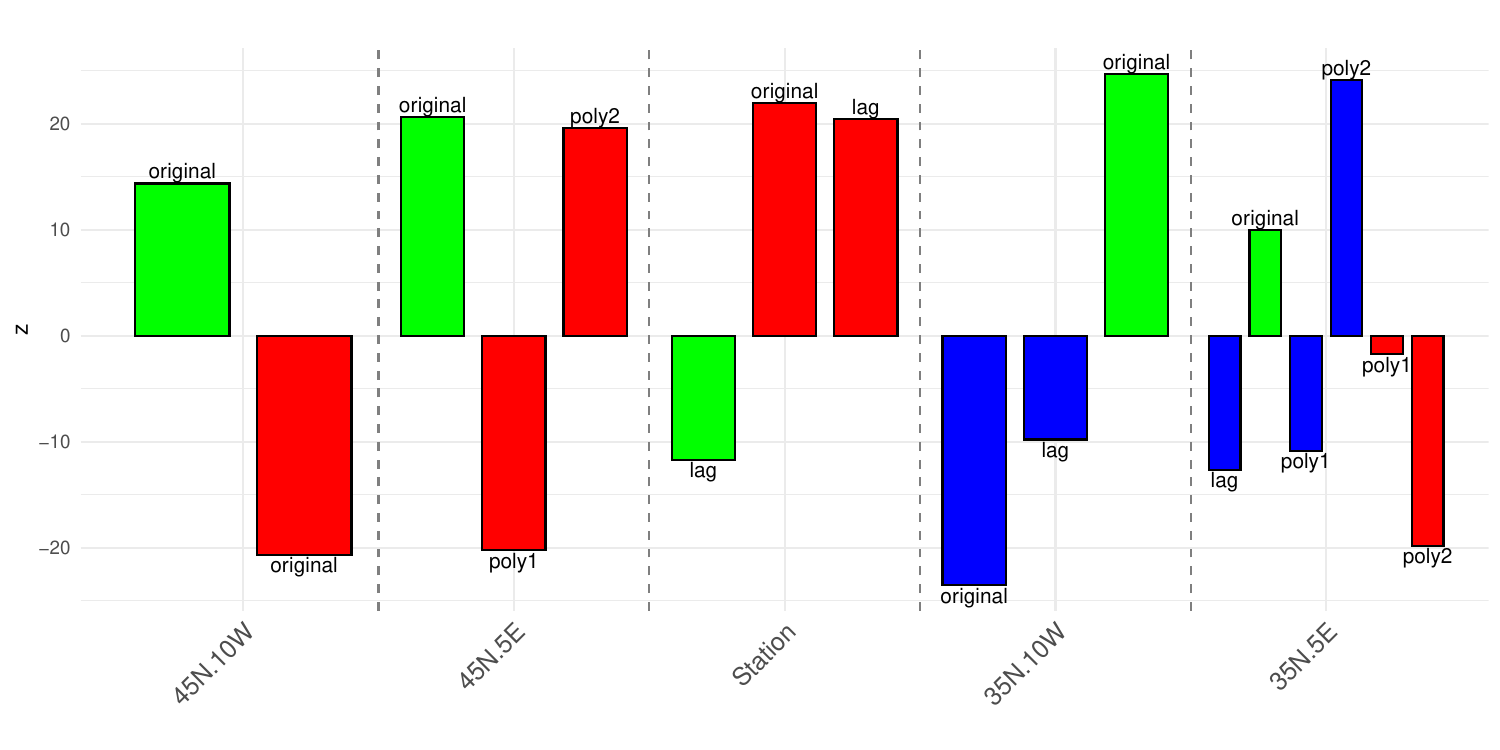}
    \caption{Diagram with the 14 geopotential variables from model M1. Variables followed by `lag' represent one-day lagged terms, while those preceded by `poly' correspond to the second-order polynomial of the corresponding variable. Top: pseudo-geographical represenation of the four corners of the grid and the nearest point to a station. Bottom: standardized estimated beta coefficients (z-scores) of the given model.}
    \label{fig:diag:var}
\end{figure}

The resulted global model M1 only accounts for atmospheric covariates. This model encompasses 14 geopotential variables which are depicted in Figure~\ref{fig:diag:var} (Supplementary Information~\ref{sup:m1:summary} includes the model summary). Variables followed by `lag' or preceded by `poly' refer to the lag term or to the second-order polynomial of that variable, respectively. Note that variables with second-order polynomial implies two terms in the model. Each variable is represented by an arrow indicating the point where it is considered, showing the four corners of the grid of the Peninsula. The label `STATION' refers to the variable at the closest grid point to each station analyzed. We can observe a higher representation of the variables at 300hPa in the lower latitudes, while the northern and local nodes only include predictors at 700 and 500hPa. This indicates a patent gradient between pressure levels and latitude. Regarding the variable relevance in the models, geopotentials at 700hPa in the nearest grid point, the single ($t=21.99$) and its 1-lagged term ($t=20.47$) showed a positive effect for the probability of $T_x$ records, while at the North-West (single: $t=-20.68$) and South-East (poly-1: $t=-1.74$; poly-2: $t=-19.8$) points showed negative effects. At 500hPa, the near grid point showed negative effect in the 1-lagged term ($t=-11.72$), while positive in the four corners of the grid (single terms: NW $t=14.39$; NE $t=20.68$; SW $t=24.73$; SE $t=9.96$) evincing the relevance of this pressure level at every input point grid. At 300hPa, we can observe a general negative effect in the South points (Single terms: SW $t=-23.53$; and 1-lagged terms: SW $t=-9.77$; SE $t=-12.66$) plus a convex effect at the South-East (poly-1: $t=-10.84$; poly-2: $t=24.15$).

Afterwards, the geopotential covariates of M1 were interacted with additional variables: geodetic (M2), climatic (M3), and spatial (M4), plus a model that sums all of them (M5). Table~\ref{tab:modelos:globales} shows performance metrics of the five global models. It presents the total $AUC$, the number of parameters ($k$), the $AUC$ for stations located less than 50km from the coast ($AUC<50km$), the AUC for stations situated more than 50km from the coast ($AUC>50km$), and the Akaike Information Criterion ($AIC$). We observe that the model with the highest overall AUC and lowest AIC is M2. Also, the most complex model M5 after the global stepwise regression equals M2, suggesting that climatic nor spatial variables improve the geodetic information added to the atmospheric model. For stations located within 50 km of the coast, the model with the highest AUC is M3 but M2 achieves a near score, whereas for stations farther than 50 km inland M2 achieves the highest AUC.

\begin{table}[htb]

    \caption{Performance metrics of the five global models. The total $AUC$, the number of parameters ($k$), the $AUC$ for stations located less than 50 km from the coast ($AUC_{<50km}$), the $AUC$ for stations situated more than 50 km from the coast ($AUC_{>50km}$), and the Akaike Information Criterion ($AIC$) are presented. The $AUC$ values were obtained during the validation period. Stations were sorted in the X-axis by distance to coast in ascending order, and the station name was abbreviated until the 6th character.}
    \medskip
    \centering
    \begin{tabular}{|c|c|c|c|c|c|}
        \hline
        Model & $k$ & $AUC$ & $AUC_{<50km}$ & $AUC_{>50km}$ & $AIC$\\
        \hline
        \textbf{M1} & 18 & 0.8575 & 0.7760 & 0.9062 & 97808.39 \\ 
        \textbf{M2} & \textbf{41} & \textbf{0.8787} & \textbf{0.7951} & \textbf{0.9253} & \textbf{95606.40} \\ 
        \textbf{M3} & 37 & 0.8759 & 0.7965 & 0.9139 & 95995.68 \\ 
        \textbf{M4} & 30 & 0.8677 & 0.7802 & 0.9105 & 96856.47 \\ 
        \textbf{M5} & 41 & 0.8787 & 0.7951 & 0.9253 & 95606.40 \\
        \hline
    \end{tabular}

    \label{tab:modelos:globales}
\end{table}

\begin{figure}[htb]
    \centering
    \includegraphics[width=1\linewidth]{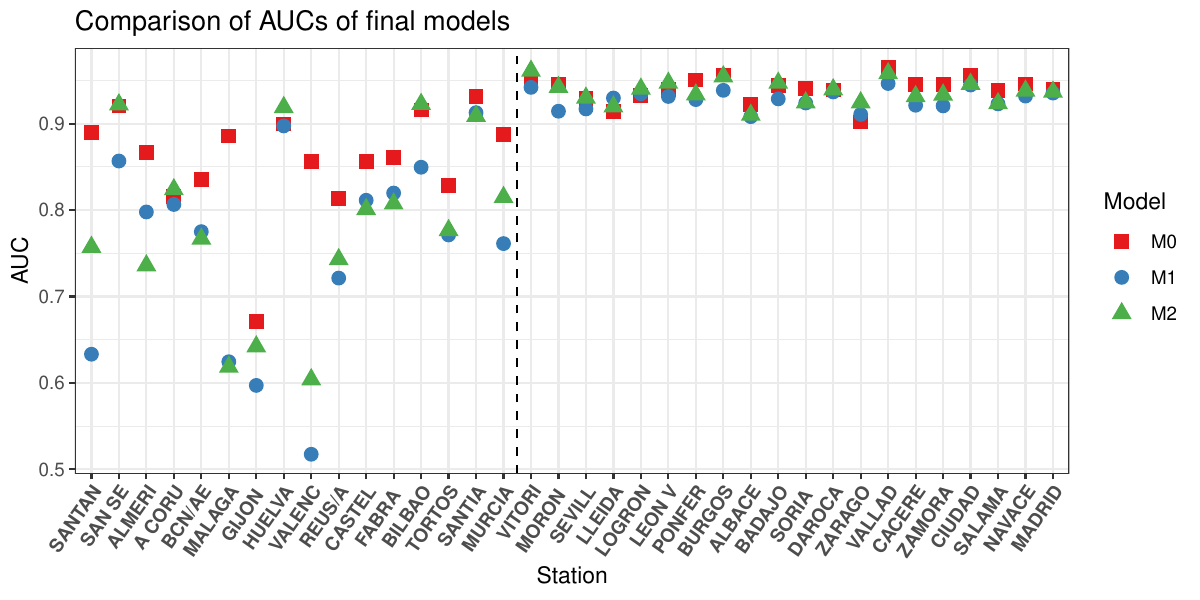}
    \caption{AUCs by station of models M0, M1, and M2. M0 is a local model for each of the stations with complete freedom. M1 is a global model with exclusively geopotential variables. M2 is a global model where M1 is interacted with Latitude and Longitude.}
    \label{fig:auc}
\end{figure}

Considering both performance and simplicity, model M2 demonstrates the best overall performance, and we therefore select it as best model. Note that this model includes latitude-longitude interactions. Figure~\ref{fig:auc} shows the AUC performance of M2 for each station which are sorted by distance to the coast, along with the AUC for individual local models (denoted as M0) and the global model with only atmospheric data M1. We can consider the AUC achieved by the M0 models as the highest possible using geopotential variables without the need of spatial information, and M1 as a ``rigid'' or too simplistic model. Then, in this framework, we can observe how is the improvement of M2 related to M1 and how close is to achieve the local modeling scores. The models exhibit better performance at stations located more than 50km from the coast. In local models, it had previously been observed that AUC values were lower for stations within 50km of the coast. A possible explanation is that global models are better suited for inland stations, where environmental conditions are more homogeneous. This suggests that global models capture broad-scale patterns prevalent in regions farther from the coast, while coastal stations exhibit higher spatial and temporal variability. Furthermore, coastal regions are inherently more challenging to model due to factors such as high atmospheric variability, the influence of ocean currents, and the presence of ``micro-climates'' affecting near-shore stations.

\subsection{Persistence validation}

After selecting M2 as the best-performing model, we assessed its effectiveness in capturing the persistence of the response variable by comparing the observed frequencies of record runs with those estimated from model simulations. This analysis used data from all monitoring stations over the 2011–2023 test period and considered runs of length one, two, three, four, and five or more (i.e., $\geq$ 5). To determine the observed frequencies, we first identified sequences of record-breaking events at each station throughout the test period and computed their relative frequencies. The results were then averaged across all stations, with variability expressed as the 95\% percentile interval. The model's performance was evaluated using 10,000 simulations. For each simulation, we generated a simulated binary time series by comparing the model's predicted probabilities to a random number uniformly distributed between 0 and 1. A value of 1 was assigned where the prediction exceeded the random number, indicating a simulated record event. From these simulated sequences, the relative frequencies of runs of various lengths were determined.

\begin{figure}[htb]
    \centering
    \includegraphics[width=0.7\textwidth]{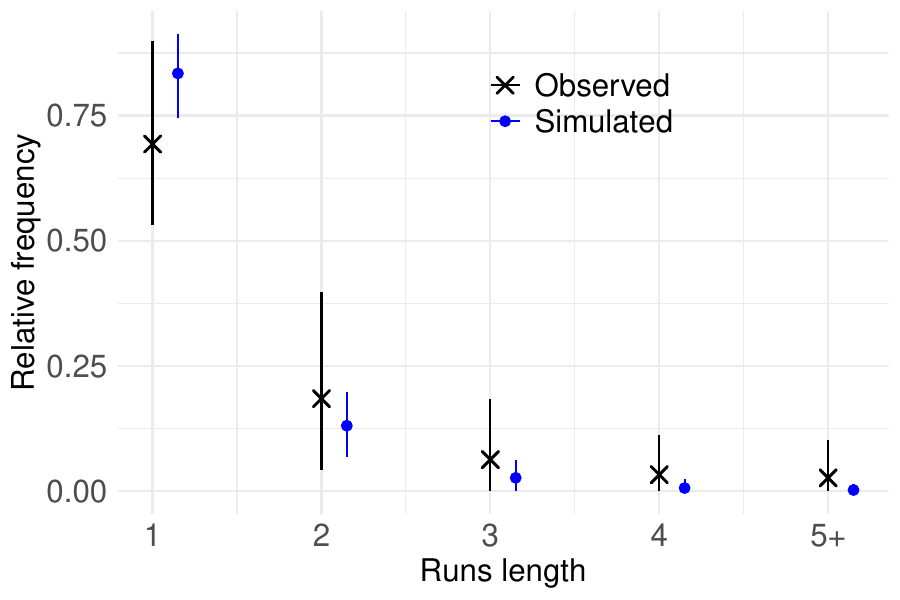}
    \caption{Average and 95\% percentile interval across stations of the relative frequencies of streaks of length one, two, three, four, and five or more for the observed and simulated M2 predictions in the 2011-2023 test period.}
    \label{fig:streaks}
\end{figure}

Table~\ref{tab:streaks} and Figure~\ref{fig:streaks} present the relative frequencies of observed runs of various lengths alongside those generated by model M2 during the 2011--2023 test period, including their associated percentile intervals. The results indicate that model M2 tends to slightly overpredict the frequency of single-day runs and underpredict the occurrence of longer runs ($\geq$5 days). Nevertheless, the 95\% percentile intervals for the simulated and observed values are largely comparable, suggesting the model captures the general distribution of runs lengths reasonably well. This pattern implies a minor bias in the model toward forecasting more isolated events than observed. Despite this, the overall agreement supports the conclusion that model M2 provides a reliable representation of the persistence of record-breaking maximum temperature events.

\begin{table}[htb]

\caption{Average (95\% percentile interval) across stations of the observed and M2-simulated record streaks relative frequencies in the 2011-2023 validation periods.}
\medskip
\centering
\begin{tabular}{lcc}
\hline
\textbf{Streak} & \textbf{Observed} & \textbf{M2-simulated} \\
\hline
Length 1 & 0.69 (0.53 - 0.90) & 0.84 (0.72 - 0.92) \\
Length 2 & 0.18 (0.04 - 0.40) & 0.12 (0.06 - 0.20) \\
Length 3 & 0.06 (0.00 - 0.18) & 0.03 (0.00 - 0.07) \\
Length 4 & 0.03 (0.00 - 0.11) & 0.01 (0.00 - 0.03) \\
Length 5+ & 0.03 (0.00 - 0.10) & 0.00 (0.00 - 0.02) \\
\hline
\end{tabular}
\label{tab:streaks}

\end{table}

\begin{figure}[htb]
    \centering
    \includegraphics[width=0.45\textwidth]{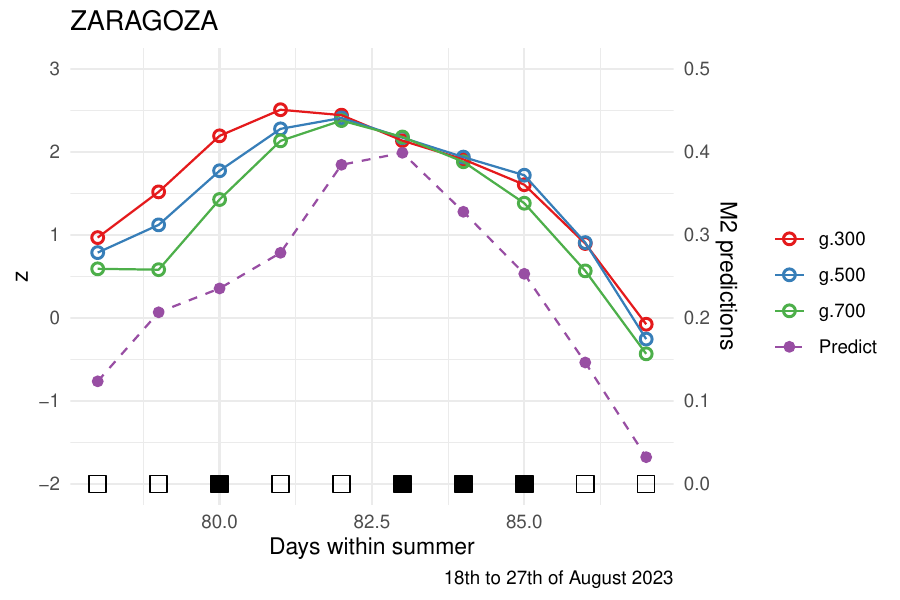}
    \includegraphics[width=0.45\textwidth]{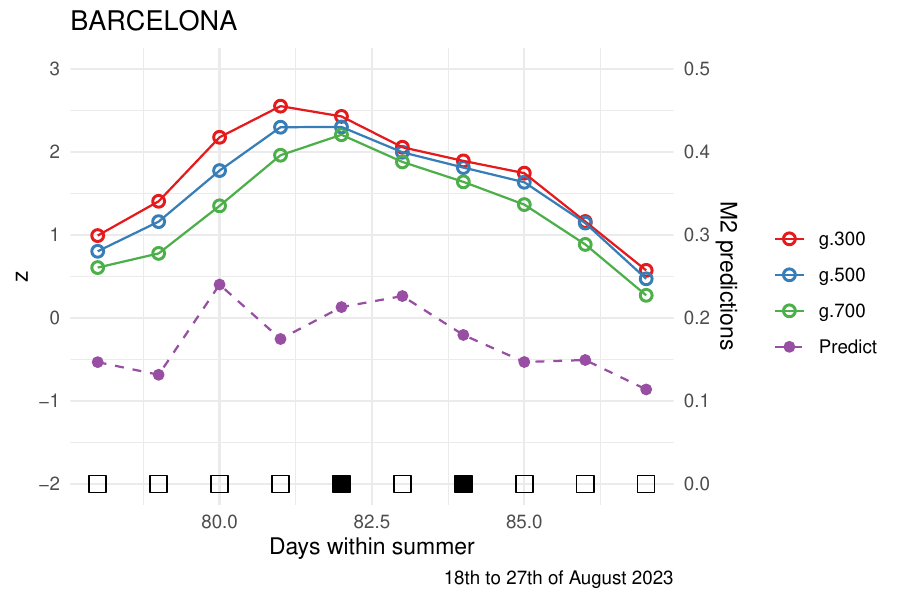}
    \includegraphics[width=0.45\textwidth]{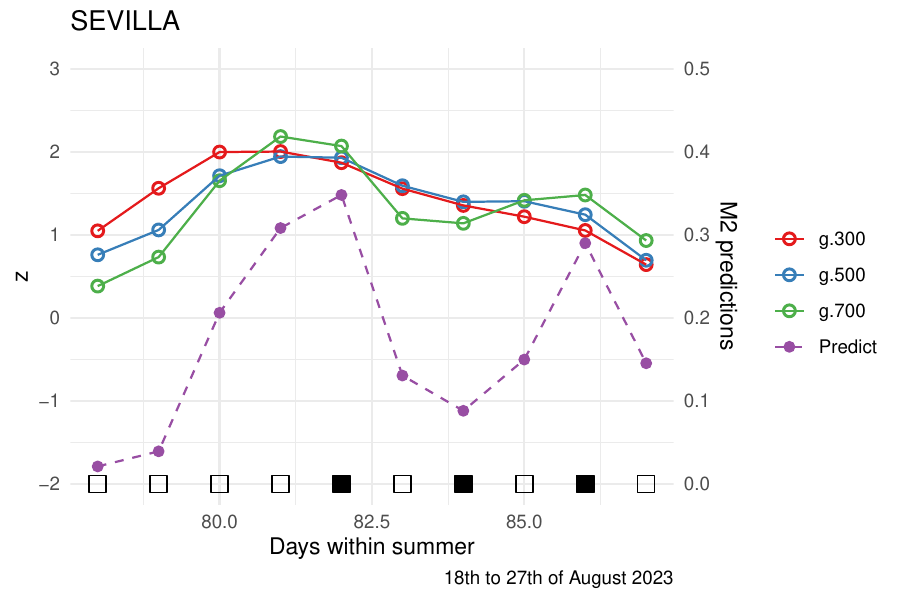}
    \includegraphics[width=0.45\textwidth]{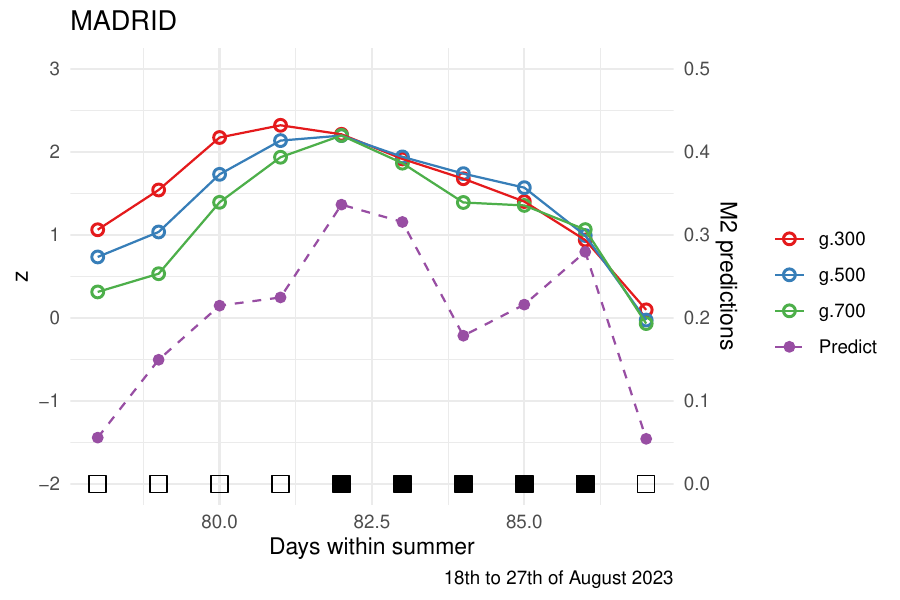}
    \caption{The plot displays the situation of the atmosphere in the days of the August 18th to the 27th of 2023. The boxes at the bottom of the plot are painted in black if the day was a record, and in white if not. Left y-axis represents the geopotentials from the nearest grid point at 700, 500, and 300 hPa scaled using the period of 1980-2010 and right y-axis represents the M2 probability of record prediction.}
    \label{fig:temporal:4stations}
\end{figure}

More specifically, Figure~\ref{fig:temporal:4stations} shows the validated M2 predictions for a period of extreme heat events (August 18--27, 2023) in the Iberian Peninsula, illustrated for four stations from distinct geographical areas. The predictions are depicted alongside the observed $T_x$ records and the scaled geopotential height at the nearest grid point. The model captures the increase in geopotential height, albeit with a slight temporal delay. Furthermore, M2 successfully reproduces discrete extreme events, such as the separated runs observed at the Sevilla station.

\subsection{Concurrence validation}

\begin{figure}
    \centering
    \includegraphics[width=0.65\linewidth]{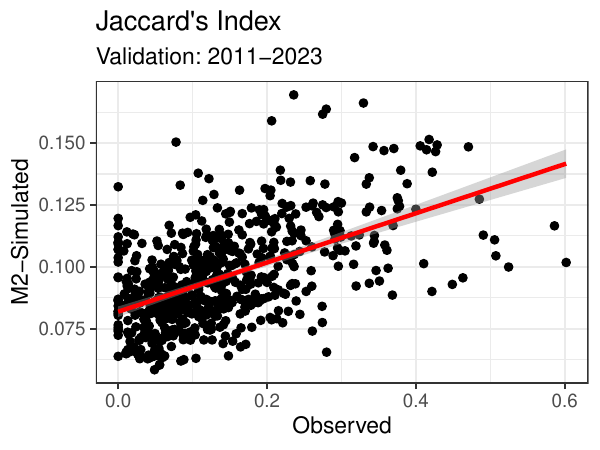}
    \caption{Scatterplot of the observed vs. M2-simulated Jaccard Indices for each pair of stations.}
    \label{fig:concurrence}
\end{figure}

The concurrence of $T_{\text{max}}$ records for the selected model M2 was assessed using the Jaccard index 
between pairs of stations, computed from 10,000 simulation runs of predicted records. The average simulated scores were compared with the observed Jaccard index scores during the 2011--2023 validation period. Figure~\ref{fig:concurrence} shows the relationship between the observed and simulated scores. The correlation between them is 0.54,
indicating a strong positive relationship. Although the magnitudes of the simulated scores are systematically lower, the model M2 correctly identifies which station pairs exhibit higher observed record concurrence.

\begin{figure}[htb]
    \centering
    \includegraphics[width=0.55\textwidth]{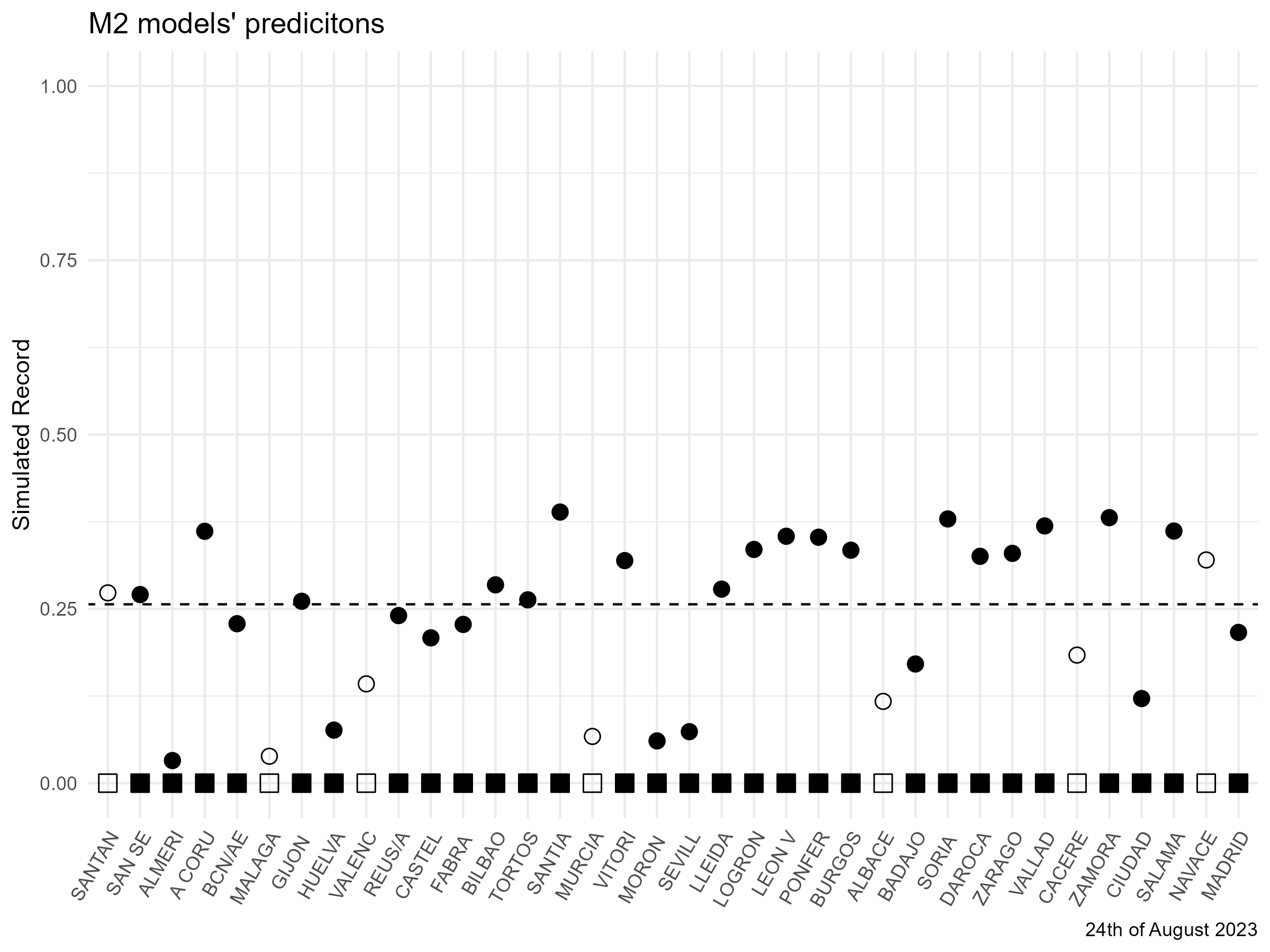}
    \includegraphics[width=0.45\textwidth]{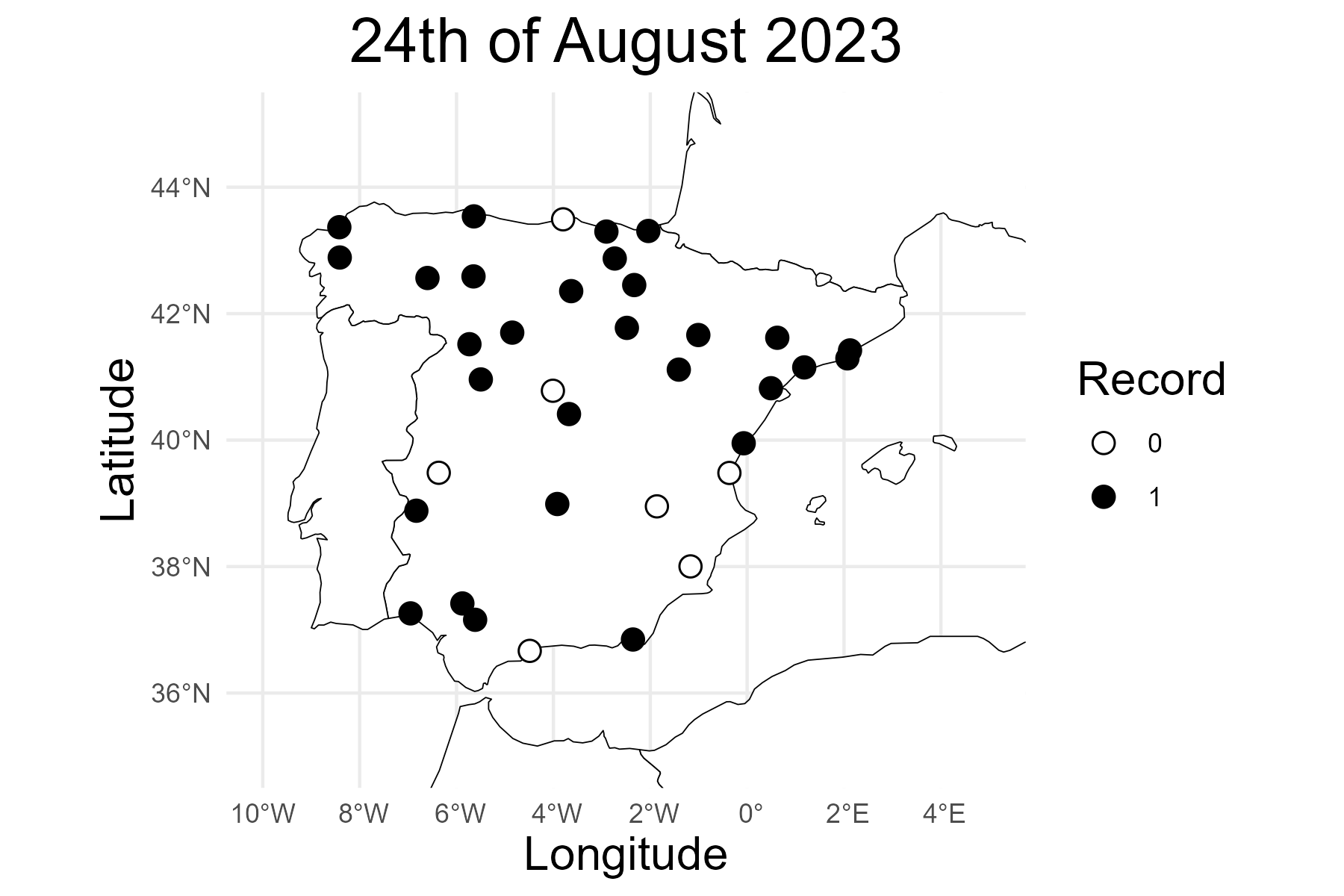}
    \includegraphics[width=0.45\textwidth]{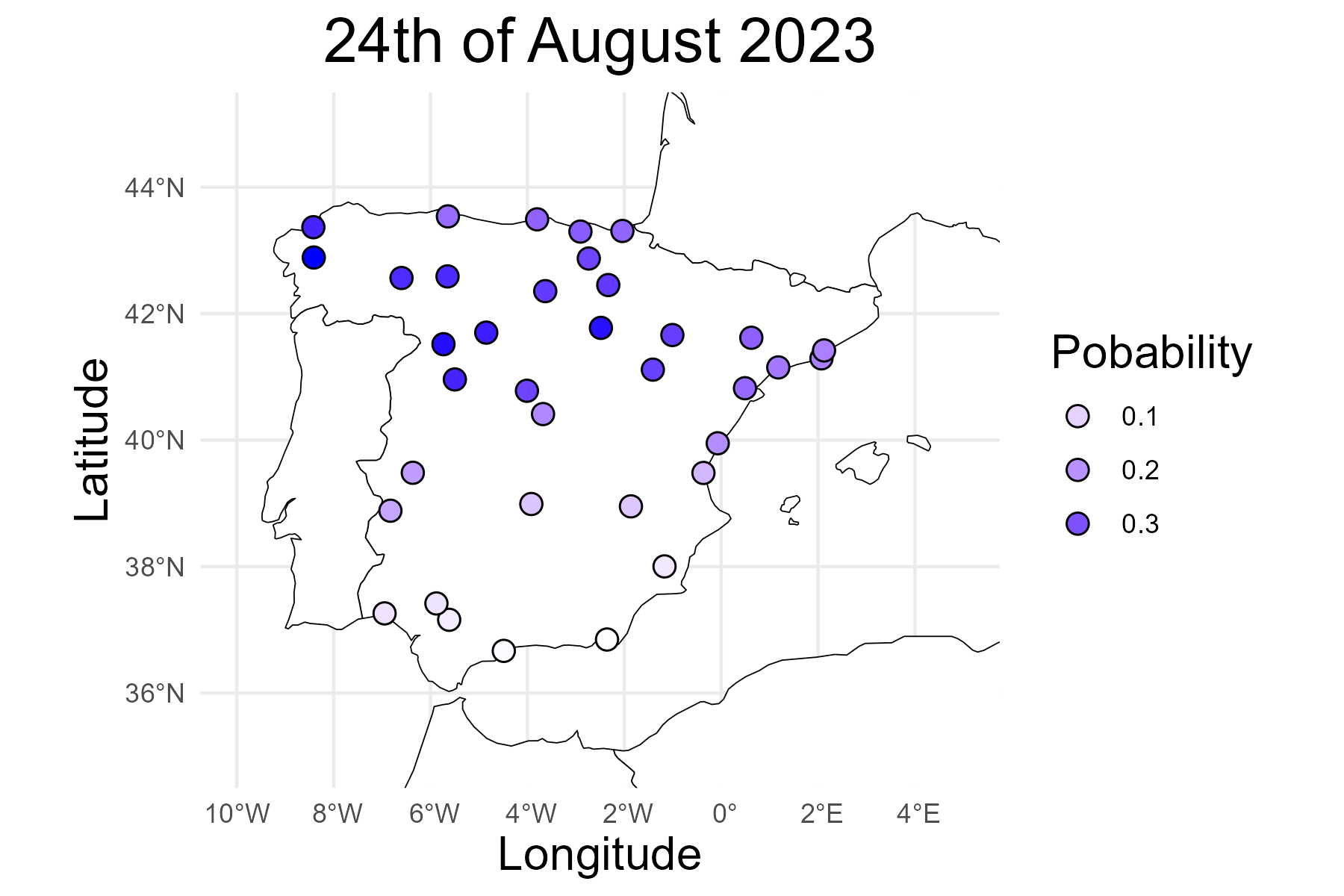}
    \caption{Spatial extreme heat event during the August 24th of 2023. Top: scatterplot with the M2 predict for each station for that day. Dashed line represents the threshold at the true negative rate of 0.95 form the M2 ROC curve. Bottom left: observed records (black) and non-records (white) for that day. Bottom right: M2 predict probabilities for that day.}
    \label{fig:spatial:ehe}
\end{figure}

Additionally, we evaluated how the selected M2 predicts extreme heat events that affects a great spatial proportion. For instance, Figure~\ref{fig:spatial:ehe} shows the M2 predictions for August 24th of 2023 which was a day where most of the stations showed a $T_x$ record. Taking as reference a true negative rate of 0.95, most of the stations with a record surpassed that prediction threshold. Also, in the right-bottom plot we can observe that the M2 predictions captures a North-South gradient, which agrees with the observed values.  

\subsection{Spatial application}

\begin{figure}[htb]
    \centering
    \includegraphics[width=0.26\linewidth]{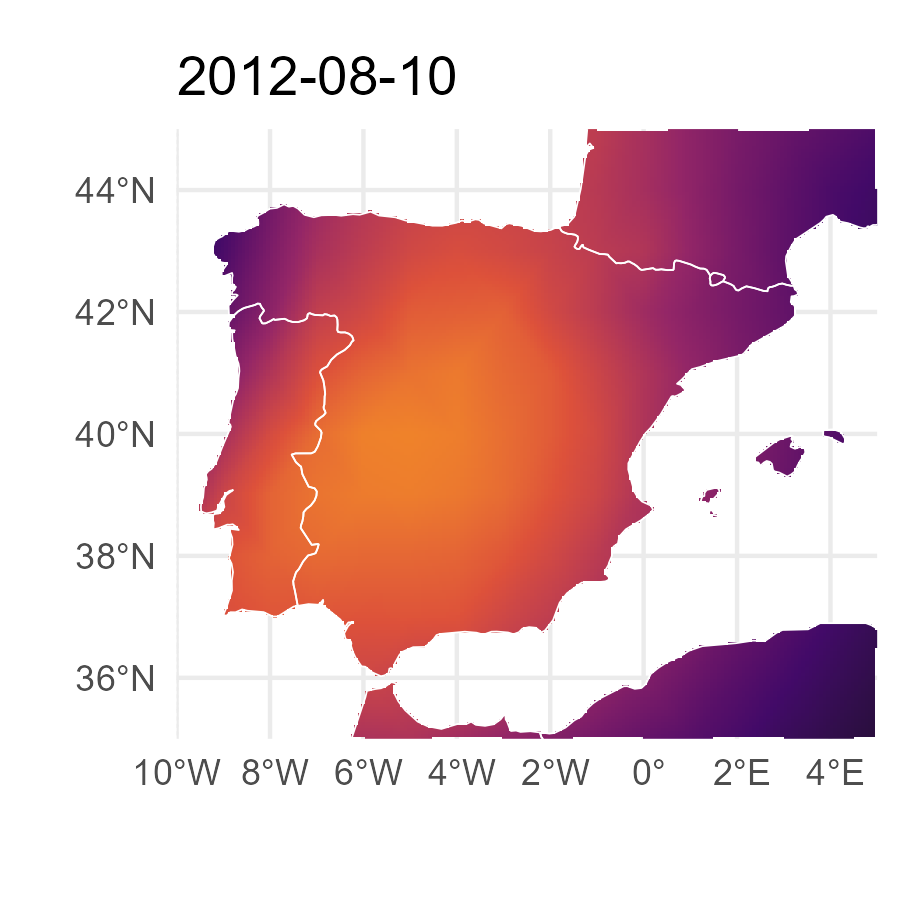}
     \includegraphics[width=0.26\linewidth]{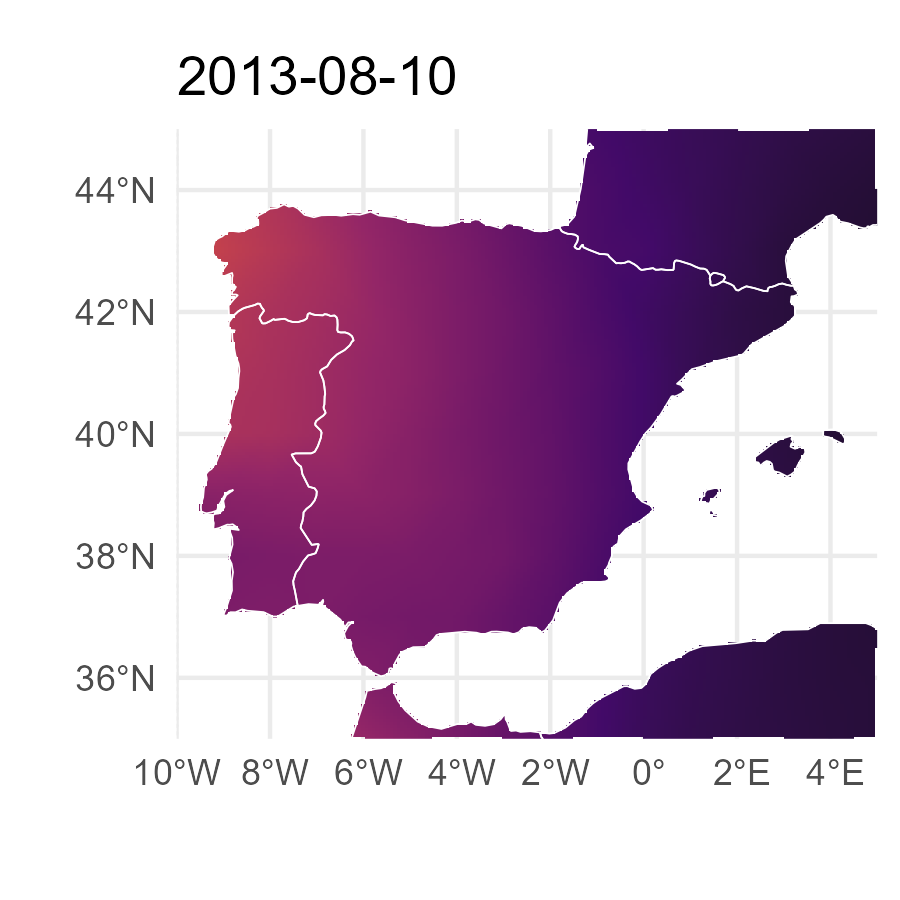}
      \includegraphics[width=0.26\linewidth]{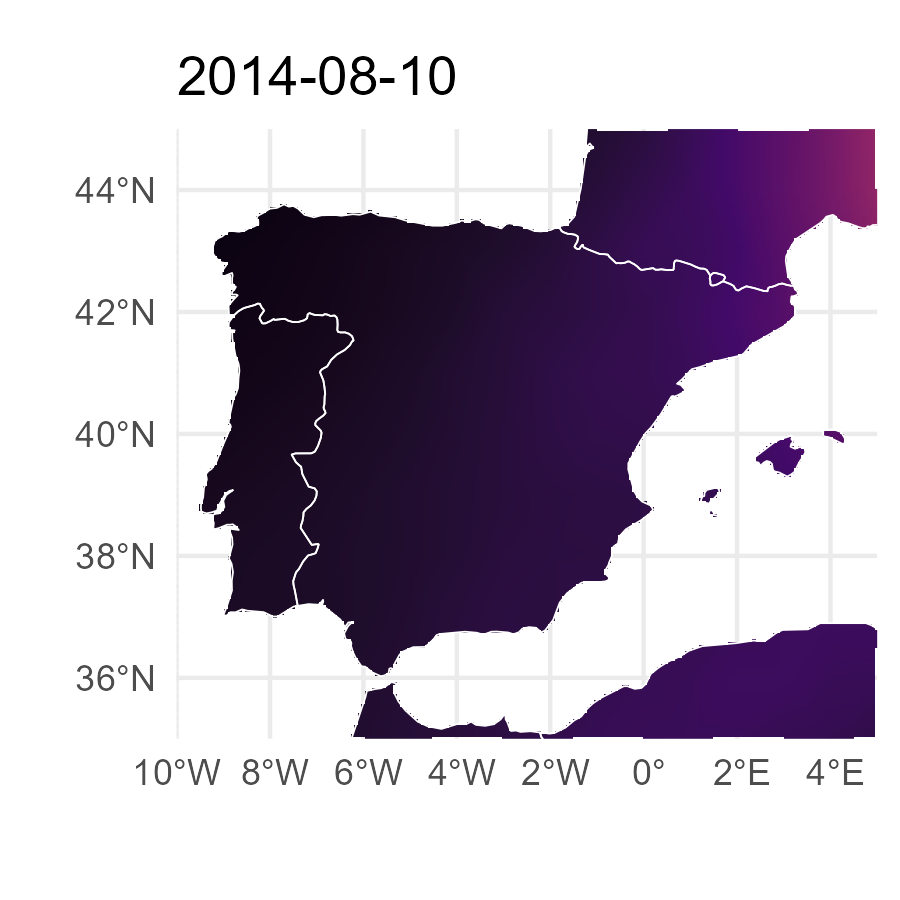}
       \includegraphics[width=0.12\linewidth]{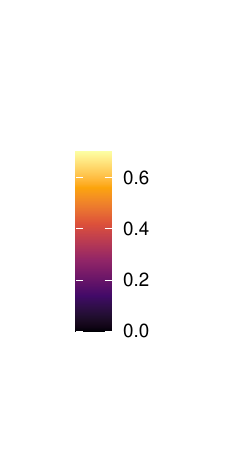}

       \includegraphics[width=0.26\linewidth]{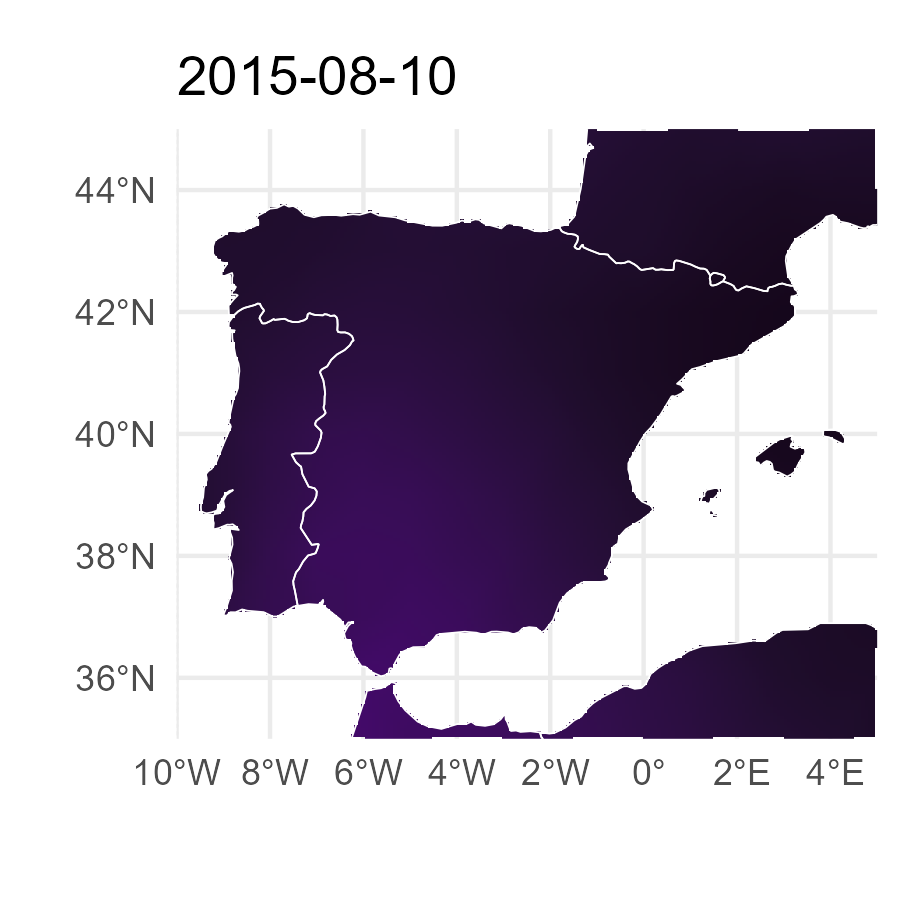}
        \includegraphics[width=0.26\linewidth]{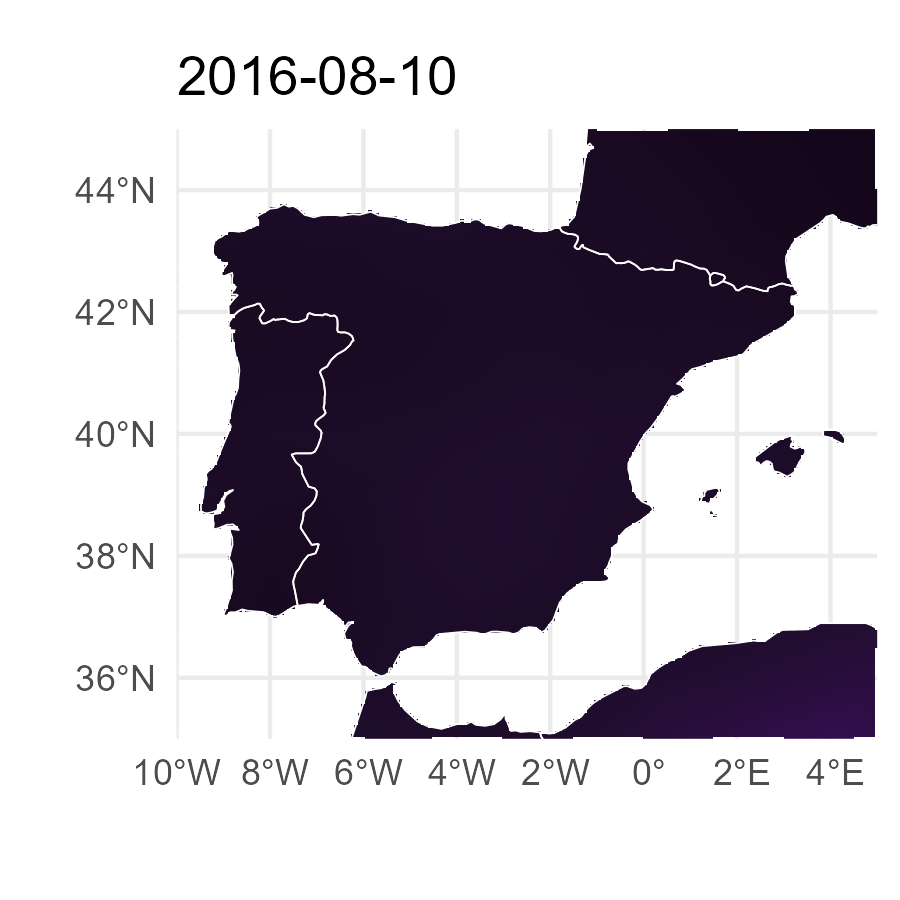}
     \includegraphics[width=0.26\linewidth]{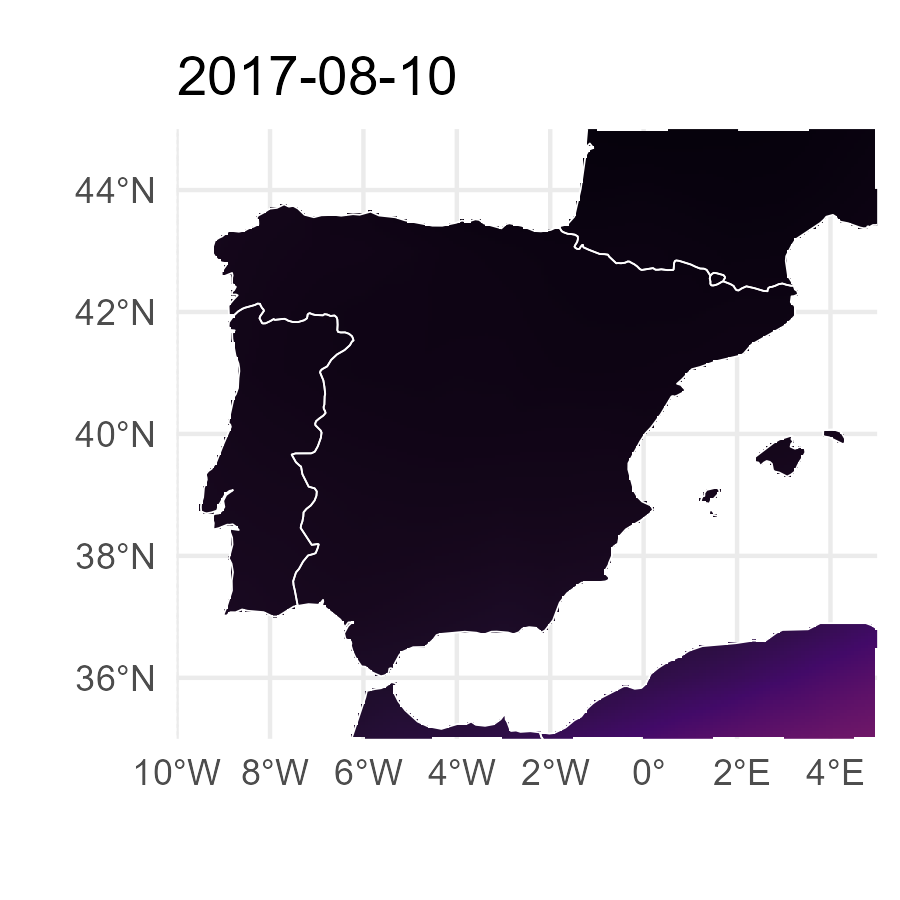}
            \includegraphics[width=0.12\linewidth]{figs/model.maps/legend.pdf}

       \includegraphics[width=0.26\linewidth]{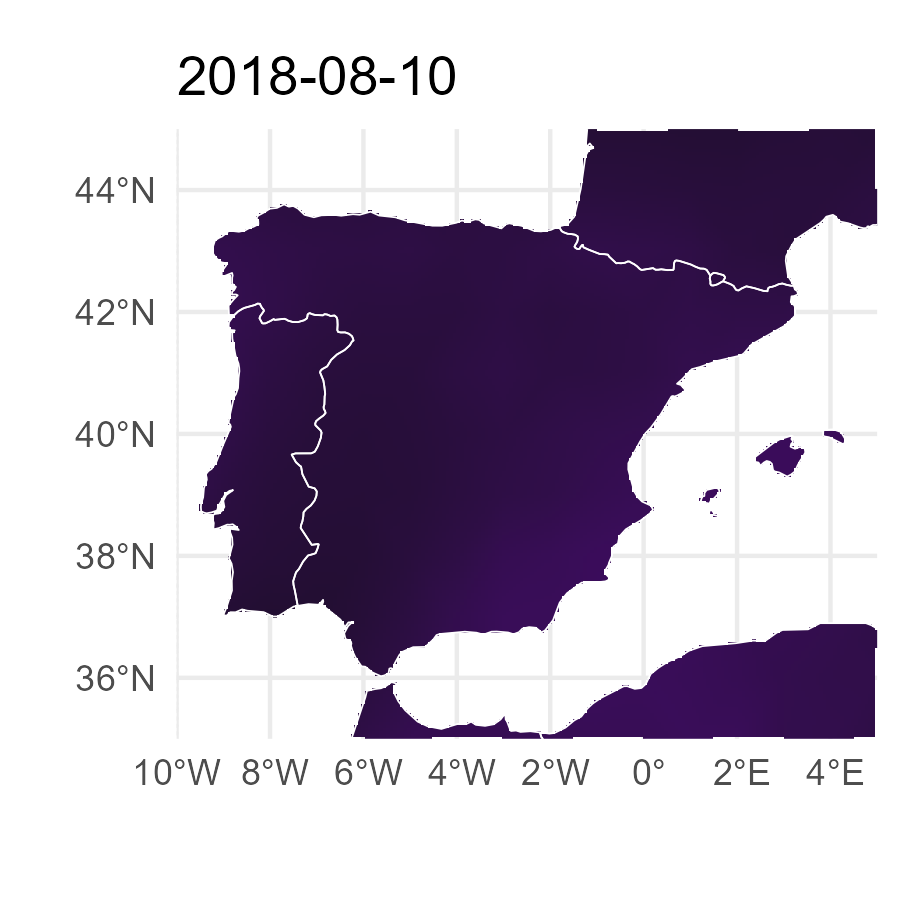}
        \includegraphics[width=0.26\linewidth]{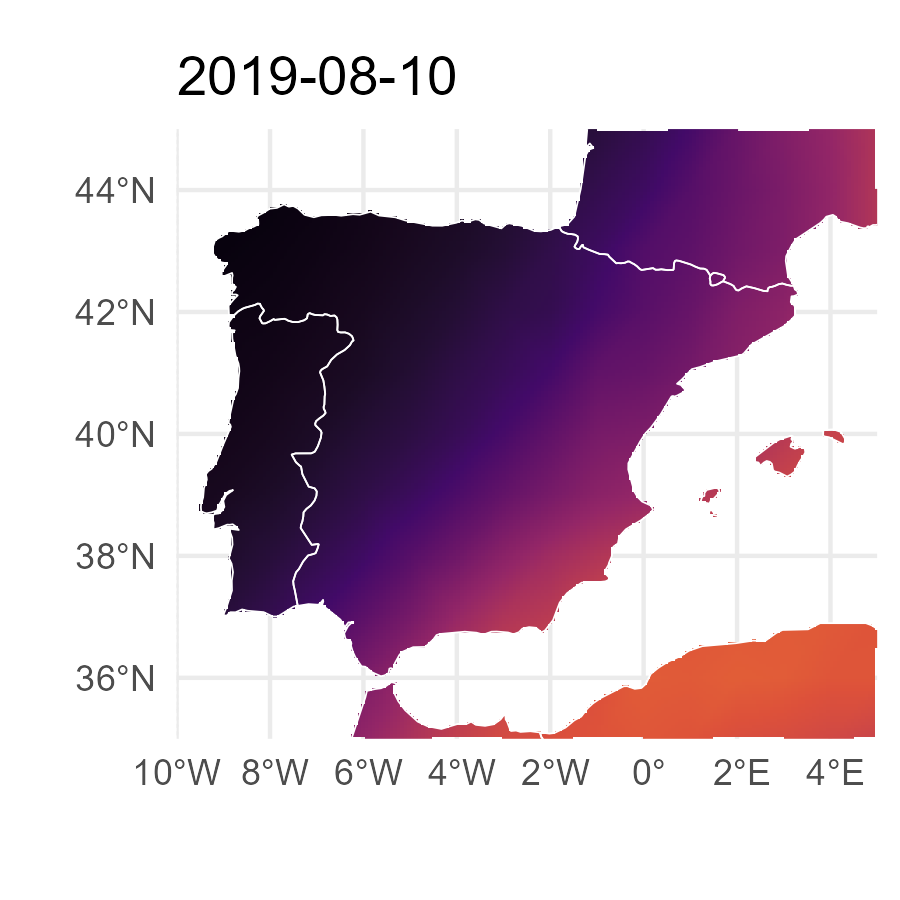}
     \includegraphics[width=0.26\linewidth]{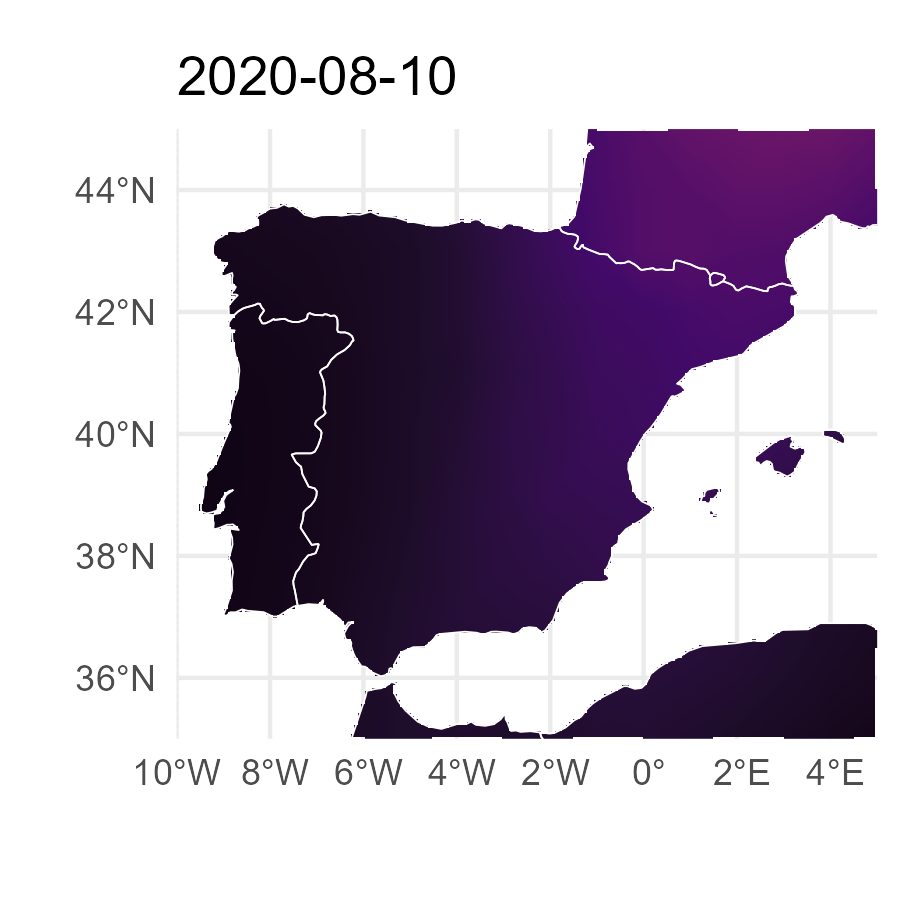}
            \includegraphics[width=0.12\linewidth]{figs/model.maps/legend.pdf}

       \includegraphics[width=0.26\linewidth]{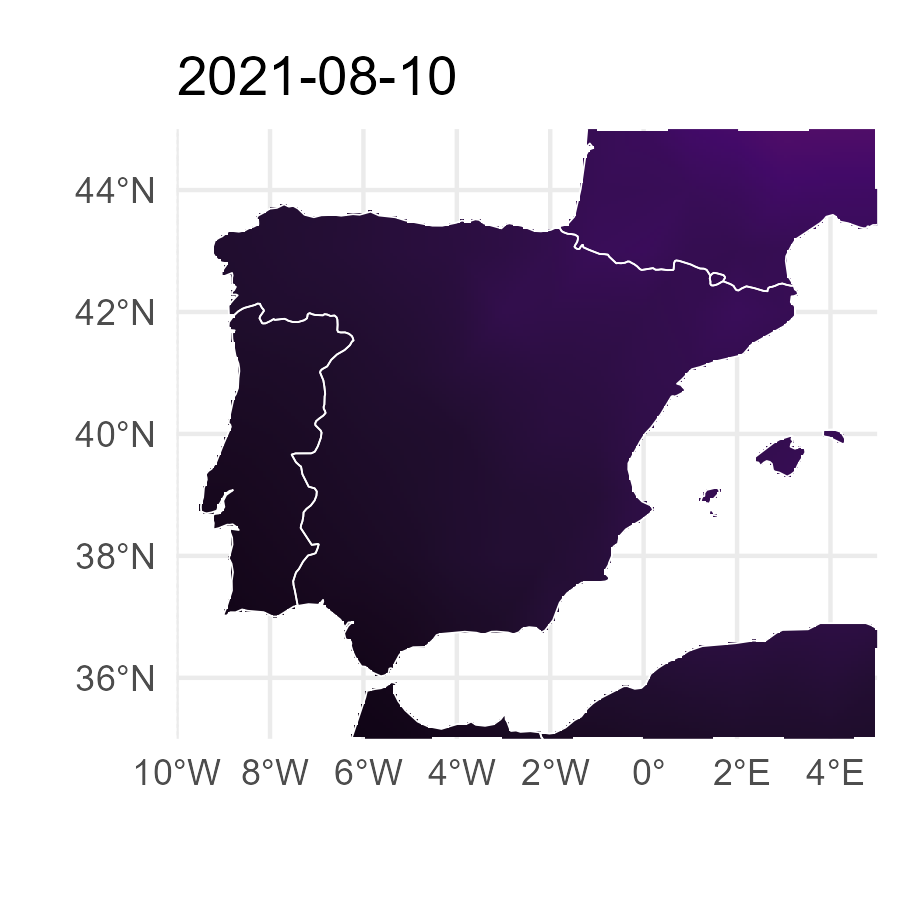}
        \includegraphics[width=0.26\linewidth]{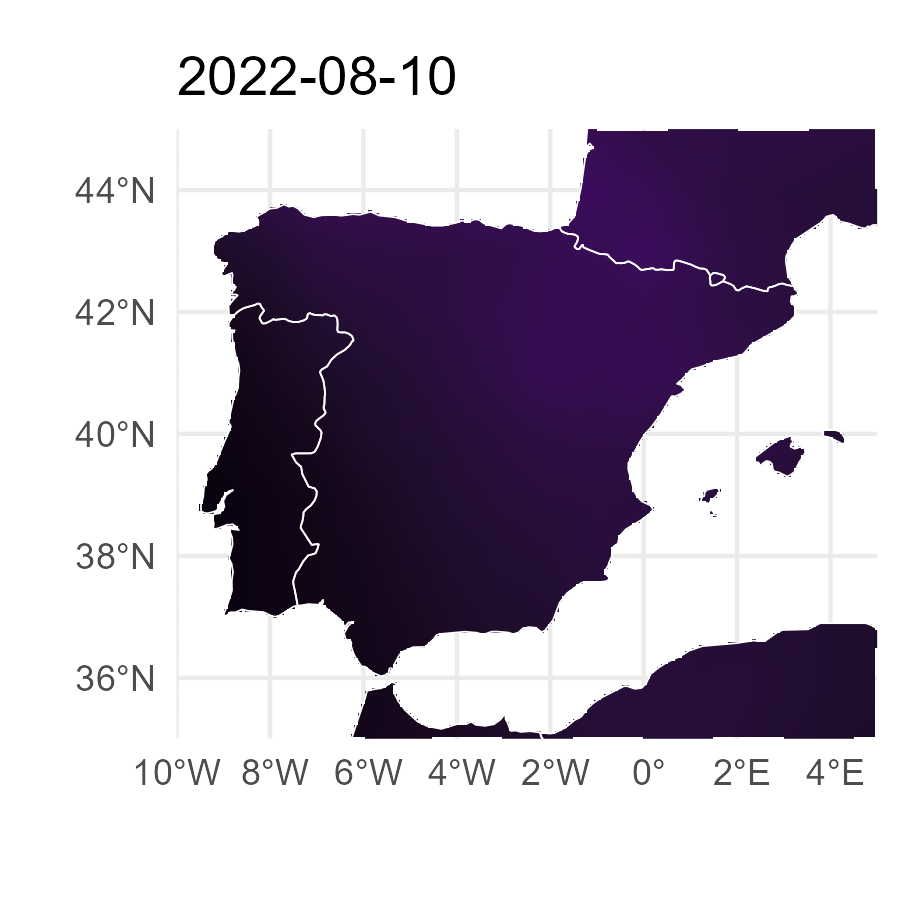}
     \includegraphics[width=0.26\linewidth]{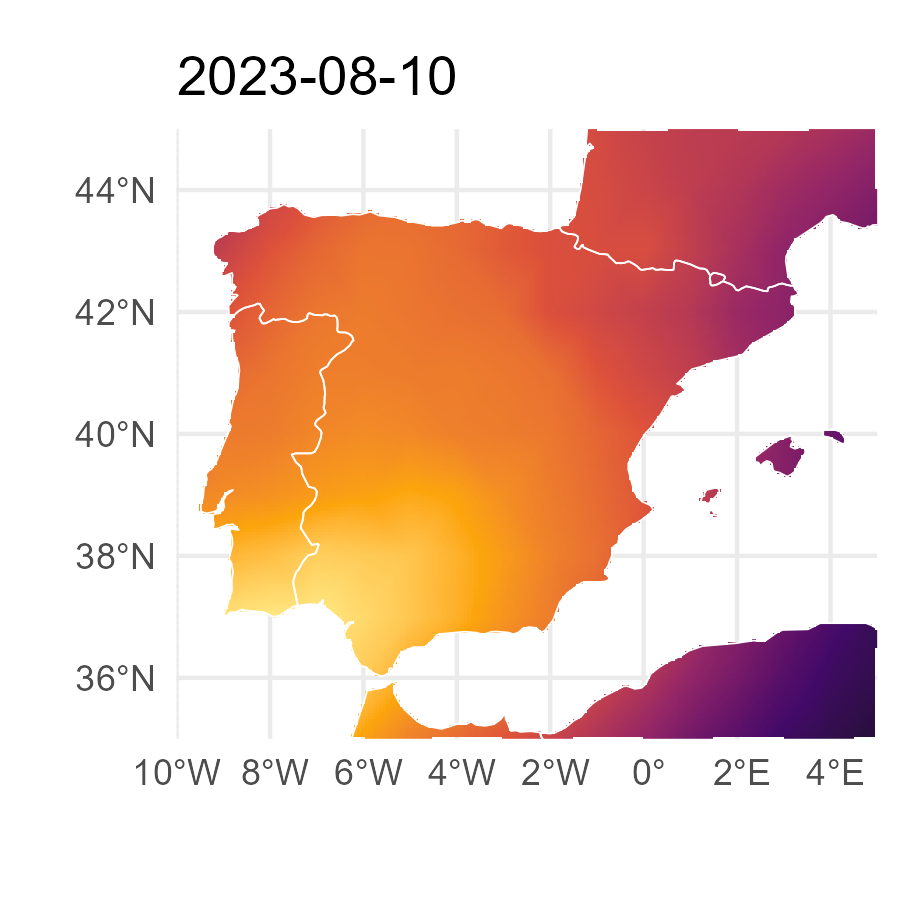}
            \includegraphics[width=0.12\linewidth]{figs/model.maps/legend.pdf}
    \caption{Predicted probabilities by model M2 of maximum temperature record occurrences on August 10 for the years 2012–2023. Each map shows the spatial distribution of probabilities across the grid covering the Iberian Peninsula.}
    \label{fig:model:spatial}
\end{figure}

The predicted probabilities of record-breaking events generated by the selected model are analyzed spatially across a grid covering the Iberian Peninsula, focusing on specific dates across multiple years. To illustrate this, a sequence of maps is produced, each representing the estimated probability of a record occurring on the same calendar day in different years. These visualizations serve to highlight the model’s effectiveness as a tool for spatial prediction of extreme temperature events. As a representative example, Figure~\ref{fig:model:spatial} shows the M2 predictions for August 10, and maps are created for each year from 2012 to 2023, all within the validation period.

Figure \ref{fig:model:spatial} presents these maps, arranged chronologically from left to right and top to bottom. A striking observation is seen in the map for August 10, 2023, where the model assigns probabilities exceeding 0.6 in certain areas. Under the assumption of a stationary climate, the probability of a record in year 2023 (i.e., $t = 64$) would be expected to be approximately $p = \frac{1}{64} \approx 0.016$. The much higher values predicted by the model suggest a significant deviation from this stationary baseline.

These spatial forecasts underscore the model’s utility in identifying regions with elevated risk of record-breaking events. For instance, in 2012, the highest predicted probabilities are concentrated in the central part of the Peninsula; in 2013, they appear predominantly in the North-West; in 2019, in the South-East; and in 2023, high probabilities are widespread across the Peninsula, with the South-West showing particularly elevated values.

\FloatBarrier

\section{Conclusion}
\label{sec:discussion}

This work describes an algorithm for building a model to predict $T_x$ record events using atmospheric fields (geopotential height) and spatial covariates. The method employs a local-to-global approach to reduce the dimensionality of the atmospheric predictors, condensing several hundred variables into fewer than twenty while maintaining comparable prediction performance. The algorithm further interacts the selected geopotential predictors with spatial and climatic covariates to improve prediction accuracy, particularly in coastal zones. This methodology yielded its best performance when geopotential covariates were interacted with geodetic information derived from the latitude and longitude of the target stations. The model achieved high predictive accuracy on out-of-sample data (AUC = 0.8787), with excellent performance at inland stations (AUC = 0.9253) and fair accuracy at coastal stations (AUC = 0.7951). Additional validation confirmed the model's ability to reproduce the observed persistence and spatial dependence of record events. In summary, we have developed a methodology for constructing a global spatio-temporal model that successfully predicts $T_x$ records across the Iberian Peninsula using a parsimonious subset of predictors, thereby enhancing both performance and interpretability.



\subsection{Future work}

The proposed model could be modified in order to generate projections for records, as a statistical downscaling tool. The estimated model ability to downscaling will be analyzed, in order to use AR6 model outputs as input in the model and to establish projections over Iberian peninsula to horizon 20 years. 
This type of tools is able to produce projections for future heatwaves or extreme temperatures in Spanish cities \citep{abaurrea2007modeling,abaurrea2018modelling}.


\newpage
\printbibliography{}

@article{abaurrea2007modeling,
  title={Modeling and forecasting extreme hot events in the central Ebro valley, a continental-Mediterranean area},
  author={Abaurrea, J and As{\'\i}n, J and Cebri{\'a}n, AC and Centelles, A},
  journal={Global and Planetary Change},
  volume={57},
  number={1-2},
  pages={43--58},
  year={2007},
  publisher={Elsevier}
}

@article{abaurrea2018modelling,
  title={Modelling the occurrence of heat waves in maximum and minimum temperatures over Spain and projections for the period 2031-60},
  author={Abaurrea, Jes{\'u}s and As{\'\i}n, J and Cebri{\'a}n, AC},
  journal={Global and planetary change},
  volume={161},
  pages={244--260},
  year={2018},
  publisher={Elsevier}
}

@article{capozzi2025,
  title={Changes in large-scale circulation behind the increase in extreme heat events in the Apennines (Italy)},
  author={Capozzi, Vincenzo and Di Bernardino, Annalisa and Budillon, Giorgio},
  journal={Atmospheric Research},
  volume={319},
  pages={108013},
  year={2025}
}

@article{castillo2023recordtest,
  title={RecordTest: An R Package to Analyze Non-Stationarity in the Extremes Based on Record-Breaking Events},
  author={Castillo-Mateo, Jorge and Cebri{\'a}n, Ana C and As{\'\i}n, Jes{\'u}s},
  journal={Journal of Statistical Software},
  volume={106},
  pages={1--28},
  year={2023}
}

@article{castillo2023statistical,
  title={Statistical analysis of extreme and record-breaking daily maximum temperatures in peninsular Spain during 1960--2021},
  author={Castillo-Mateo, Jorge and Cebri{\'a}n, Ana C and As{\'\i}n, Jes{\'u}s},
  journal={Atmospheric Research},
  pages={106934},
  year={2023},
  publisher={Elsevier}
}

@article{castillo2024spatio,
  title={Spatio-temporal modeling for record-breaking temperature events in Spain},
  author={Castillo-Mateo, Jorge and Gelfand, Alan E and Gracia-Tabuenca, Zeus and As{\'\i}n, Jes{\'u}s and Cebri{\'a}n, Ana C},
  journal={arXiv preprint arXiv:2403.00080},
  year={2024}
}

@article{cebrian2022record,
  title={Record tests to detect non-stationarity in the tails with an application to climate change},
  author={Cebri{\'a}n, Ana C and Castillo-Mateo, Jorge and As{\'\i}n, Jes{\'u}s},
  journal={Stochastic Environmental Research and Risk Assessment},
  volume={36},
  number={2},
  pages={313--330},
  year={2022},
  publisher={Springer}
}

@article{di2020contribution,
  title={Contribution of mean climate to hot temperature extremes for present and future climates},
  author={Di Luca, Alejandro and de El{\'\i}a, Ram{\'o}n and Bador, Margot and Arg{\"u}eso, Daniel},
  journal={Weather and Climate Extremes},
  volume={28},
  pages={100255},
  year={2020},
  publisher={Elsevier}
}

@article{dupuy2020climate,
  title={Climate change impact on future wildfire danger and activity in southern Europe: a review},
  author={Dupuy, Jean-luc and Fargeon, H{\'e}l{\`e}ne and Martin-StPaul, Nicolas and Pimont, Fran{\c{c}}ois and Ruffault, Julien and Guijarro, Mercedes and Hernando, Carmen and Madrigal, Javier and Fernandes, Paulo},
  journal={Annals of Forest Science},
  volume={77},
  pages={1--24},
  year={2020},
  publisher={Springer}
}

@article{garcia2015attributing,
  title={Attributing trends in extremely hot days to changes in atmospheric dynamics},
  author={Garc{\'\i}a-Valero, Juan Andr{\'e}s and Mont{\'a}vez, JP and G{\'o}mez-Navarro, JJ and Jim{\'e}nez-Guerrero, Pedro},
  journal={Natural Hazards and Earth System Sciences},
  volume={15},
  number={9},
  pages={2143--2159},
  year={2015},
  publisher={Copernicus GmbH}
}

@article{gonzalez2019potential,
  title={Potential increase in hazard from Mediterranean hurricane activity with global warming},
  author={Gonz{\'a}lez-Alem{\'a}n, Juan J and Pascale, Salvatore and Gutierrez-Fernandez, Jes{\'u}s and Murakami, Hiroyuki and Gaertner, Miguel A and Vecchi, Gabriel A},
  journal={Geophysical Research Letters},
  volume={46},
  number={3},
  pages={1754--1764},
  year={2019},
  publisher={Wiley Online Library}
}

@article{huang2025,
  title={Two oceanic origins for the record-breaking extreme high temperature event over northern China in October 2023},
  author={Huang, Hongjie and Zhu, Zhiwei and Ma, Qianrong and Hu, Shujuan and Wang, Lijuan},
  journal={Climate Dynamics},
  volume={63},
  number={3},
  pages={1--16},
  year={2025}
}

@article{klok2009updated,
  title={Updated and extended European dataset of daily climate observations},
  author={Klok, EJ and Klein Tank, AMG},
  journal={International Journal of Climatology: A Journal of the Royal Meteorological Society},
  volume={29},
  number={8},
  pages={1182--1191},
  year={2009},
  publisher={Wiley Online Library}
}

@article{qian2024rapid,
  title={Rapid attribution of the record-breaking heatwave event in North China in June 2023 and future risks},
  author={Qian, Cheng and Ye, Yangbo and Jiang, Jiacheng and Zhong, Yangyang and Zhang, Yuting and Pinto, Izidine and Huang, Cunrui and Li, Sihan and Wei, Ke},
  journal={Environmental Research Letters},
  volume={19},
  number={1},
  pages={014028},
  year={2024},
  publisher={IOP Publishing}
}

@article{ryu2023teleconnections,
  title={Teleconnections link to summer heat extremes in the south-central US: Insights from CMIP5 and CMIP6 simulations},
  author={Ryu, Jung-Hee and Kang, Song-Lak},
  journal={Weather and Climate Extremes},
  volume={42},
  pages={100635},
  year={2023},
  publisher={Elsevier}
}

@article{twardosz2021warming,
  title={Warming in Europe: recent trends in annual and seasonal temperatures},
  author={Twardosz, Robert and Walanus, Adam and Guzik, Izabela},
  journal={Pure and Applied Geophysics},
  volume={178},
  number={10},
  pages={4021--4032},
  year={2021},
  publisher={Springer}
}

@online{ESOTC2019,
    author = "ESOTC",
    title = "European State of the Climate 2019",
    year={2019},
    url  = "https://climate.copernicus.eu/ESOTC/2019",
    doi = "https://doi.org/10.24381/zw9t-hx58"
}

@online{aemet2023,
    author = "AEMET",
    title = "Informe anual 2022",
    year={2023},
    url  = "http://hdl.handle.net/20.500.11765/15937"
}

@Book{MASS4,
    title = {Modern Applied Statistics with S},
    author = {W. N. Venables and B. D. Ripley},
    publisher = {Springer},
    edition = {Fourth},
    address = {New York},
    year = {2002},
    note = {ISBN 0-387-95457-0},
    url = {https://www.stats.ox.ac.uk/pub/MASS4/},
}

@article{buntgen2024recent,
  title={Recent summer warming over the western Mediterranean region is unprecedented since medieval times},
  author={B{\"u}ntgen, Ulf and Reinig, Frederick and Verstege, Anne and Piermattei, Alma and Kunz, Marcel and Krusic, Paul and Slavin, Philip and {\v{S}}t{\v{e}}p{\'a}nek, Petr and Torbenson, Max and del Castillo, Edurne Martinez and others},
  journal={Global and Planetary Change},
  volume={232},
  pages={104336},
  year={2024},
  publisher={Elsevier},
  doi = "https://doi.org/10.1016/j.gloplacha.2023.104336"
}

@article{roye2021effects,
  title={Effects of hot nights on mortality in Southern Europe},
  author={Roy{\'e}, Dominic and Sera, Francesco and Tob{\'\i}as, Aurelio and Lowe, Rachel and Gasparrini, Antonio and Pascal, Mathilde and de’Donato, Francesca and Nunes, Baltazar and Teixeira, Joao Paulo},
  journal={Epidemiology},
  volume={32},
  number={4},
  pages={487--498},
  year={2021},
  publisher={LWW},
  doi = "https://doi.org/10.1097/EDE.0000000000001359"
}

@article{breshears2021underappreciated,
  title={Underappreciated plant vulnerabilities to heat waves},
  author={Breshears, David D and Fontaine, Joseph B and Ruthrof, Katinka X and Field, Jason P and Feng, Xiao and Burger, Joseph R and Law, Darin J and Kala, Jatin and Hardy, Giles E St J},
  journal={New Phytologist},
  volume={231},
  number={1},
  pages={32--39},
  year={2021},
  publisher={Wiley Online Library},
  doi = "https://doi.org/10.1111/nph.17348"
}

@misc{hersbach2023era5,
  title={ERA5 hourly data on single levels from 1940 to present, Copernicus Climate Change Service (C3S) Climate Data Store (CDS)[data set]},
  author={Hersbach, Hans and Bell, Bill and Berrisford, Paul and Biavati, Gionata and Hor{\'a}nyi, Andr{\'a}s and Mu{\~n}oz Sabater, Joaqu{\'\i}n and Nicolas, Julien and Peubey, Carole and Radu, Raluca and Rozum, Iryna and others},
  year={2023},
  doi={https://doi.org/10.24381/cds.adbb2d47}
}

@article{om2022,
  title={Observed trends in extreme temperature events over northern part of the Korean Peninsula during 1960--2019 and a comparative overview},
  author={Om, Kum-Chol and Ren, Guoyu and Kim, Kwang-Hyon and Pak, Yon-I and Jong, Sang-Il and Kil, Hyon-Nam},
  journal={Atmospheric Research},
  volume={270},
  pages={106061},
  year={2022}
}

@article{paredes2023understanding,
  title={Understanding the Magnification of Heatwaves over Spain: Relevant changes in the most extreme events},
  author={Paredes-Fortuny, L and Khodayar, S},
  journal={Weather and Climate Extremes},
  volume={42},
  pages={100631},
  year={2023},
  publisher={Elsevier},
  doi={https://doi.org/10.1016/j.wace.2023.100631}
}

@article{saddique2020,
  title={Trends in temperature and precipitation extremes in historical (1961--1990) and projected (2061--2090) periods in a data scarce mountain basin, northern Pakistan},
  author={Saddique, Naeem and Khaliq, Abdul and Bernhofer, Christian},
  journal={Stochastic Environmental Research and Risk Assessment},
  volume={34},
  pages={1441--1455},
  year={2020}
}

@article{zhang2025,
  title={Attribution of historical extreme heat events in different climate zones of China},
  author={Zhang, Yuxia and Sun, Ying},
  journal={Environmental Research Letters},
  year={2025}
}

\newpage

\begin{flushleft}
{\Large
\textbf\newline{Supplememtary information:\\Prediction of Maximum Temperature record in Spain 1960-2023 by the means of ERA5 atmospheric geopotentials}
}
\newline
\\
Elsa Barrio 1\textsuperscript{1,*},
Jesús Abaurrea \textsuperscript{1},
Jesús Asín \textsuperscript{1},
Jorge Castillo-Mateo \textsuperscript{1},
Ana Carmen Cebrián \textsuperscript{1},
Zeus Gracia \textsuperscript{1},
\\
\bigskip
\bf{1} Department of Statistical Methods, University of Zaragoza.
\\
\bigskip
* e.barrio@unizar.es
\end{flushleft}


\section*{Extended Exploratory data analysis}

\subsection*{Relationship between geopotential variables and daily temperature}

\begin{table}[htb]
    \centering
    \caption{$N_t$ record test z-statistic (p-values) for the geopotentials at three pressure levels (700, 500, and 300 hPa) at the four grid corners (45N10W, 45N5E, 35N10W, 35N5E) and for the $T_x$ records at its closest station.}
    \medskip
    \begin{tabular}{|l|c|c|c|c|}
        \hline
        Station \& Grid point & $T_x$ & g700 & g500 & g300\\
        \hline
        \hline
        A Coruña \& g.45N.10W & 1.4510 (0.145) & -0.912 (0.682) & -0.2031 (0.5512) & 0.3288 (0.4385)\\
        \hline
        Barcelona-Fabra \& g.45N.5E & 7.2405 (0.0027) & 2.7507 (0.0835) & 2.5144 (0.1205) & 3.1643 (0.0796)\\
        \hline
        Huelva \& g.35N.10W & 1.3329 (0.183) & 2.16 (0.1259) & 3.5187 (0.0613) & 6.2953 (0.0031)\\
        \hline
        Murcia \& g.35N.5E & 2.1009 (0.099) & 4.0504 (0.0327) & 3.4596 (0.0802) & 5.291 (0.0161)\\
        \hline
    \end{tabular}
    \label{sup:tab:nttest}
\end{table}
\subsection*{Trends of atmospheric variables}



\begin{figure}[htb]
    \centering
    \begin{subfigure}[b]{0.33\textwidth}
    \includegraphics[width=\textwidth]{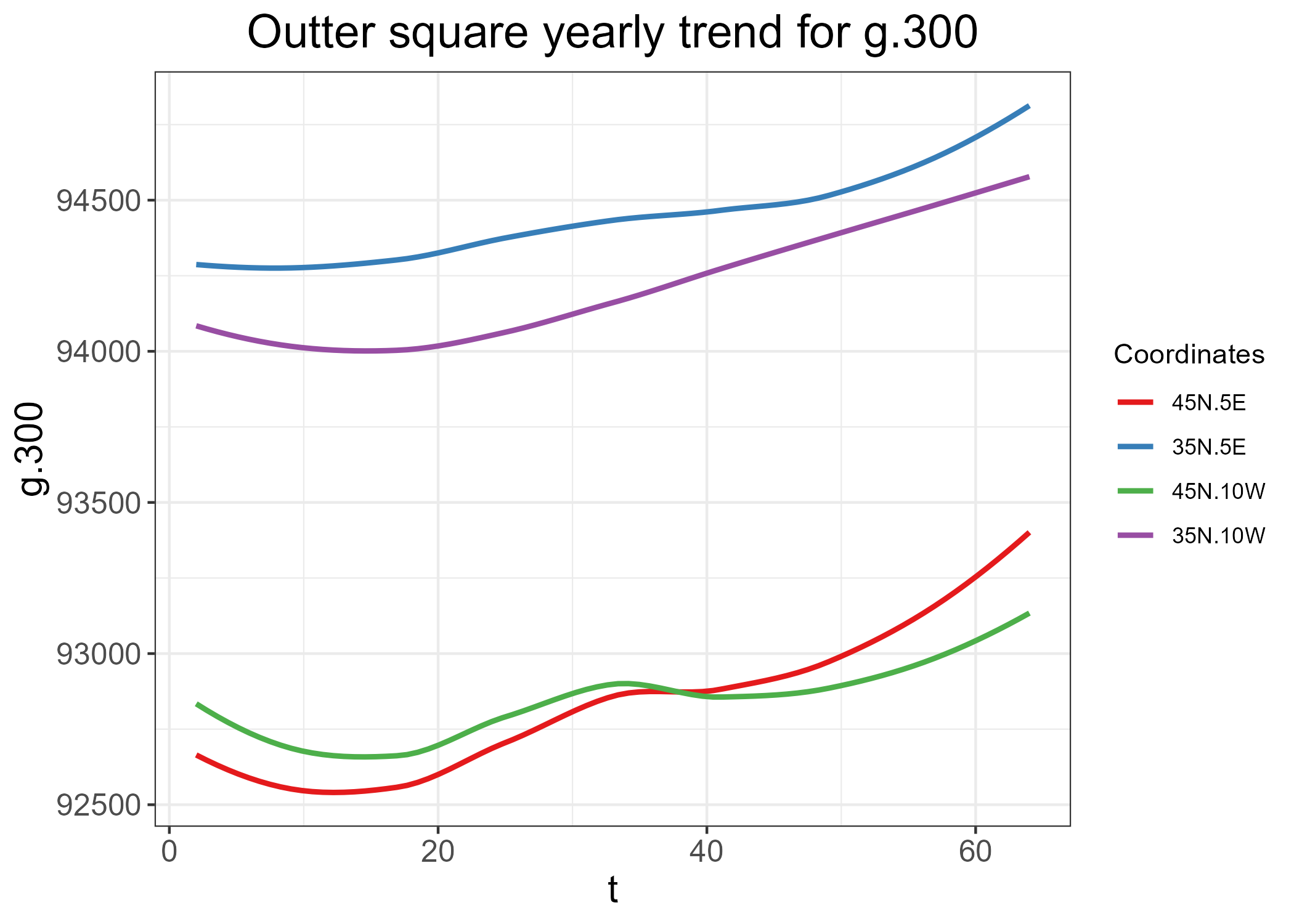}
        \caption{g.300}
        \label{fig:plot1}
    \end{subfigure}
    \begin{subfigure}[b]{0.33\textwidth}
    \includegraphics[width=\textwidth]{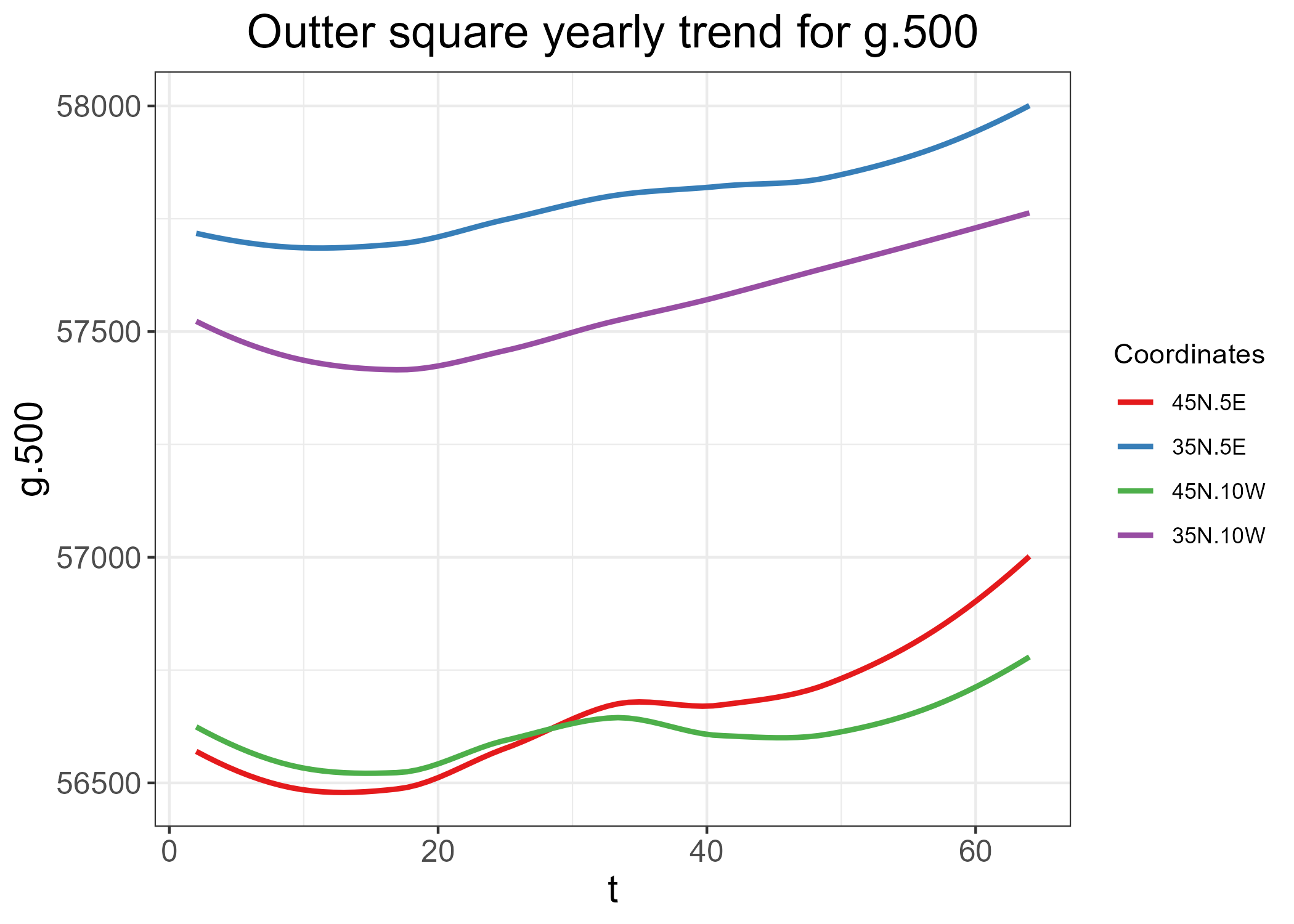}
        \caption{g.500}
        \label{fig:plot2}
    \end{subfigure}
    \begin{subfigure}[b]{0.33\textwidth}
    \includegraphics[width=\textwidth]{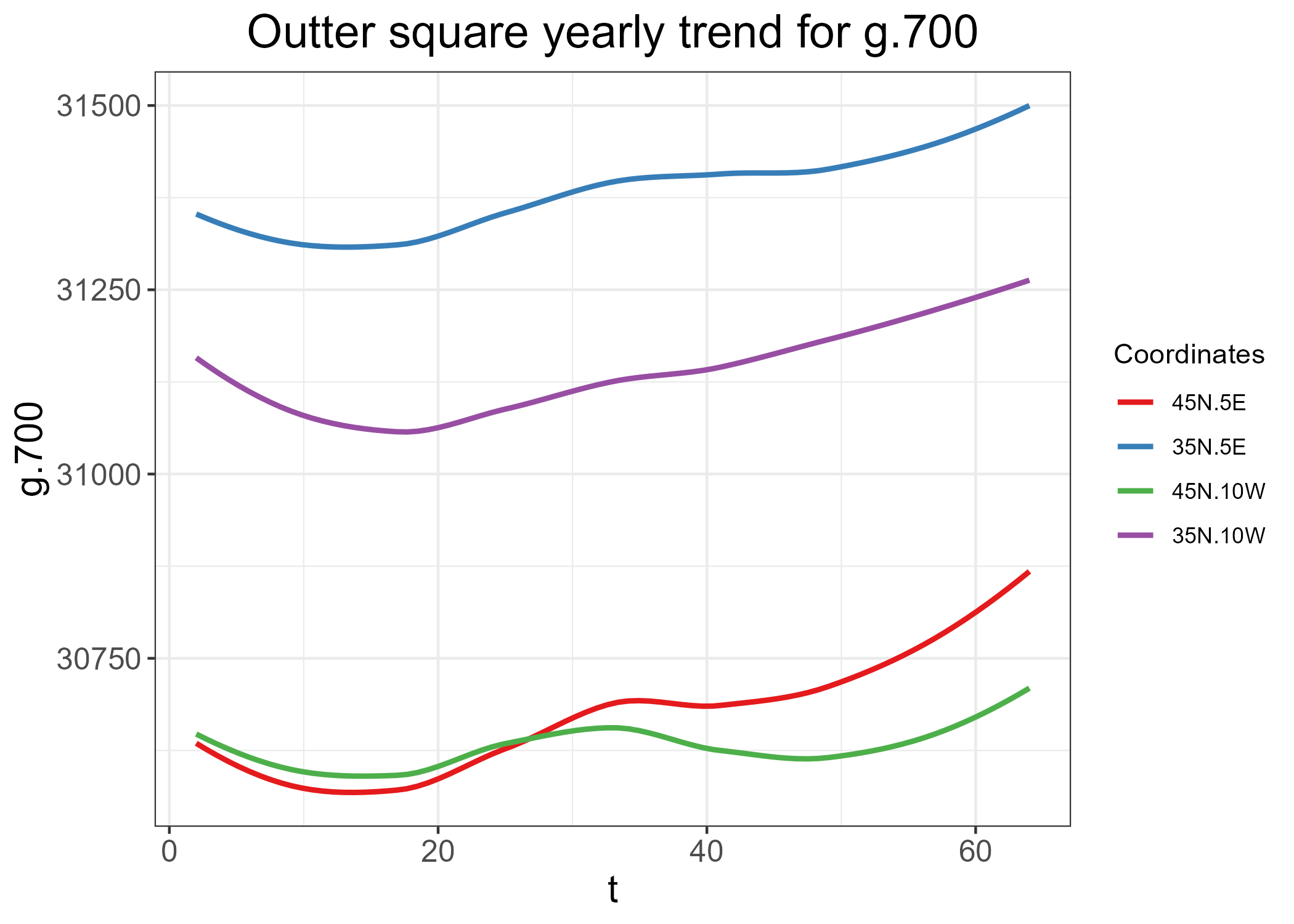}
        \caption{g.700}
        \label{fig:outter.yearly.trends}
    \end{subfigure}
    \caption{Yearly trends of Geopotential variables at pressure levels of 300hPa, 500hPa, and 700 hPa, at four fourthest grid points 45N-10W, 45N-5E, 35N-10W, and 35N-5E.}
    \label{fig:both_plots}
\end{figure}


\FloatBarrier
\clearpage
\section*{Extended Modeling Results}

\subsection*{Local models}


\begin{table}[ht]
    \caption{Number of parameters ($k$) and performance metrics ($AUC$, $AIC$) for each local model corresponding to each station. $AIC$ was computed in the train period (1960-2010) and the $AUC$ in the test period (2011-2023). Stations are sorted by distance to the coast in ascending order.}
    \medskip
    \centering
    \begin{tabular}{|l|c|c|c|}
        \hline
        Station & $k$ & $AUC$ & $AIC$ \\ 
        \hline
        Santander & 20 & 0.89 & 2033 \\ 
        San Sebastián & 17 & 0.92 & 1881 \\ 
        Almería & 21 & 0.87 & 2325 \\ 
        A Coruña & 25 & 0.82 & 2041 \\ 
        Barcelona-Airport & 28 & 0.84 & 2352 \\ 
        Málaga & 27 & 0.89 & 1961 \\ 
        Gijón & 16 & 0.67 & 2160 \\ 
        Huelva & 22 & 0.90 & 1873 \\ 
        Valencia & 25 & 0.86 & 2227 \\ 
        Reus & 21 & 0.81 & 2208 \\ 
        Castellón & 29 & 0.86 & 2564 \\ 
        Barcelona-Fabra & 20 & 0.86 & 2365 \\ 
        Bilbao & 18 & 0.92 & 1967 \\ 
        Tortosa & 22 & 0.83 & 2679 \\ 
        Santiago & 15 & 0.93 & 1962 \\ 
        Murcia & 26 & 0.89 & 1914 \\ 
        Vitoria & 25 & 0.95 & 1871 \\ 
        Moron & 19 & 0.95 & 1902 \\ 
        Sevilla & 27 & 0.93 & 1711 \\ 
        Lleida & 22 & 0.91 & 1976 \\ 
        Logroño & 26 & 0.93 & 1950 \\ 
        León & 23 & 0.94 & 1791 \\ 
        Ponferrada & 15 & 0.95 & 1803 \\ 
        Burgos & 22 & 0.96 & 2011 \\ 
        Albacete & 17 & 0.92 & 1878 \\ 
        Badajoz & 20 & 0.94 & 1971 \\ 
        Soria & 34 & 0.94 & 1748 \\ 
        Daroca & 23 & 0.94 & 1859 \\ 
        Zaragoza & 28 & 0.90 & 2001 \\ 
        Valladolid & 22 & 0.96 & 1812 \\ 
        Cáceres & 29 & 0.95 & 1790 \\ 
        Zamora & 23 & 0.95 & 1871 \\ 
        Ciudad Real & 34 & 0.96 & 1763 \\ 
        Salamanca & 25 & 0.94 & 1820 \\ 
        Navacerrada & 26 & 0.95 & 1946 \\ 
        Madrid & 27 & 0.94 & 1998 \\ 
        \hline
    \end{tabular}
    \label{sup:tab:local}
\end{table}


\begin{table}[htb]
\caption{Number of surviving covariates with different restrictions.}
\centering
\begin{tabular}{lrrr}
  \hline
 & $n\geq9$ & $n \geq 12$ & $n \geq 18$ \\ 
  \hline
  $z \geq 1.6$ & 37 & 30 & 19 \\ 
  $z \geq 2$ & 31 & 23 & 15 \\ 
  $z \geq 2.6$ & 19 & 16 & 8 \\ 
  $z \geq 3.2$ & 12 & 6 & 4 \\ 
   \hline
\end{tabular}
\label{sup:tab:local:thresholds}
\end{table}

\newpage
\FloatBarrier
\clearpage

\subsection*{Summary of only-atmospheric model (M1)}\label{sup:m1:summary}

\begin{verbatim}

Call:
glm(formula = m1.frm, family = binomial(link = "logit"), data = global.df[idx.train, 
    ])

Coefficients:
                        Estimate Std. Error z value Pr(>|z|)    
(Intercept)           -1.335e+02  8.634e+00 -15.467   <2e-16 ***
g700.                  1.101e-03  5.009e-05  21.988   <2e-16 ***
g700.45N.10W          -1.426e-03  6.898e-05 -20.675   <2e-16 ***
poly(g300.35N.5E, 2)1 -1.818e+02  1.677e+01 -10.842   <2e-16 ***
poly(g300.35N.5E, 2)2  1.301e+02  5.389e+00  24.148   <2e-16 ***
g700..lag1             1.838e-03  8.980e-05  20.472   <2e-16 ***
g300.35N.10W.lag1     -2.238e-04  2.291e-05  -9.769   <2e-16 ***
poly(g700.35N.5E, 2)1 -3.142e+01  1.811e+01  -1.735   0.0827 .  
poly(g700.35N.5E, 2)2 -1.358e+02  6.858e+00 -19.799   <2e-16 ***
g500.45N.10W           6.964e-04  4.840e-05  14.389   <2e-16 ***
g300.35N.10W          -9.787e-04  4.158e-05 -23.534   <2e-16 ***
poly(g700.45N.5E, 2)1 -3.675e+02  1.818e+01 -20.211   <2e-16 ***
poly(g700.45N.5E, 2)2  7.976e+01  4.065e+00  19.619   <2e-16 ***
g500.45N.5E            1.186e-03  5.736e-05  20.677   <2e-16 ***
g300.35N.5E.lag1      -3.767e-04  2.976e-05 -12.656   <2e-16 ***
g500.35N.10W           1.551e-03  6.273e-05  24.730   <2e-16 ***
g500..lag1            -7.235e-04  6.176e-05 -11.715   <2e-16 ***
g500.35N.5E            1.353e-03  1.359e-04   9.956   <2e-16 ***
---
Signif. codes:  0 ‘***’ 0.001 ‘**’ 0.01 ‘*’ 0.05 ‘.’ 0.1 ‘ ’ 1

(Dispersion parameter for binomial family taken to be 1)

    Null deviance: 107882  on 168911  degrees of freedom
Residual deviance:  97772  on 168894  degrees of freedom
AIC: 97808

Number of Fisher Scoring iterations: 6

\end{verbatim}

\FloatBarrier
\clearpage

\subsection*{Summary of atmospheric-geodetic model (M2)}

\begin{verbatim}

Call:
glm(formula = m2.frm, family = binomial(link = "logit"), data = global.df[idx.train, 
    ])

Coefficients:
                            Estimate Std. Error z value Pr(>|z|)    
(Intercept)               -4.854e+02  5.869e+01  -8.270  < 2e-16 ***
g700.                      1.960e-03  8.825e-05  22.204  < 2e-16 ***
g700.45N.10W               1.866e-02  1.372e-03  13.606  < 2e-16 ***
LAT                        7.824e+00  1.423e+00   5.498 3.84e-08 ***
poly(g300.35N.5E, 2)1     -2.044e+02  1.695e+01 -12.062  < 2e-16 ***
poly(g300.35N.5E, 2)2      1.273e+02  5.469e+00  23.278  < 2e-16 ***
g700..lag1                 1.928e-03  1.190e-04  16.209  < 2e-16 ***
g300.35N.10W.lag1         -2.440e-04  2.354e-05 -10.367  < 2e-16 ***
poly(g700.35N.5E, 2)1     -6.168e+01  1.860e+01  -3.316 0.000913 ***
poly(g700.35N.5E, 2)2     -1.281e+02  6.933e+00 -18.470  < 2e-16 ***
poly(g700.45N.5E, 2)1     -3.372e+03  3.349e+02 -10.067  < 2e-16 ***
poly(g700.45N.5E, 2)2     -1.478e+02  7.839e+01  -1.885 0.059435 .  
g500.45N.5E                5.405e-03  1.129e-03   4.788 1.68e-06 ***
g300.35N.5E.lag1          -3.207e-04  3.035e-05 -10.567  < 2e-16 ***
g500.35N.5E               -2.162e-03  5.093e-04  -4.246 2.17e-05 ***
g300.35N.10W              -6.987e-04  4.595e-05 -15.206  < 2e-16 ***
g500.45N.10W              -4.946e-03  9.753e-04  -5.071 3.96e-07 ***
g500.35N.10W               1.071e-03  6.564e-05  16.319  < 2e-16 ***
LON                       -9.311e+00  1.074e+00  -8.667  < 2e-16 ***
g700.45N.10W.lag1         -6.364e-04  8.876e-05  -7.170 7.48e-13 ***
g500..lag1                -8.183e-04  7.556e-05 -10.830  < 2e-16 ***
g500.45N.10W.lag1         -3.435e-03  4.198e-04  -8.182 2.79e-16 ***
g500.45N.5E.lag1           3.703e-03  4.080e-04   9.077  < 2e-16 ***
g700.45N.5E.lag1          -2.655e-04  1.015e-04  -2.616 0.008889 ** 
g700.45N.10W:LAT          -4.756e-04  3.347e-05 -14.208  < 2e-16 ***
LAT:poly(g700.45N.5E, 2)1  6.913e+01  8.198e+00   8.433  < 2e-16 ***
LAT:poly(g700.45N.5E, 2)2  5.644e+00  1.919e+00   2.941 0.003268 ** 
LAT:g500.45N.10W           1.251e-04  2.382e-05   5.252 1.50e-07 ***
poly(g700.45N.5E, 2)1:LON -7.367e+01  6.321e+00 -11.655  < 2e-16 ***
poly(g700.45N.5E, 2)2:LON  5.979e+00  1.349e+00   4.431 9.38e-06 ***
g300.35N.10W:LON           3.533e-05  4.845e-06   7.292 3.06e-13 ***
g700.:LON                 -1.475e-04  1.477e-05  -9.987  < 2e-16 ***
g700.45N.10W:LON           2.233e-04  2.466e-05   9.054  < 2e-16 ***
LON:g500.45N.10W.lag1     -6.900e-05  7.356e-06  -9.380  < 2e-16 ***
LAT:g500.45N.10W.lag1      8.997e-05  1.021e-05   8.812  < 2e-16 ***
g500.45N.5E:LON            1.318e-04  1.951e-05   6.753 1.45e-11 ***
LAT:g500.35N.5E            8.697e-05  1.200e-05   7.249 4.20e-13 ***
LAT:g500.45N.5E           -1.048e-04  2.766e-05  -3.787 0.000152 ***
LON:g700.45N.5E.lag1       1.424e-04  1.165e-05  12.217  < 2e-16 ***
LAT:g500.45N.5E.lag1      -7.714e-05  9.939e-06  -7.761 8.41e-15 ***
g500.45N.10W:LON          -7.420e-05  1.747e-05  -4.247 2.17e-05 ***
---
Signif. codes:  0 ‘***’ 0.001 ‘**’ 0.01 ‘*’ 0.05 ‘.’ 0.1 ‘ ’ 1

(Dispersion parameter for binomial family taken to be 1)

    Null deviance: 107882  on 168911  degrees of freedom
Residual deviance:  95524  on 168871  degrees of freedom
AIC: 95606

Number of Fisher Scoring iterations: 6

\end{verbatim}

\subsection*{Alternative global model}


\begin{table}[htb]

    \caption{Performance metrics of the alternative global models. M6 is a global model using the stringent stepwise regression without the Step 1, 
    M1 is also included as reference. The total $AUC$, the number of parameters ($k$), the $AUC$ for stations located less than 50 km from the coast ($AUC_{<50km}$), the $AUC$ for stations situated more than 50 km from the coast ($AUC_{>50km}$), and the Akaike Information Criterion ($AIC$) are presented. The $AUC$ values were obtained during the validation period.}
    \medskip
    \centering
    \begin{tabular}{|c|c|c|c|c|c|}
    \hline
    Model & $AUC$ & $k$ & $AUC_{<50km}$ & $AUC_{>50km}$ & $AIC$\\
    \hline
    \textbf{M1} & 0.8575 & 18 & 0.7760 & 0.9062 & 97808.39\\ 
    \textbf{M6} & 0.8561 & 34 & 0.7733 & 0.9055 & 97136.57\\ 
    \hline
    
    \end{tabular}

    \label{sup:tab:model_alt}
\end{table}

\subsection*{Supplementary Concurrence validation results}

\begin{figure}[htb]
    \centering
    \includegraphics[width=0.45\textwidth]{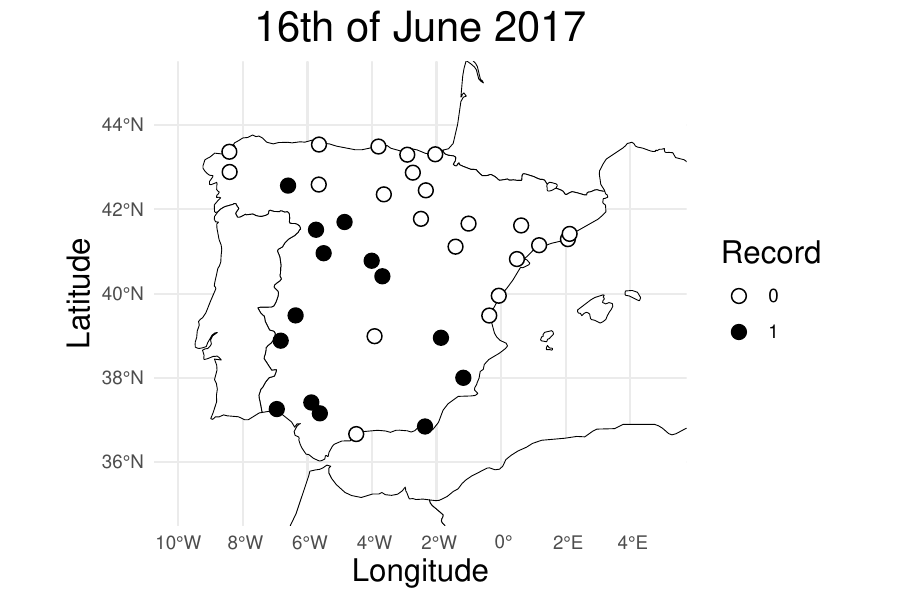}
    \includegraphics[width=0.45\textwidth]{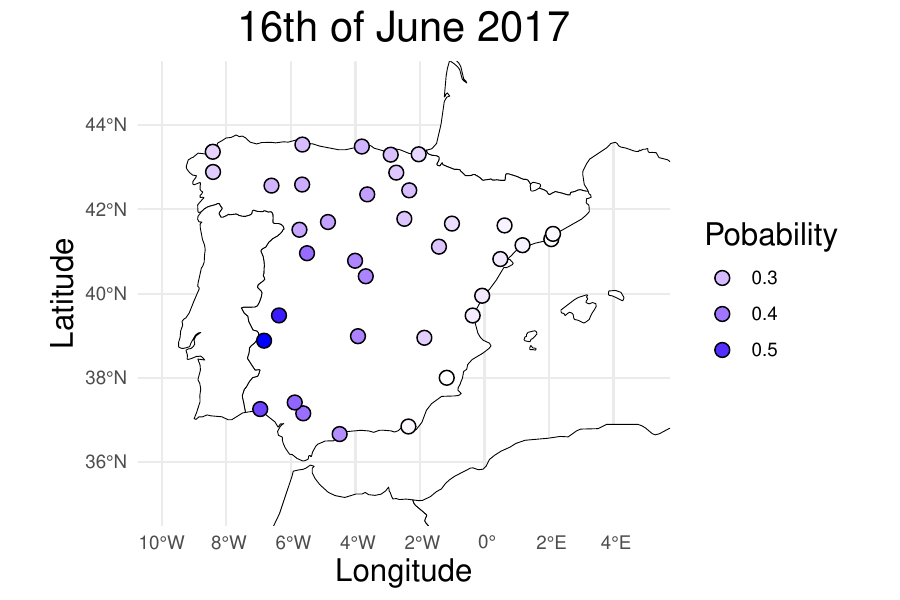}
    \includegraphics[width=0.45\textwidth]{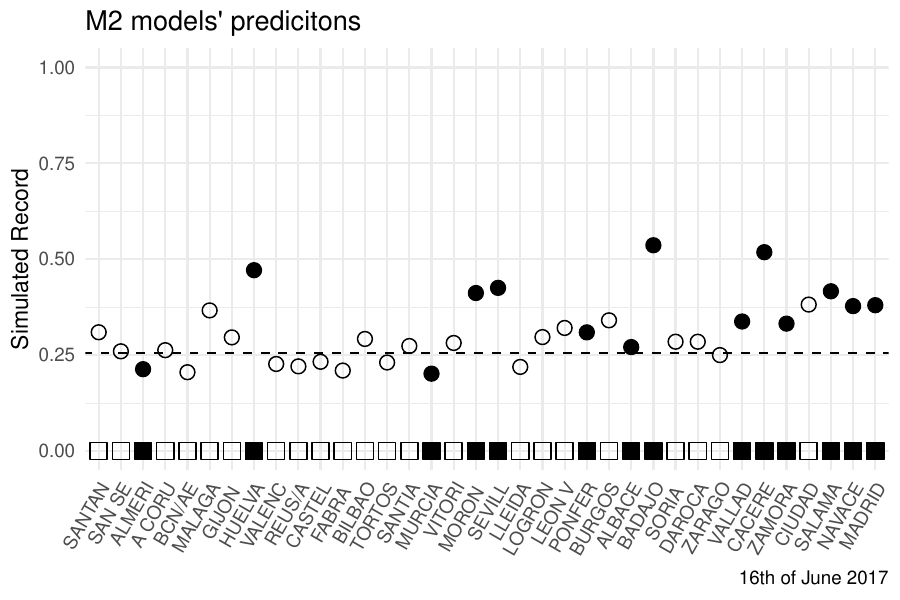}
    \caption{Top left: observed records (black) and non-records (white) for the 17th of June 2017. Top right: M2 predict probabilities for that day. Bottom: scatterplot with the M2 predict for each station for that day. Dashed line represents the threshold at the true negative rate of 0.95 form the M2 ROC curve.}
    \label{fig:enter-label}
\end{figure}

\newpage
\subsection*{State of the atmosphere}

\begin{figure}[htb]
    \centering
    \includegraphics[width=0.49\textwidth]{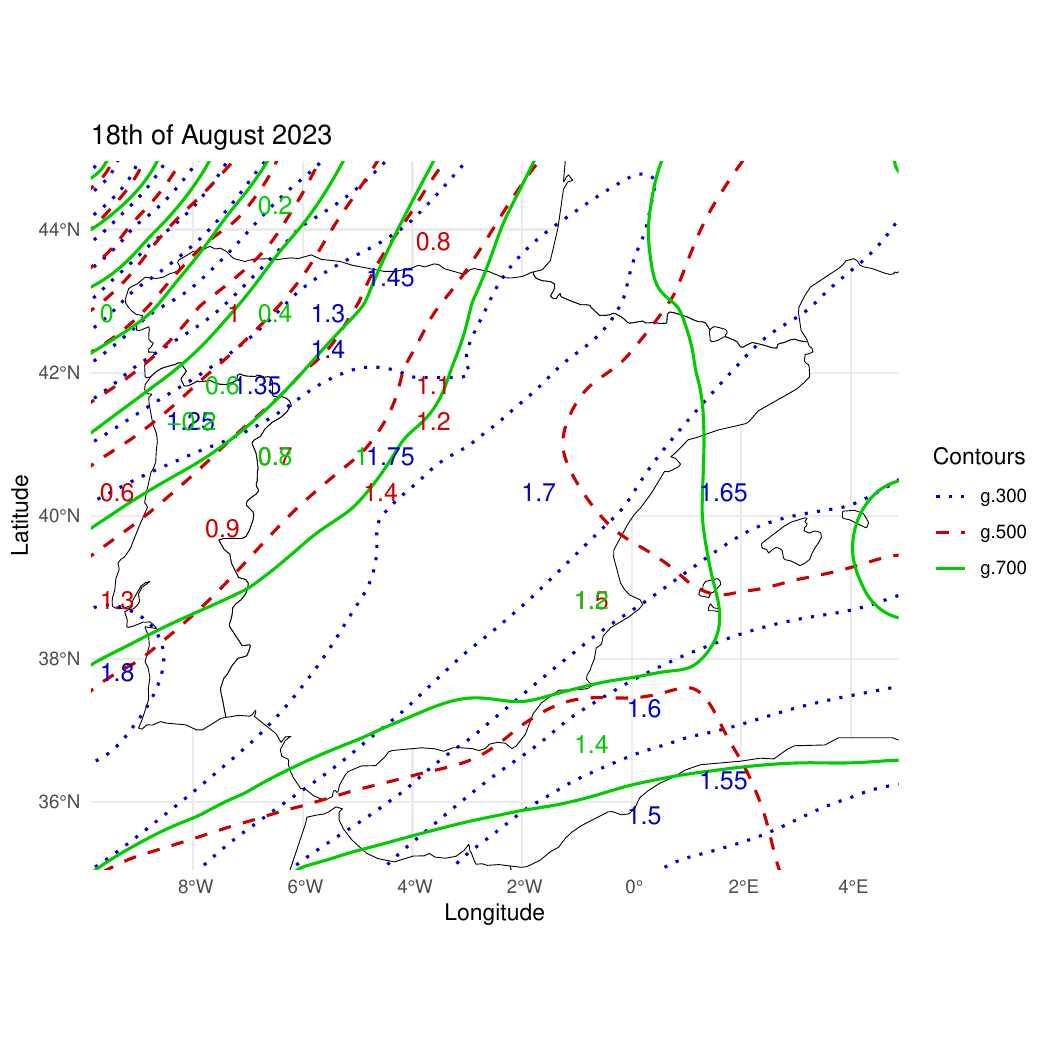}
    \includegraphics[width=0.49\textwidth]{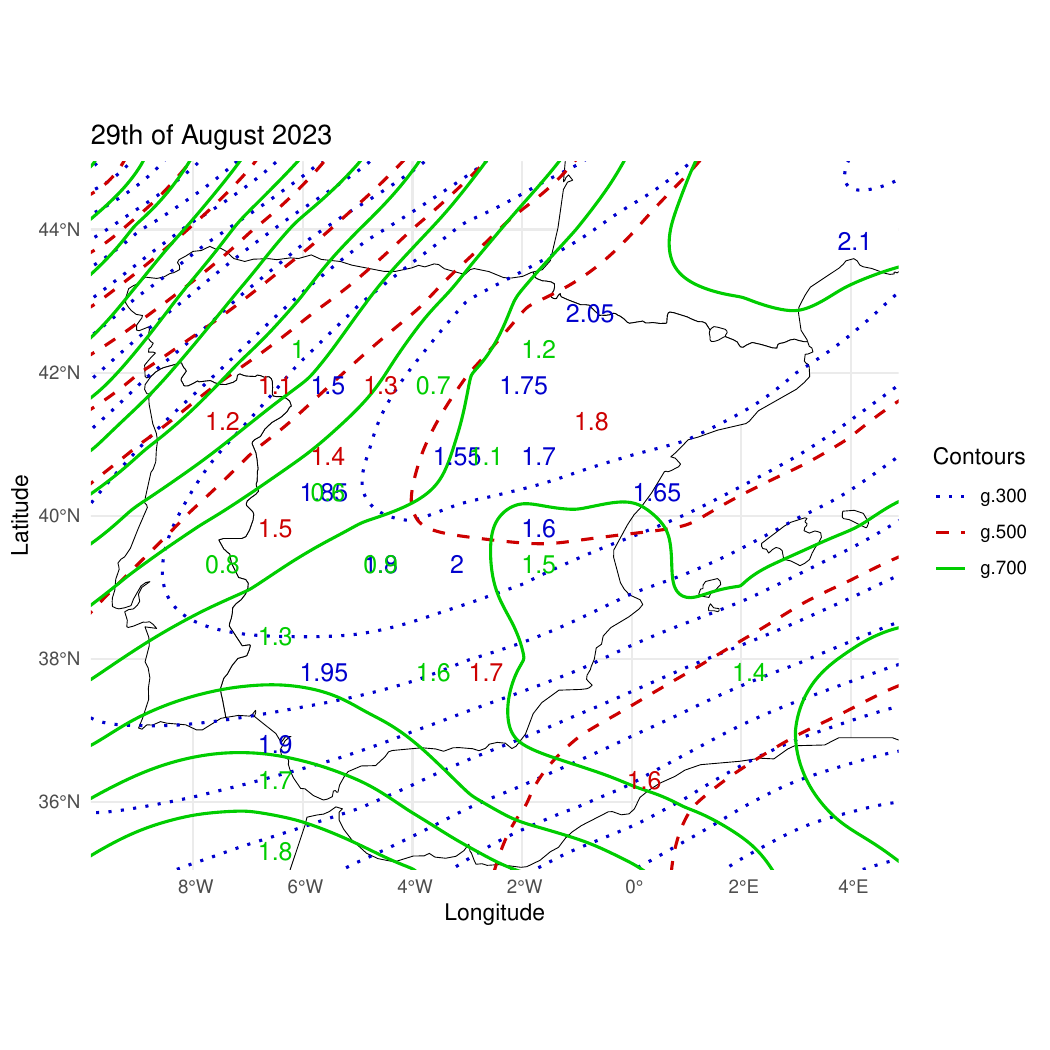}
    \includegraphics[width=0.49\textwidth]{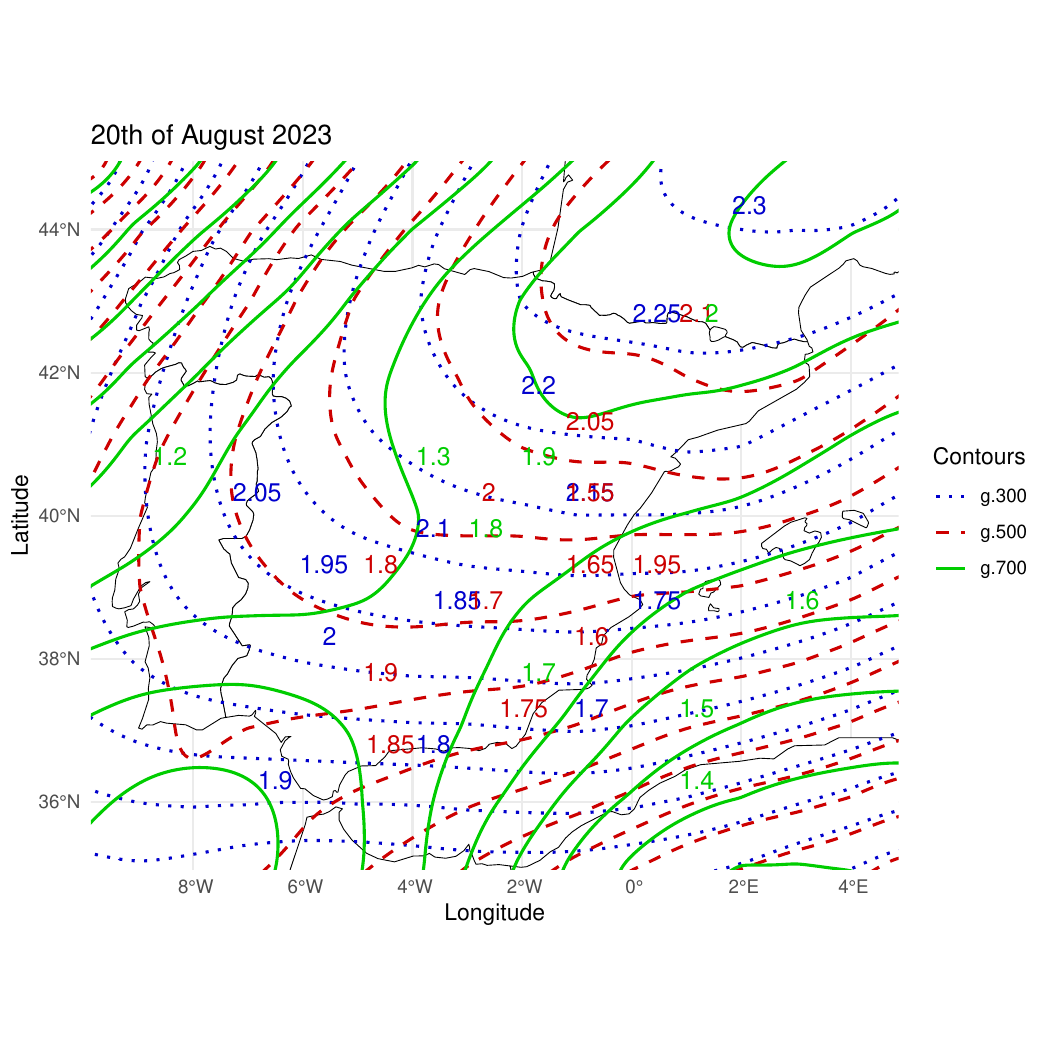}
    \includegraphics[width=0.49\textwidth]{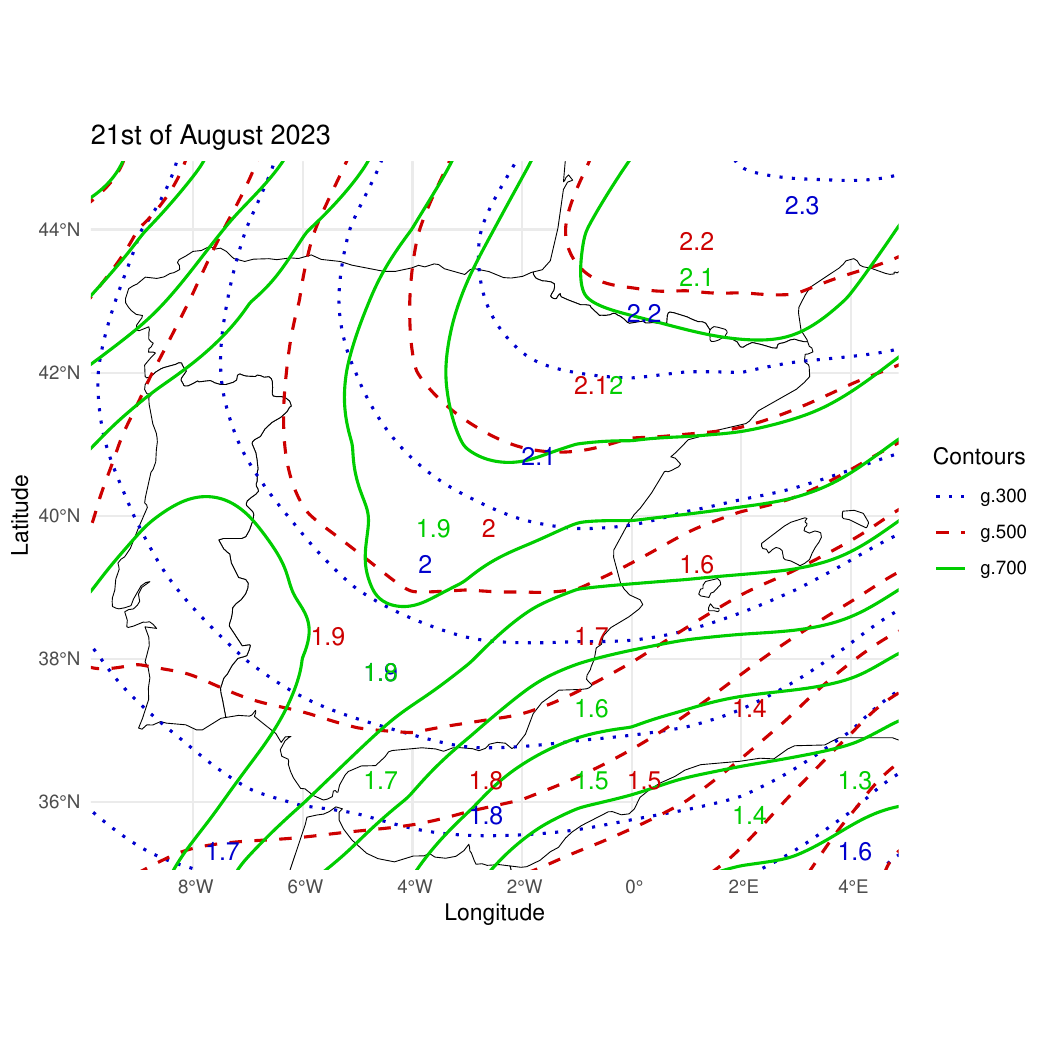}
    \includegraphics[width=0.49\textwidth]{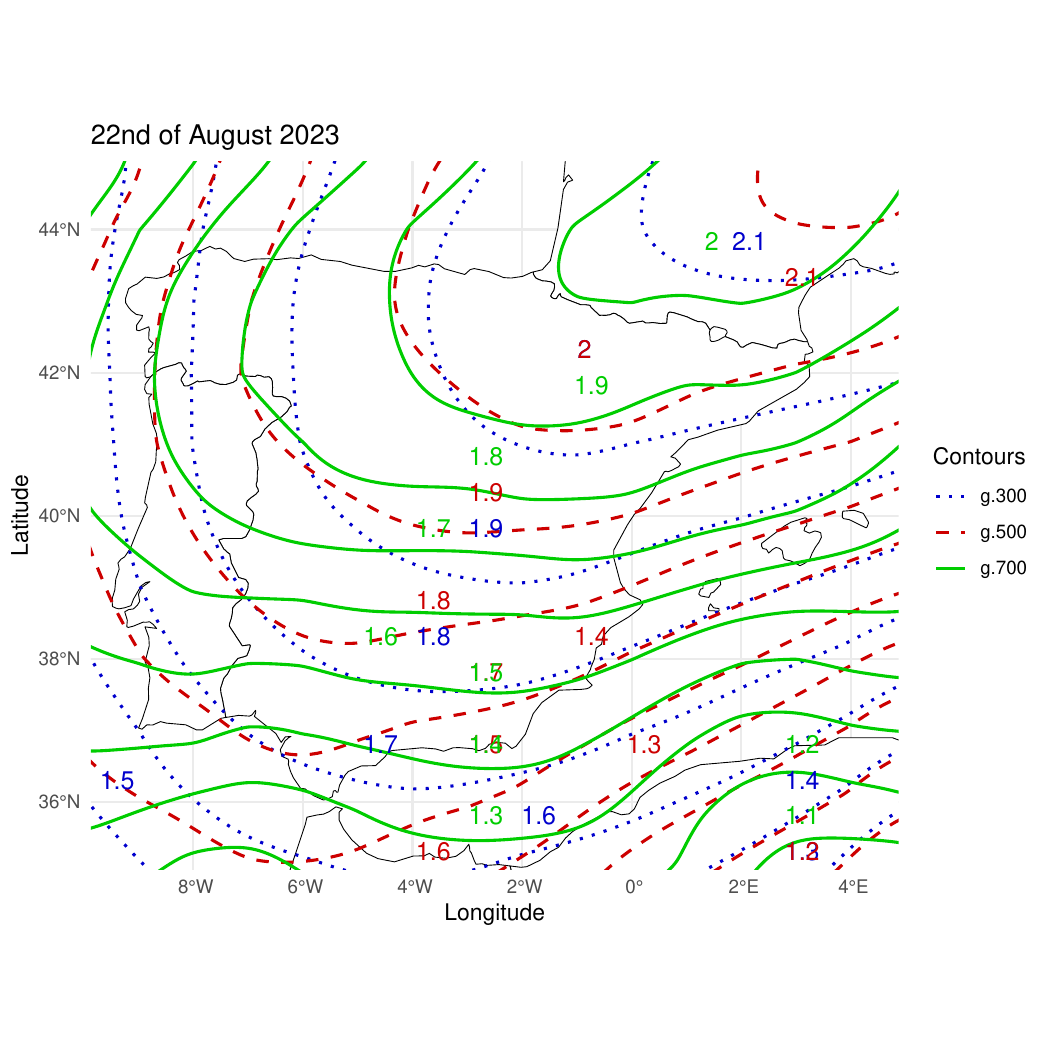}
    \includegraphics[width=0.49\textwidth]{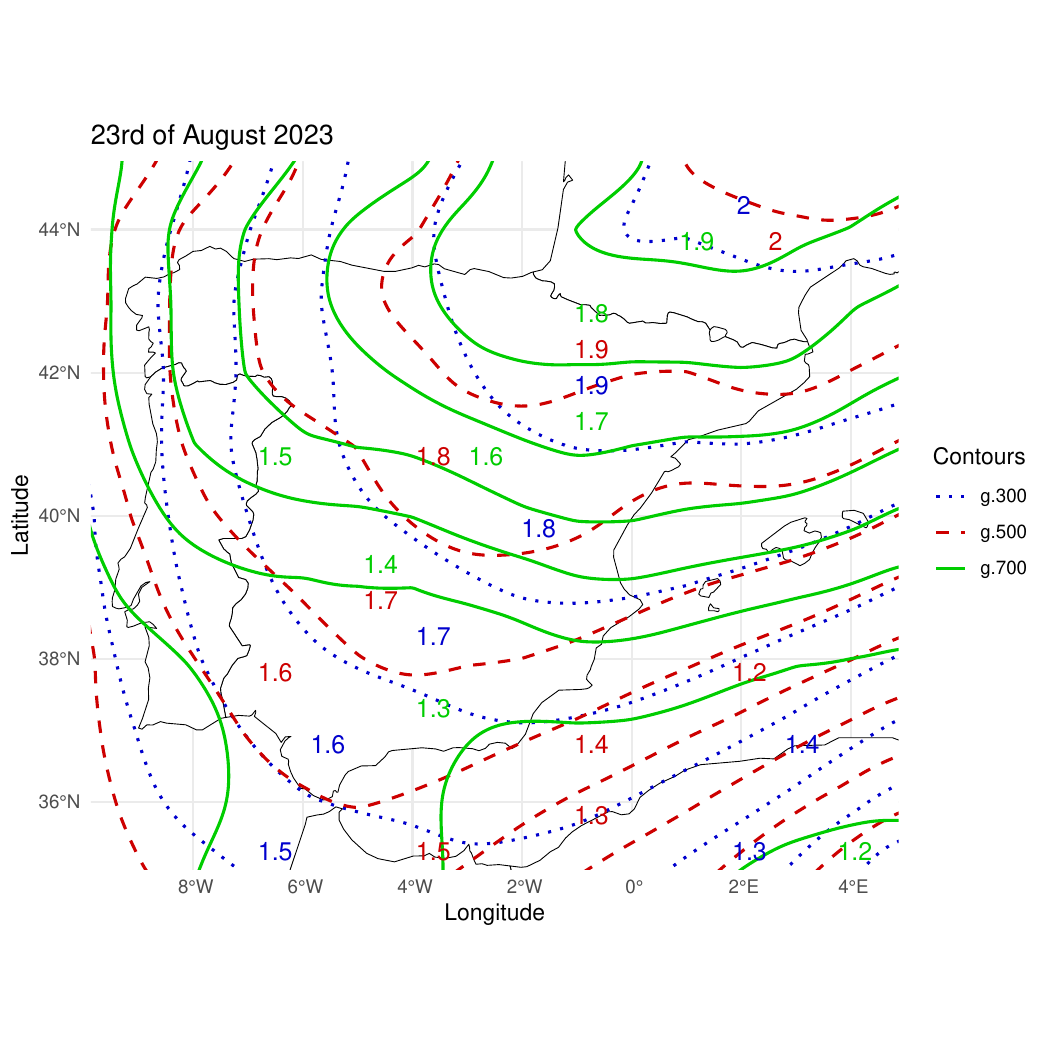}
    \label{fig:GEO-MAPS}
\end{figure}

\begin{figure}[htb]
    \centering
    \includegraphics[width=0.49\textwidth]{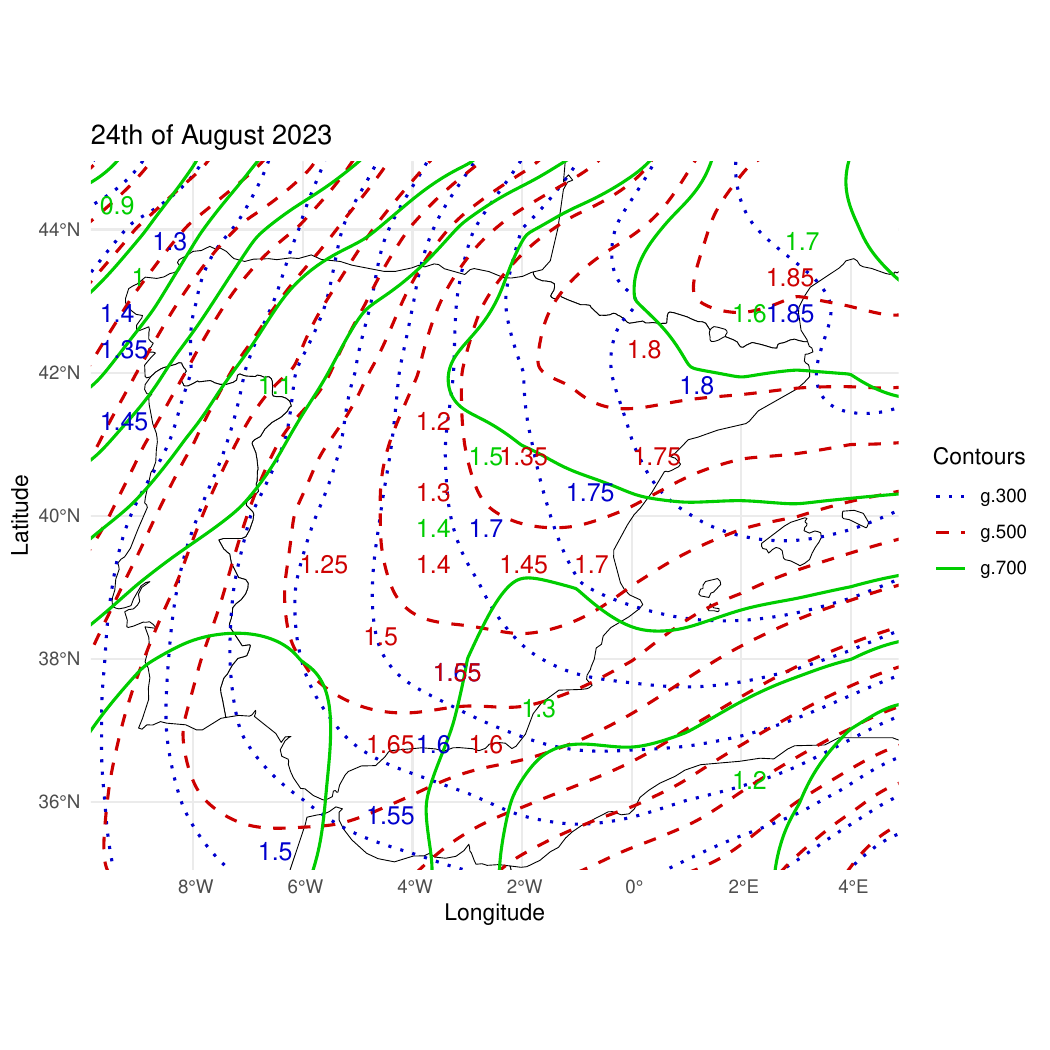}
    \includegraphics[width=0.49\textwidth]{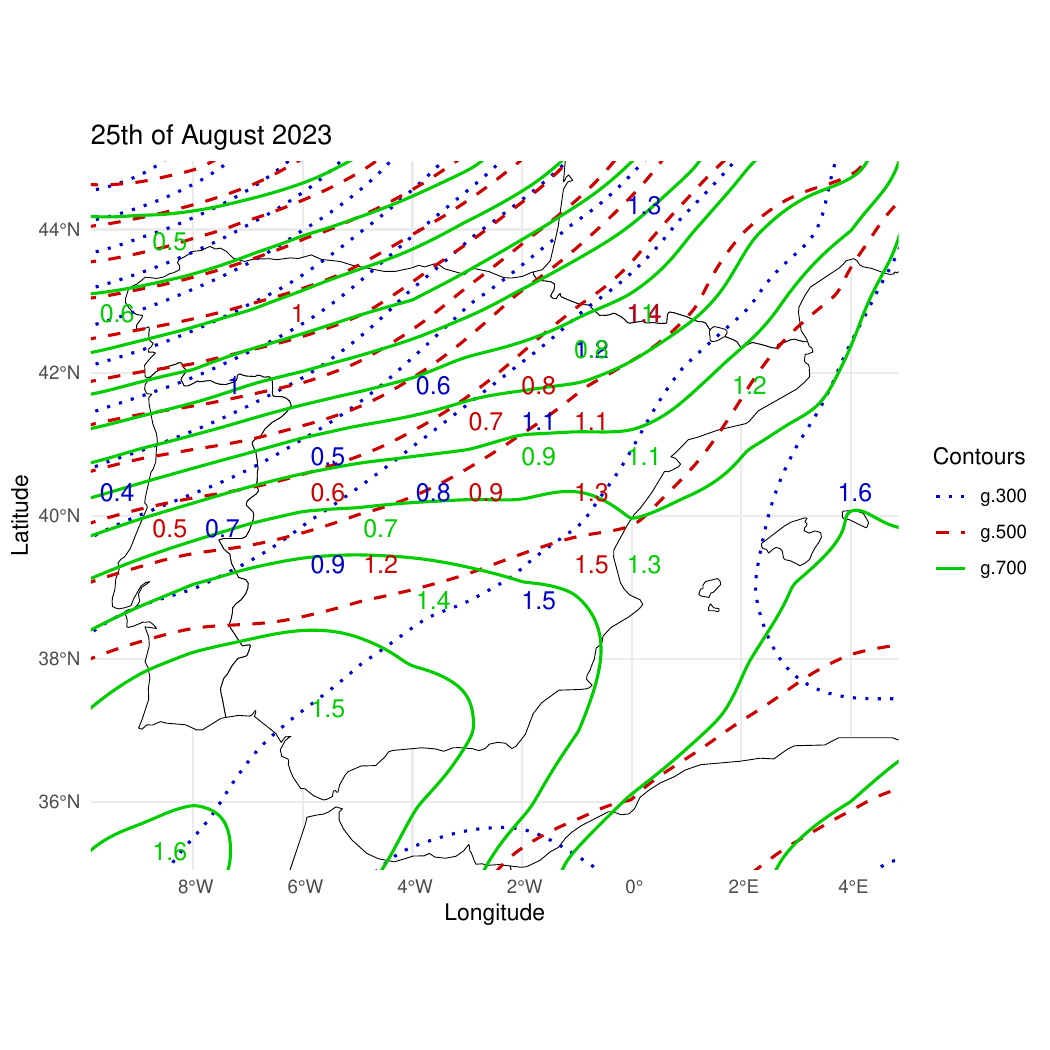}
    \includegraphics[width=0.49\textwidth]{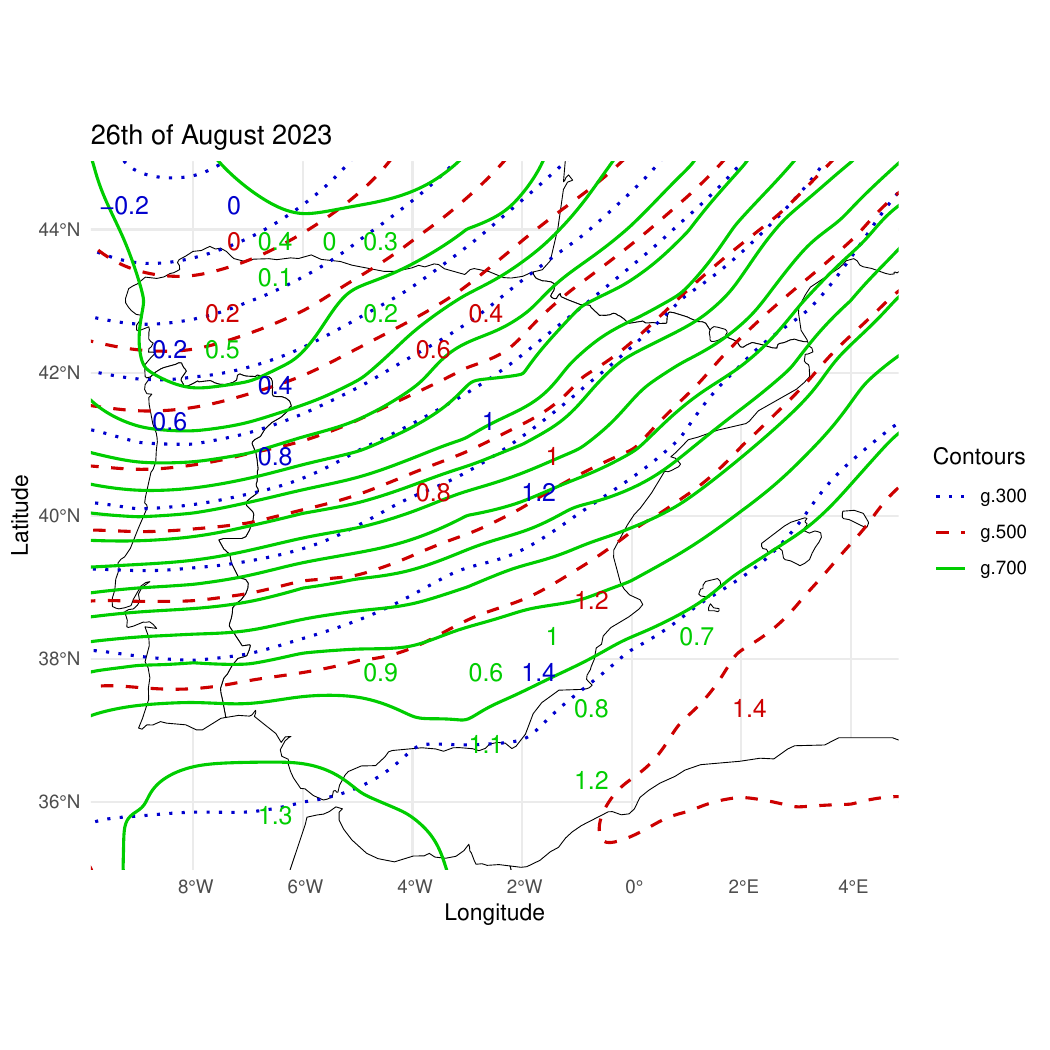}
    \includegraphics[width=0.49\textwidth]{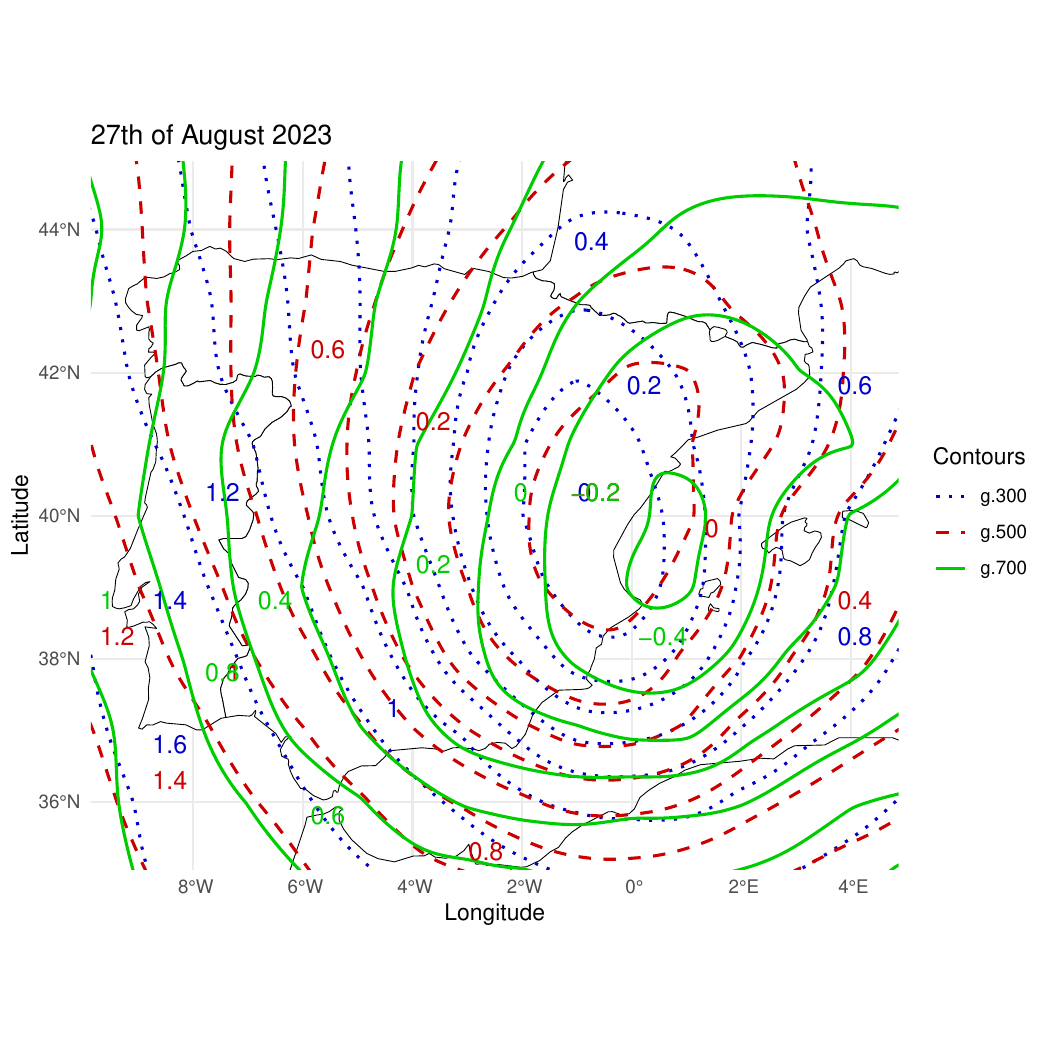}       
    \caption{Contour maps of standardized geopotentials, calculated relative to the 1980–2010 reference period. Each panel corresponds to a different day between 18 and 27 August 2023, read left to right and top to bottom. Contour lines indicate different geopotential levels: 300 hPa (blue dotted), 500 hPa (red dashed), and 700 hPa (green solid).}
    \label{fig:GEO-MAPS}
\end{figure}

\end{document}